\documentclass{aa}  
\usepackage{graphicx}
\usepackage{txfonts}
\usepackage{hyperref}
\usepackage[utf8]{inputenc}
\usepackage{subcaption}
\usepackage{placeins}
\usepackage{amsmath}
\usepackage{caption}

\newcommand*\diff{\mathop{}\!\mathrm{d}}

\begin{document}

    \title{Evidence for a large off-centered galactic outflow and its connection to the extraplanar diffuse ionized gas in IC\,1553}

   \author{L. Dirks
          \inst{1}
          \and
          R.-J. Dettmar
          \inst{1,2,3}
          \and
          D. J. Bomans
          \inst{1,2,3}
          \and
          P. Kamphuis
          \inst{1}
          \and
          U. Schilling
          \inst{1}
          }

   \institute{Ruhr  University  Bochum,  Faculty  of  Physics  and  Astronomy,  Astronomical  Institute  (AIRUB),  Universitätsstrasse  150,  44801 Bochum, Germany
   \and
   Ruhr University Bochum, Research Department "Plasmas with Complex Interaction"
   \and
    Ruhr Astroparticle and Plasma Physics Center (RAPP Center)
    }

   \date{Received December 13, 2022; Accepted August 1, 2023}

 
  \abstract
   {}
   {We analyze a MUSE optical integral field spectrum of the star-forming edge-on galaxy IC\,1553 in order to study its extraplanar diffuse ionized gas (eDIG) and the processes shaping its disk-halo interface.}
   {We extracted the optical emission line properties from the integral field spectrum and generated the commonly used emission line diagnostic diagrams in order to analyze the ionization conditions and the distribution of the eDIG. Furthermore, we performed gravitational potential fitting to investigate the kinematics of a suspected galactic outflow.}
   {We find that the eDIG scale height has a maximum value of approximately 1.0\,kpc and decreases roughly linearly with the radial distance from the galactic center in projection. The ionization state of the eDIG is not consistent with a pure photoionization scenario and instead requires a significant contribution from shock ionization. This, in addition to the gas kinematics, strongly suggests the presence of a galactic scale outflow, the origin of which lies at least 1.4\,kpc away from the galactic center. The inferred shock velocity in the eDIG of approximately 225\,km\,s$^{-1}$ is comparable to the escape velocity estimated from our potential modelling. 
   The asymmetric distribution of currently star-forming clusters produces a range of different ionization conditions in the eDIG. As a result, the vertical emission line profiles vary quantitatively and qualitatively along the major axis of the galaxy. This analysis illustrates that it is crucial in studies of the eDIG to use observations that take the spatial and kinematical distributions into account, such as those done with integral field units, to form an accurate picture of the relevant physical properties.
   }
   {}

\titlerunning {The extraplanar diffuse ionized gas of IC\,1553 }
   \maketitle
%

\section{Introduction}

While models accounting for the star-formation-driven dynamical interstellar medium (ISM) successfully describe the overall picture of the various ISM components observed in nearby galaxies, details of the processes governing the distribution and physical conditions of the different phases remain an open issue. The warm ionized medium (WIM) or diffuse ionized gas (DIG) observed in the halos of star-forming disk galaxies (for a review see \cite{haffner09} and \cite{putman12}) poses questions regarding its origin, ionization, and heating. The existence of an extraplanar layer of low-density ionized gas was first proposed by \cite{hoyle63} to explain an absorption feature in the Galactic diffuse radio continuum background. The Galactic DIG has since been detected through the dispersion of pulsar signals \citep[][]{guelin74, reynolds91}, in emission of strong optical spectral lines in the direction of high-velocity \ion{H}{i} clouds \citep[][]{weiner96, tufte98}, and from absorption lines of intermediate ions in spectra of background sources \citep[][]{collins09}. The DIG in external galaxies was originally studied using H$\alpha$ imaging \citep{dettmar90, rand90} or long-slit spectroscopy \citep{dettmar92two}, and more recently using integral field spectroscopy \citep{ho16, jones17, lopez-coba17, levy19, rautio22}.

The DIG in the Milky Way occupies the temperature range between 6000\,K and 10000\,K and densities between 0.03\,cm$^{-3}$ and 0.08\,cm$^{-3}$ with a filling factor around $f \approx$ 0.4 -- 0.2 in the volume 2 -- 3\,kpc above the galactic midplane \citep{reynolds91two, hill08, haffner09}. For the external galaxy NGC\,891 \cite{rand90} arrived at a similar density estimate. More commonly, however, the DIG is characterized in terms of the behavior of its optical emission lines rather than its thermodynamical conditions. Generally, the DIG displays elevated levels of the line ratios $[\ion{S}{ii}]\lambda 6717$/H$\alpha$ and $[\ion{N}{ii}]\lambda 6583$/H$\alpha$ and lower $[\ion{O}{iii}]\lambda 5007$/H$\alpha$ values than classical planar \ion{H}{ii} regions \citep[e.g.,][]{mathis86,reynolds99}. Additionally, $[\ion{S}{ii}]\lambda 6717$/H$\alpha$ and $[\ion{N}{ii}]\lambda 6583$/H$\alpha$ tend to rise with increasing altitude above the midplane \citep[e.g., ][]{golla96, levy19}. These varying line ratios are contrasted by a nearly constant $[\ion{S}{ii}]\lambda 6717$/$[\ion{N}{ii}]\lambda 6583$ ratio \citep[e.g., ][]{haffner99, reynolds99}. While the abnormal ratios of forbidden lines to H$\alpha$ relative to \ion{H}{ii} regions can be explained by photoionization at low ionization parameters, $q$ \citep{domgoergen94}, the constancy of $[\ion{S}{ii}]\lambda 6717$/$[\ion{N}{ii}]\lambda 6583$ requires an additional source of heating \citep{reynolds92, rand98, reynolds99, collins01}. \cite{wood04} successfully modeled this additional heating source with the spectral hardening of the ionizing photon field from O- and B-type stars as it propagates from the midplane into the halo. These general properties of the optical DIG emission lines are commonly observed, but not ubiquitous. \cite{golla96}, for example, observed a rising trend for $[\ion{S}{ii}]\lambda 6717$/$[\ion{N}{ii}]\lambda 6583$ with increasing altitude in NGC\,4631 rather than the usual constancy, but at a lower rate than predicted by the \cite{domgoergen94} models. We note, however, that the interaction between  NGC\,4631, NGC\,4656, and NGC\,4627 may introduce complications that are not considered in general models of the DIG.

The necessary energy budget for the production and maintenance of the DIG implies that the radiation field from young massive stars in the disk is the primary ionizing source \citep{reynolds84, mathis86, domgoergen94, haffner09}. Additional sources that have been proposed to contribute at a significant level include, among others, shocks \citep{martin97}, radiation from hot-gas-cool-gas interfaces \citep{haffner09}, and hot low-mass evolved stars \citep[HOLMES,][]{sokolowski91, flores11}.

The extraplanar H$\alpha$ emission in DIG-halos is found to correlate with tracers of star formation in the disk \citep{dettmar92one, Rossa2003a, Rossa2003b, jones17, rautio22} on local and galactic scales. This coupling to the star formation indicates that the ionized gas originates in the disk and is transported into the lower halo by feedback processes. The necessary gas transport process in this paradigm connects to the study of galactic winds, which are defined as gas flows with a velocity component that exceeds the escape velocity of the galactic potential, and galactic outflows, which remain bound to their host galaxy. Many questions regarding the detailed workings of galactic outflows and winds remain unsolved (see reviews by \cite{Veilleux2005}, \cite{Rupke2018}, and \cite{Zhang2018}), but it is clear that they are important mechanisms regulating the evolution of galaxies, the circumgalactic medium, and the intergalactic medium \citep[e.g.,][]{Heckman1990}, and consequently also the disk-halo interface. In the context of this paper, the questions about the ionization mechanism(s), the structure of the base of the wind, and the kinematics of the flow are especially noteworthy. The ionization of the gas in diffuse ionized halos and in galactic winds is a longstanding, not fully resolved question, as discussed above; photoionization from OB stars is clearly an important contributor \citep[e.g.,][]{Heckman1990}, but shocks can also play a role \citep[e.g.,][]{Rich2010}, as well as ionization by photons from the hot gas inside the outflow or wind \citep[e.g.,][]{Sarkar2022}. Even non-equilibrium effects can play a role \citep{Gray2019a, Gray2019b, Sarkar2022}. To understand the structure of the outflow or wind and its energetics, knowing the main driving mechanism is critical; current models include contributions from overpressure due to the hot gas and its mechanical feedback on the surrounding gas, radiation pressure, and the interaction of cosmic rays with the magnetic field frozen within the ionized gas \citep[e.g.,][]{Fielding2022}. This short introduction to outflows and winds is focused on the warm, ionized outflow component that we discuss in this paper but it is important to keep in mind that this is only one part of a multiphase phenomenon, with hot gas inside the core of the outflow, cool gas at the compressed shell, clouds inside the flow, and cosmic rays interacting with the magnetic field frozen into the ionized gas.


\section{Data}
We used archival Integral Field Unit (IFU) data observed with the Multi Unit Spectroscopic Explorer (MUSE, \citeauthor{bacon10} \citeyear{bacon10}) originally intended by \citet{comeron19} to investigate the stellar kinematics in the thick disks of star-forming edge-on galaxies. \cite{rautio22} used this same data set to analyze the DIG of the majority of galaxies in this sample. In this work we picked IC\,1553 from this sample to study the DIG of a single galaxy in more detail.

The MUSE spectrograph has a channel width of $1.25$\,{\AA} spanning the spectral range of 4750 -- $9351$\,{\AA}, and spaxels of size $0.2\arcsec \times 0.2\arcsec$ over a total $1\arcmin \times 1\arcmin$ field of view. The data cube was pre-reduced with the MUSE pipeline \citep{weilbacher12} and \texttt{ZAP} \citep{soto16}. For a detailed description of the reduction process see \citet{comeron19}.

We used the fitting code 3-Dimensional Emission Line Analysis \citep[\texttt{3DELA},][]{schilling} to extract the emission line properties. \texttt{3DELA} is a collection of Python scripts designed for the fast evaluation of IFU data cubes. The treatment of spectra with a small signal-to-noise ratio has been of special importance in detecting features of the gas dynamics far away from the galaxy disk, such as the presence of outflows. The spectral line fitting requires an initial estimate of the systemic velocity, preferably obtained by measuring the wavelength of the shifted H$\alpha$ line in the spectral cube in a region near the suspected center of mass. The spectral range for the fit of the line under consideration is known therefrom.

Primarily the code has been developed for the study of the galaxies of the Calar Alto Legacy Integral Field Area Survey \citep[CALIFA,][]{sanchez12, garciabenito15}. Due to the survey's low spectral resolution of 6 Å in the V500 setup, the spectral lines were only defined by a few data points. For this reason \texttt{3DELA} fits lines with a single Gaussian component. The current version of \texttt{3DELA} allows the fitting of emission lines from data cubes from other instruments -- like MUSE -- and surveys (e.g., Sloan Digital Sky Survey \citep[SDSS,][]{jonsson20} or Mapping Nearby Galaxies at APO \citep[MaNGA,][]{bundy15}) by transposing them to an internal standard format conforming to FITS. This procedure requires some value-substitution strategies due to missing values in the FITS headers. Typical tasks of preparation or post-processing (e.g., estimation of the systemic velocity, dereddening, the plotting of specific results or the calculation of diagnostic diagrams) are performed by discrete scripts using the internal standard format of the data cubes.

\begin{figure}[t]
    \includegraphics[width=0.5\textwidth]{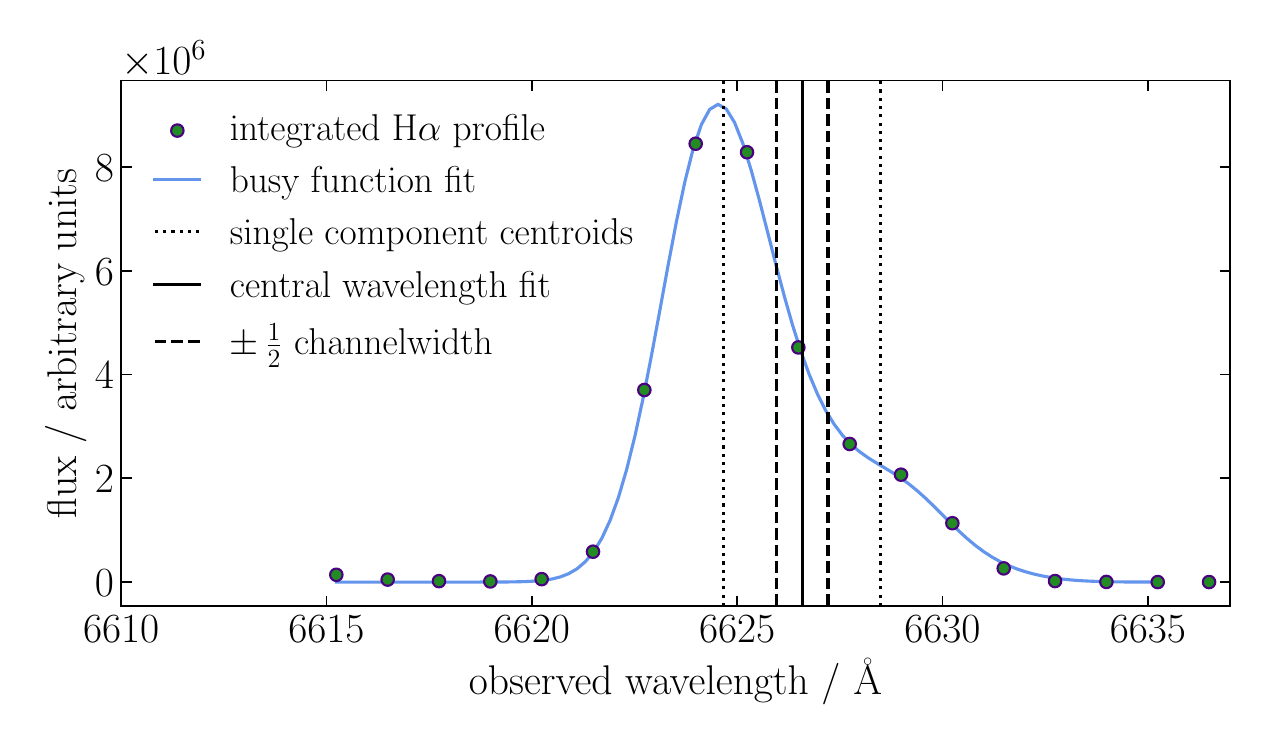}
        \caption{Visualization of the busy function fit to determine the systemic velocity of IC\,1553. The integrated H$\alpha$ profile was obtained by collapsing the data cube along both spatial axes for bins with altitude $|z|\leq 10\arcsec$. The central wavelength of the profile is taken as the mean of the centroid wavelengths of the red- and blue-shifted components in the busy function fit. We estimate the uncertainty in this determination as \mbox{$\pm 0.5 \cdot \textrm{channel width}$}.}
        \label{fig:busyfct}
\end{figure}

\begin{figure*}[t]
    \includegraphics[width=\textwidth]{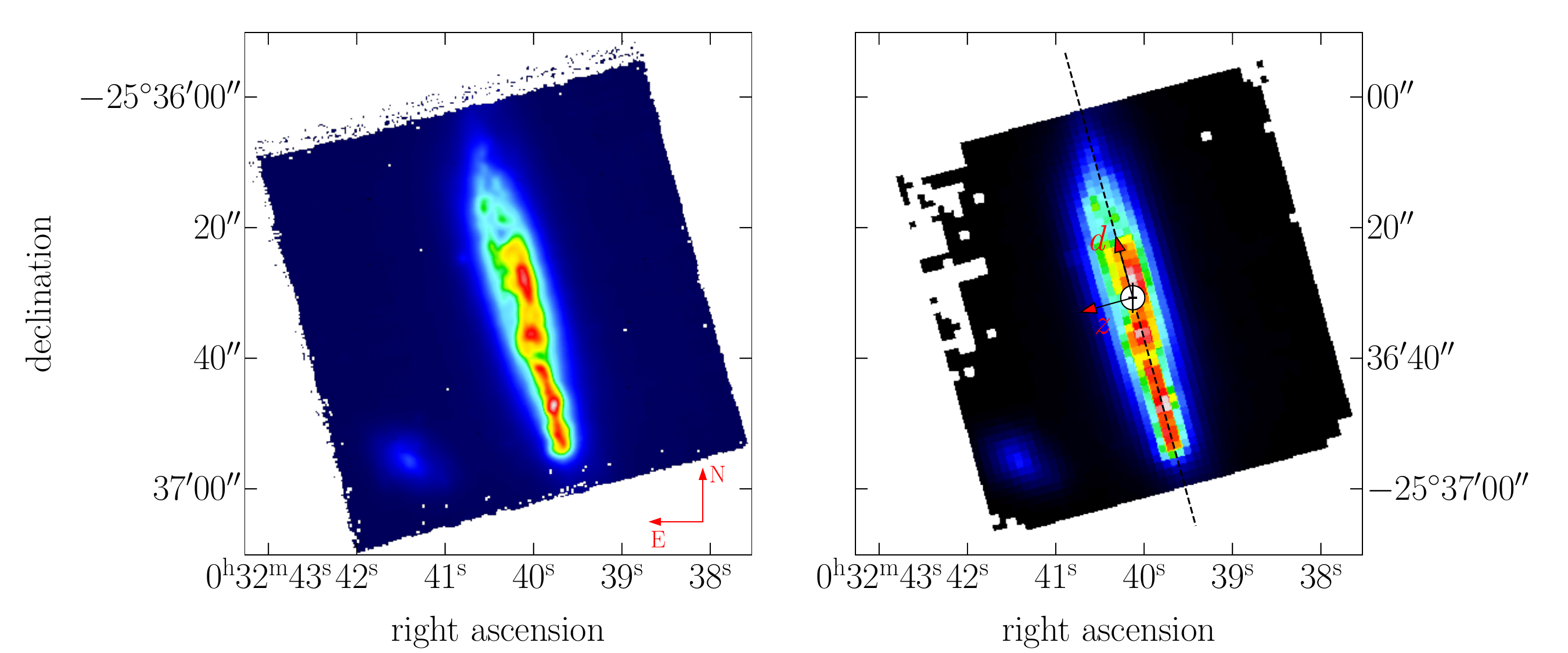}
        \caption{Synthetic $V$-band images constructed from the data cube at the original resolution \textit{(left)}, and after rebinning with a regular $50 \times 50$ grid of resolution elements \textit{(right)}. The dashed line and the white circle are the estimates for the plane of the disk and the dynamic center of IC\,1553, respectively.}
        \label{fig:continuumimage}
\end{figure*}

The data cube was rebinned to increase the signal-to-noise ratio with a regular grid of $50 \times 50$ resolution elements over the field of view (i.e., $6\times 6$ spaxels per resolution element), resulting in an effective spatial resolution of $1.2\arcsec$. We chose a regular grid over other commonly used practices like the Voronoi tessellations \citep[e.g.,][]{cappellari03} to simplify the decomposition of a bin's position into its projected radial distance from the center of IC\,1553 and its height above the disk, and to be able to sample this coordinate space in a consistent pattern. In order to have the spaxels aligned with the major and minor axes of the galaxy, the orientation of the grid was determined in the following manner. We extracted a synthetic \textit{V}-band image from the data cube and separated it into ten bins along a rough estimate of the major axis. The midplane in each bin is then taken to be the position that ensures the $V$-band flux is equally distributed above and below the plane of the disk. Finally, these positions were fit with a linear function to determine the plane of the disk. The nominal systemic velocity of IC\,1533 was ascertained by fitting the H$\alpha$ profile in the spatially integrated spectrum of the disk of IC\,1553 (masking all bins of altitude $|z| > 10\arcsec$) with the busy function \citep{busyfct14, westmeier14}. This fit yields a centroid wavelength for each of the approaching and receding halves of the galaxy (see Fig. \ref{fig:busyfct}), which can then be averaged to calculate the systemic velocity. In what follows we take the center of IC\,1553 to be the position along the midplane of the disk, where the H$\alpha$ emission line is Doppler-shifted with this systemic velocity. We assume an uncertainty in the determination of the systemic velocity equivalent to half the width of the spectral channels, in other words $\pm 0.625\,${\AA}, and obtain a systemic velocity value of $v_\mathrm{sys}=(2914\pm 29)\,\textrm{km}\,\textrm{s}^{-1}$. The results of the rebinning and the fits to the disk plane and the center of the galaxy are shown for the synthetic $V$-band image in Fig. \ref{fig:continuumimage}. Positive values for the projected radial distance, $d$, signify a position north of the center along the dashed line in Fig. \ref{fig:continuumimage}. Positive values for the height above the midplane, $z$, signify a position east of the disk. We note that the emitting extended object in the southeastern corner of the field of view in Fig. \ref{fig:continuumimage} is a background galaxy and not interacting with IC\,1553. This background galaxy contaminates the H$\alpha$ line of IC\,1553 (see Appendix \ref{appendix}); we therefore masked this area in all diagnostic maps using H$\alpha$.

In some quantitative considerations in the following sections we employ data quality cuts on the H$\alpha$ flux. In these considerations the detection limit is given as a multiple of the background level, $\sigma \approx 5.13\times 10^{-8}\,{\textrm{ergs}\,\textrm{s}^{-1}\,\textrm{cm}^{-2}\,\textrm{sr}^{-1} \equiv 0.21\,\textrm{Rayleighs}}$ (R, $1\,\textrm{R} = 10^6 / (4\pi)\,\textrm{photons}\,\textrm{s}^{-1}\,\textrm{cm}^{-2}\,\textrm{sr}^{-1}$), which we estimate with the standard deviation of non-zero H$\alpha$ flux values in bins far away from the disk of IC\,1553 ($25\arcsec \leq |z| \leq 30\arcsec$).

Within this framework we have produced maps for the emission line fluxes commonly used to characterize the warm ionized medium -- including H$\beta$, [\ion{O}{iii}]$\lambda 4959,5007$, [\ion{O}{i}]$\lambda6300$, H$\alpha$, [\ion{N}{ii}]$\lambda 6548,6583$, [\ion{S}{ii}]$\lambda 6716,6731,$ and [\ion{S}{iii}]$\lambda 9069$ -- as well as their observed velocity offsets from the rest-frame system and their full width at half maximum (FWHM).

The emission line fluxes were corrected for dust attenuation using the same methods as \citet{dominguez13}. Assuming typical gas conditions in star-forming regions for the electron temperature, $T_{\mathrm{e}} = 10^4\,\textrm{K,}$ and density, $n_e = 100\,\textrm{cm}^{-3}$, yields a largely constant Balmer decrement (H$\alpha$/H$\beta$)$_{int} = 2.86$ for Case B recombination \citep{osterbrock06}. The extinction in magnitudes at wavelength $\lambda$ is then given from the measured Balmer decrement as \citep{dominguez13}:
\begin{equation}\label{eq:dustcorrect}
    A_{\lambda} = k(\lambda) \times 1.97 \log_{10} \left[ \frac{(\mathrm{H}\alpha /\mathrm{H}\beta)}{2.86} \right],
\end{equation}
where $k(\lambda)$ can be sampled from the reddening curve of \cite{calzetti00}. The observed H$\alpha$ and H$\beta$ fluxes contain not only the emission from the ionized gas, but also some component from the emission of the stellar population along the line of sight, as do all recombination lines. Typically, the stellar population produces a strong absorption line feature in H$\beta$ and a weaker absorption line feature in H$\alpha$. This alters the measured Balmer decrement, and constitutes a form of contamination in the context of Eq. (\ref{eq:dustcorrect}). We corrected the observed Balmer decrement for this effect by separating the emission components from the stellar population and the ionized gas using the Penalized PiXel-Fitting code \citep[\texttt{pPXF},][]{cappellari04, cappellari17} with the E-MILES stellar population synthesis models \citep{vazdekis16}. The left panel of Fig. \ref{fig:ha_map} shows a map of the derived gas-phase Balmer decrements.

The FWHM was corrected for widening from the line spread function (LSF) via
\begin{equation}\label{eq:lsf_correct}
    {\rm FWHM} = \sqrt{{\rm FWHM}_{\mathrm{observed}}^2 - {\rm LSF}^2(\lambda)}.
\end{equation}
The parametrized LSF given by \citet{bacon17} for the udf10 field (their Eq. (8)) was used in this correction.

We assume the same distance to IC\,1553 as \citet{comeron19}, namely $36.5\,{\rm Mpc}$ \citep{tully16}. Some of the general properties of IC\,1553 are summarized in Table \ref{tab:general_props}.

\renewcommand{\arraystretch}{1.5}
\begin{table}
    \caption{Overview of some general properties of IC\,1553. The maximum rotational velocity and metallicity give ranges of singular measurements and are not meant as statistically significant intervals.}
    \label{tab:general_props}
    {\centering
    \begin{tabular}{r l}
    \hline \hline 
         stellar mass$^{1}$: &$13.40 \times 10^9\,\textrm{M}_{\odot}$\\
         gas disk mass$^{2}$: &$2.96 \times 10^9\,\textrm{M}_{\odot}$\\
         distance$^{3}$: &36.5\,Mpc\\
         absolute $g$-band brightness$^{4}$: &$-18.90\,\textrm{mag}$\\
         maximum rotational velocity$^{5}$: &$128\,\textrm{km\,s}^{-1}$ -- $142\,\textrm{km\,s}^{-1}$\\
         metallicity$^{6}$ (12+$\log (O/H)$): &$8.47$ -- $8.48$\,dex\\
         \hline
    \end{tabular}
    }
    \phantom{test}\\
    {$^{1}$}{Sum of bulge-, thin, and thick disk masses reported by \cite{comeron18}}\\
    {$^{2}$}{\citep{comeron18}}\\
    {$^{3}$}{\citep{tully16}}\\
    {$^{4}$}{From the Dark Energy Survey \citep{DES05, abbott18, abbott21}, assuming a distance of $36.5\,\textrm{Mpc}$}\\
    {$^{5}$}{$128\,\textrm{km\,s}^{-1}$ from \ion{H}{i} data \citep{courtois09} and $142\,\textrm{km\,s}^{-1}$ from optical nebular emission lines \citep{comeron19}}. Our fit to the optical emission lines yields a value of $134\,\textrm{km\,s}^{-1}$ (see Appendix \ref{appendixc}).\\
    {$^{6}$}{Estimated from two \ion{H}{ii} regions in the northern disk (marked with red circles in the second panel of Fig. \ref{fig:ha_map}) with the $S_{23}$ diagnostic (Eq. \ref{eq:S_23}), as outlined by \cite{kewley02}}.
\end{table}

\section{Results}

\subsection{H$\alpha$ distribution}\label{sect:Ha_dist}

\begin{figure*}
        \begin{subfigure}{.33\linewidth}
        \centering
        \includegraphics[width=\hsize]{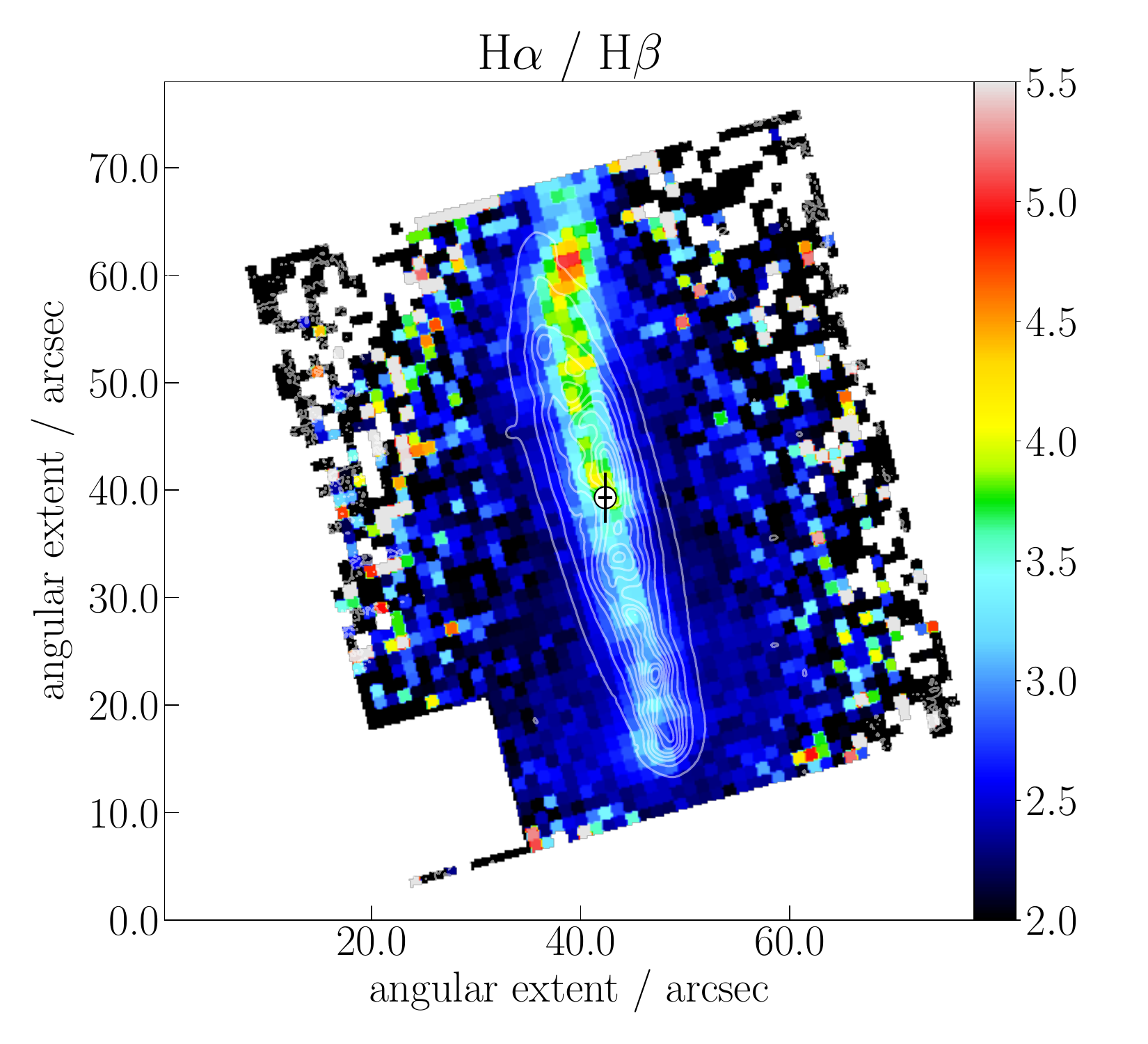}
    \end{subfigure}
        \centering
        \begin{subfigure}{.33\linewidth}
        \centering
        \includegraphics[width=\hsize]{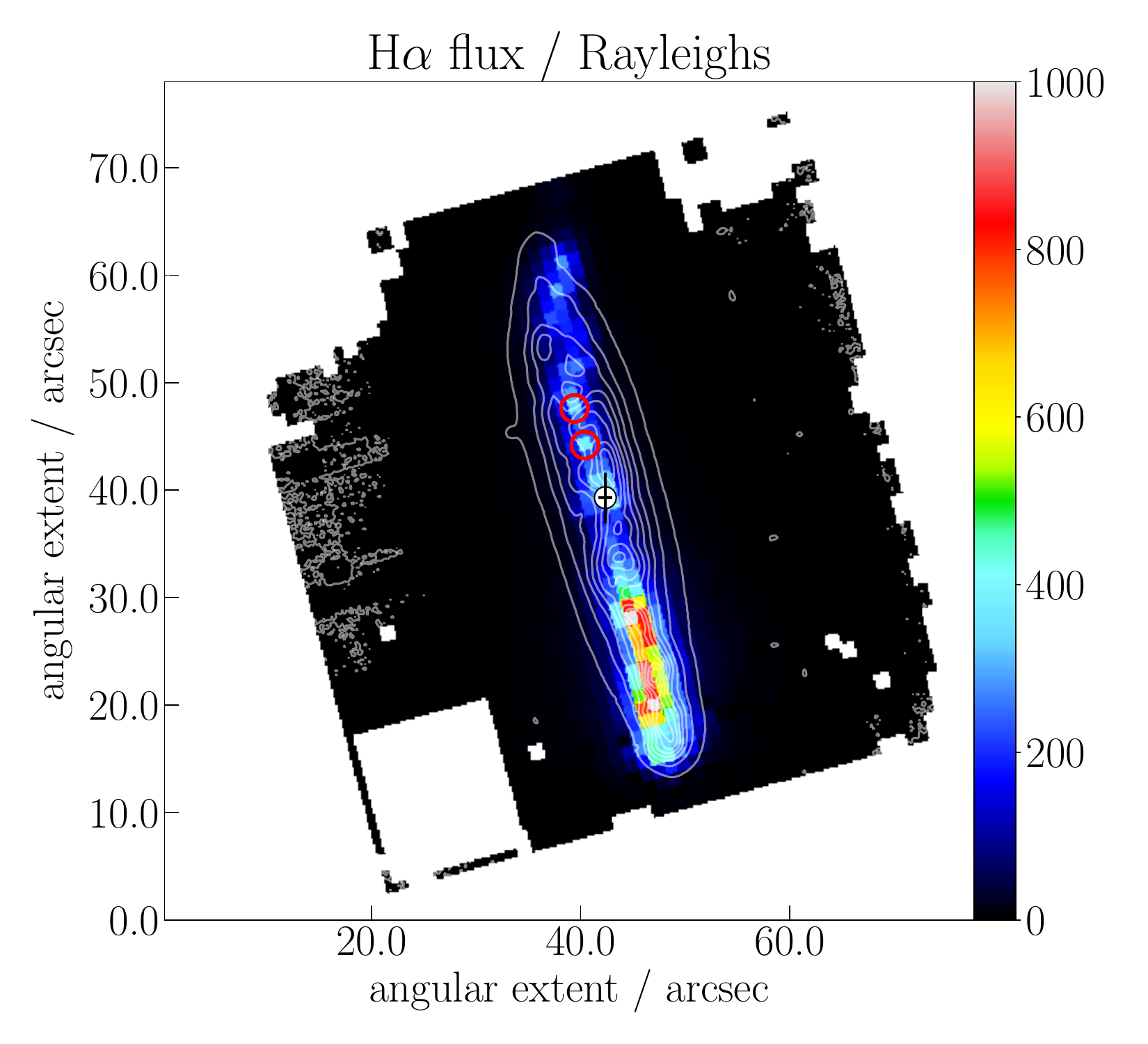}
        \end{subfigure}
        \begin{subfigure}{.33\linewidth}
        \centering
        \includegraphics[width=\hsize]{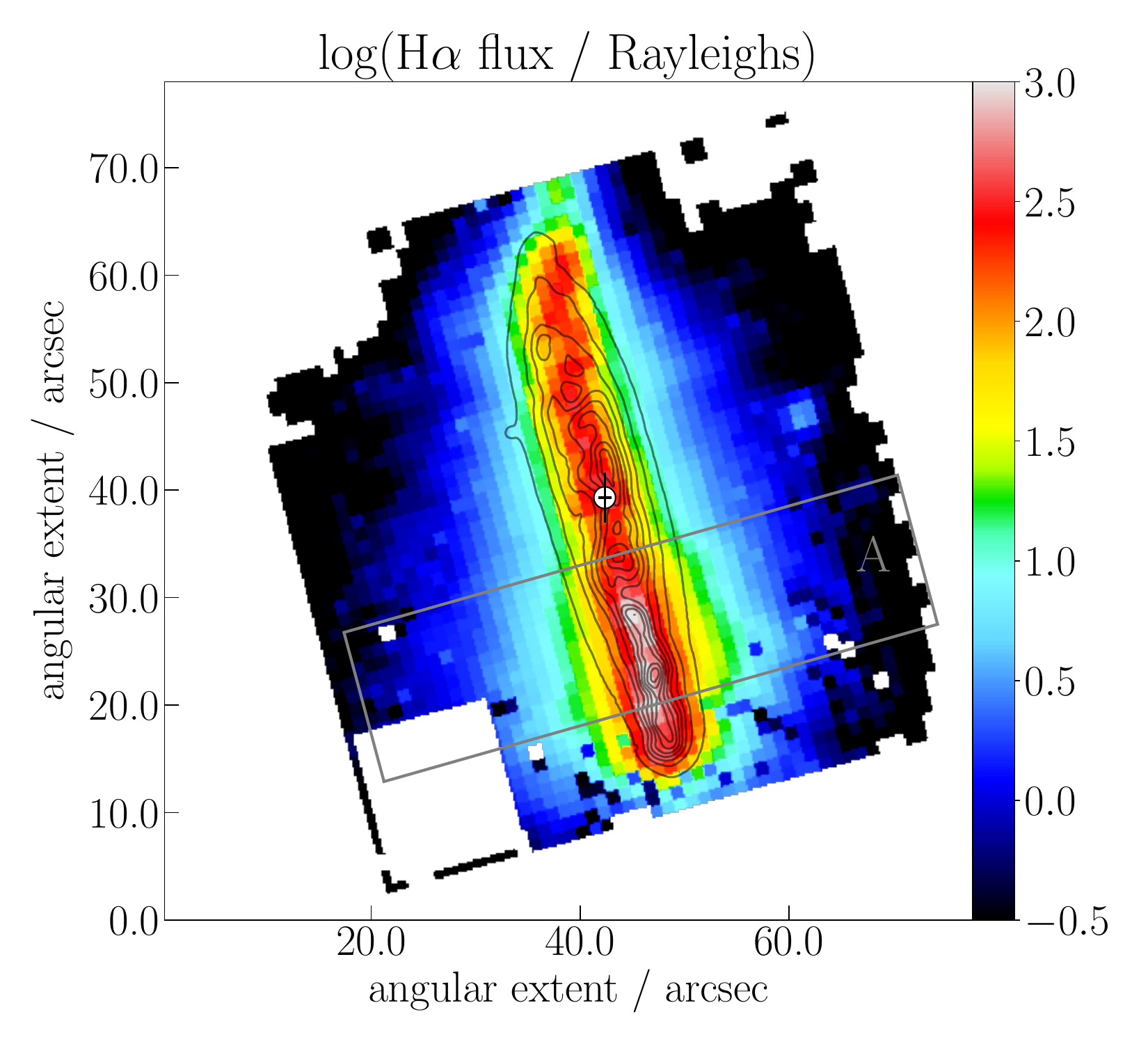}
        \end{subfigure}
    \caption{Maps of the Balmer emission in IC\,1553. \textit{Left}: Map of the Balmer decrement H$\alpha$/H$\beta$ obtained from the spectral template fitting with \texttt{pPXF}. \textit{Middle}: Map of the extinction-corrected H$\alpha$ emission line flux in Rayleighs with linear scaling. \textit{Right}: The same as the middle panel, but with logarithmic scaling. The contours are arbitrary equidistant levels from the synthetic $V$-band image meant to give orientation for the eye. The white square in the lower left corner corresponds to the masked background galaxy. The red circles in the middle panel mark the \ion{H}{ii} regions used for the oxygen abundance measurement listed in Table \ref{tab:general_props}. The boxed region marked \textbf{A} in the right-hand panel is identical to the gray shaded area in Fig. \ref{fig:scaleheights} below.}\label{fig:ha_map}
\end{figure*} 

\begin{figure*}[t]
    \includegraphics[width=0.33\textwidth]{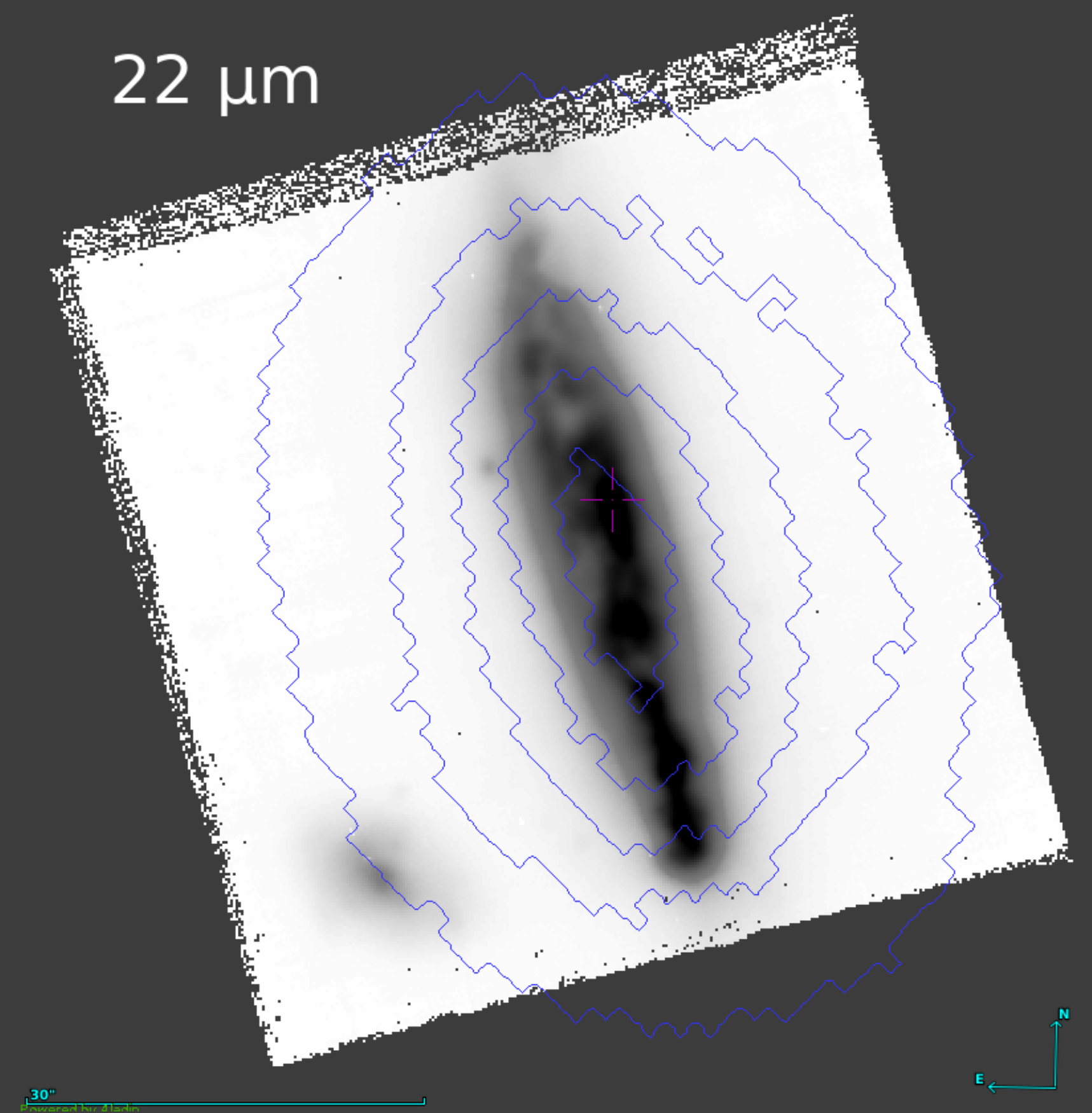}
    \includegraphics[width=0.33\textwidth]{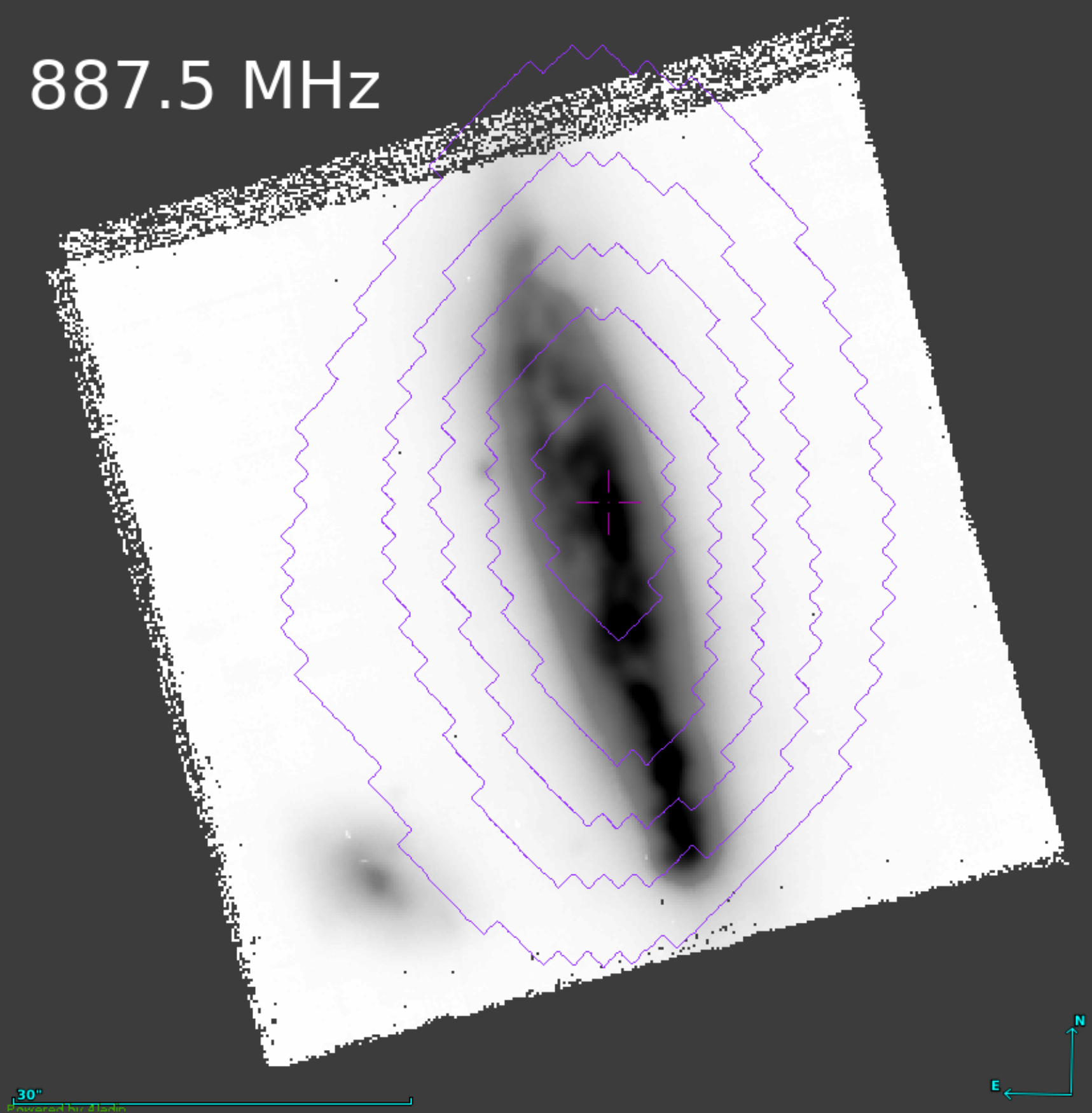}
    \includegraphics[width=0.33\textwidth]{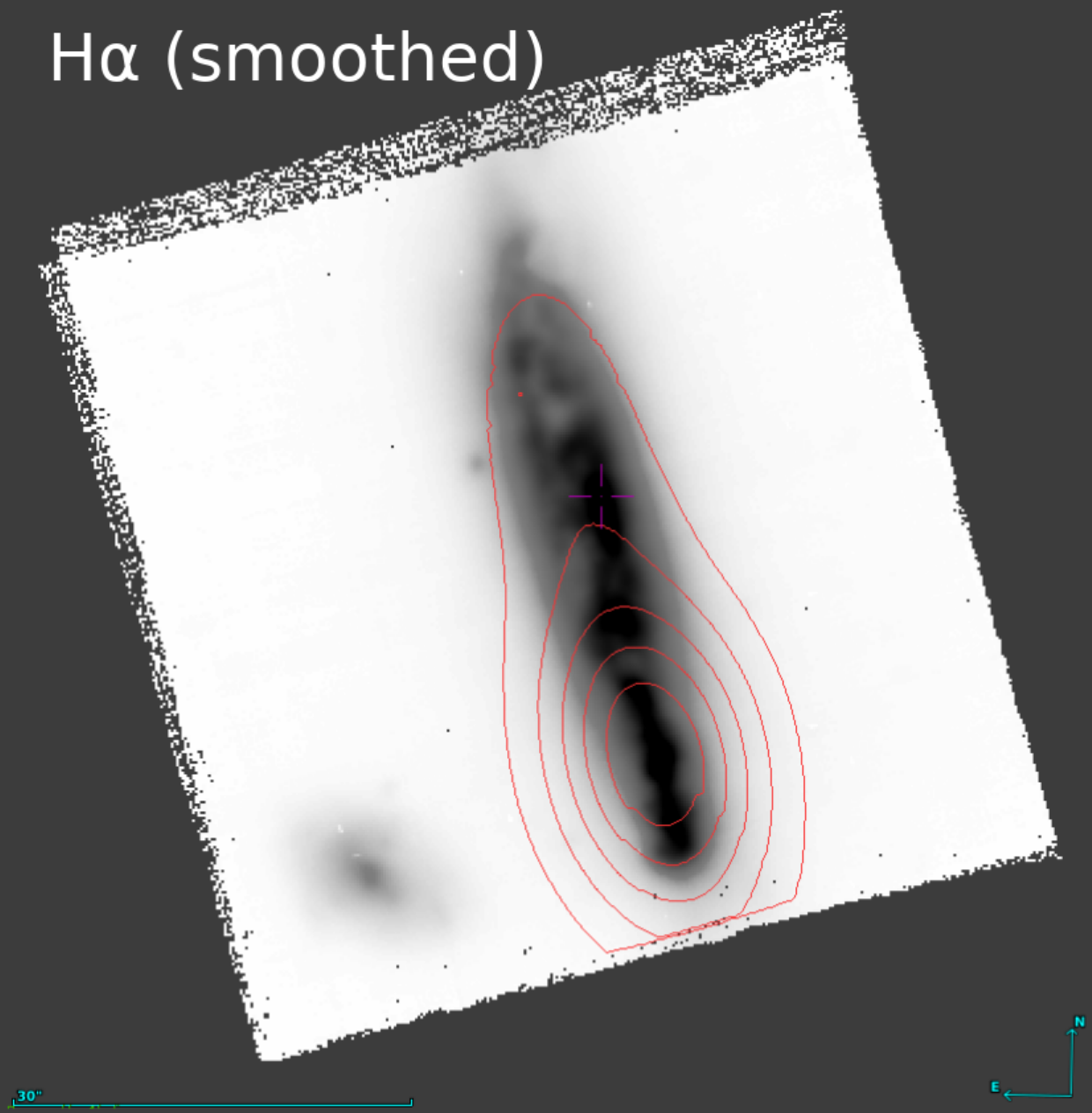}
    \caption{Synthetic $V$-band image of IC\,1553 overlaid with equidistant contours from the W4-band image from the ALLWISE Data Release at $22\,\mu$m \textit{(left)}, from the RACS-low image at $887.5\,\textrm{MHz}$ \textit{(middle),} and from the H$\alpha$ image extracted from our data cube smoothed with a Gaussian kernel \textit{(right)}. The nominal angular resolutions are $12.0\arcsec$ for the WISE W4-band, $15.0\arcsec$ for RACS-low, and the H$\alpha$ image was convolved with a $12.0\arcsec$-wide Gaussian for comparison.}
    \label{fig:sfr_tracers}
\end{figure*}

The strongest tracer of the ionized component of the warm ISM is the H$\alpha$ emission line. The middle and right panels of Fig. \ref{fig:ha_map} show maps of the extinction-corrected H$\alpha$ fluxes on a linear and logarithmic scale under contour lines derived from the $V$-band continuum image. From the linear scale map it is evident that IC\,1553 exhibits a pronounced asymmetry in H$\alpha$ brightness between its northern and southern sides, with the majority of the flux being produced in the south. This indicates that the southern half is currently undergoing significantly more star formation than the north. 

In Fig. \ref{fig:sfr_tracers} we show maps of alternative tracers for star formation as contour lines. The first panel shows the brightness in the W4-band at $22\,\mu$m from the ALLWISE Data Release \citep{cutri13} of the Wide-field Infrared Survey Explorer (WISE), which traces star formation via the emission of dust heated by stellar ultraviolet light \citep{rieke09, kennicutt12, cluver17}. The second panel of Fig. \ref{fig:sfr_tracers} shows contours from the first data release of the Rapid ASKAP Continuum Survey (RACS) \citep{mcconnell20} at $887.5\,\textrm{MHz}$. Radio continuum emission is produced by synchrotron radiation from relativistic electrons, which are accelerated either by an active galactic nucleus (AGN) or by supernovae \citep{condon92, vaddi16}. The fact that IC\,1553 does not show an AGN signature in optical emission line ratios \citep[see Sect. \ref{sect:ionization} below and][]{rautio22} suggests that here the radio continuum emission is related to star formation. Investigations of global star formation rates and radio continuum luminosities show a linear correlation, but usually use radio data at 1.4\,GHz due to the strong far-infrared/radio correlation \citep{murphy11, kennicutt12}. Examinations of local star formation rates and radio luminosities down to 150\,MHz even show a superlinear relation \citep{heesen19, heesen22}, so that we can use the RACS image for a relative comparison of the star formation in the northern and southern halves of IC\,1553. The nominal angular resolution is $12\arcsec$ for WISE W4 images \citep{wright10} and $15\arcsec$ for the RACS-low band, so both are able to separate the northern and southern halves of the galaxy's $\sim 1\arcmin$-wide disk from each other. 

As can be seen in Fig. \ref{fig:sfr_tracers}, in both the W4 and RACS-low bands the galaxy appears reasonably symmetrical across the extent of its stellar disk. For comparison, the right-hand panel in Fig. \ref{fig:sfr_tracers} shows the H$\alpha$ image extracted from the data cube convolved with a $12.0\arcsec$-wide Gaussian kernel to emulate a similar spatial resolution. The primary difference between these tracers is that H$\alpha$ is sensitive on a shorter timescale of $\lesssim$10\,Myrs while infrared and radio continuum emission are sensitive to star formation for up to 100\,Myrs \citep{kennicutt12}. This implies that the northern half of IC\,1553 underwent comparable star formation rates to what its southern half currently exhibits, but was quenched sometime within the last 10 -- 100\,Myrs.

From the right-hand panel of Fig. \ref{fig:ha_map} it can be seen that diffuse H$\alpha$ emission is detected up to altitudes of $\sim$20{\arcsec} above the midplane. The north-south asymmetry is also present in this more diffuse layer of ionized gas, where the emission above the southern star-forming cluster is generally stronger than above more quiescent regions of the disk. 

We use this extended emission to characterize the spatial distribution of the DIG. The DIG of edge-on galaxies is usually well fit in the vertical direction by a two-component exponential function \citep{reynolds91}, or a single exponential component, if the plane of the disk is masked sufficiently \citep[e.g.,][]{levy19}. \cite{rautio22} obtained a two-component fit for IC\,1553 averaged over its projected radial extent and characterized its radial variation in terms of the total H$\alpha$ flux. We aim to model the radial DIG distribution by fitting exponential functions at all projected radial positions according to our binning pattern and using all bins with H$\alpha$ fluxes above $5\sigma = 1.06\,\textrm{R}$.

\begin{figure*}[t]
    \includegraphics[width=0.5\textwidth]{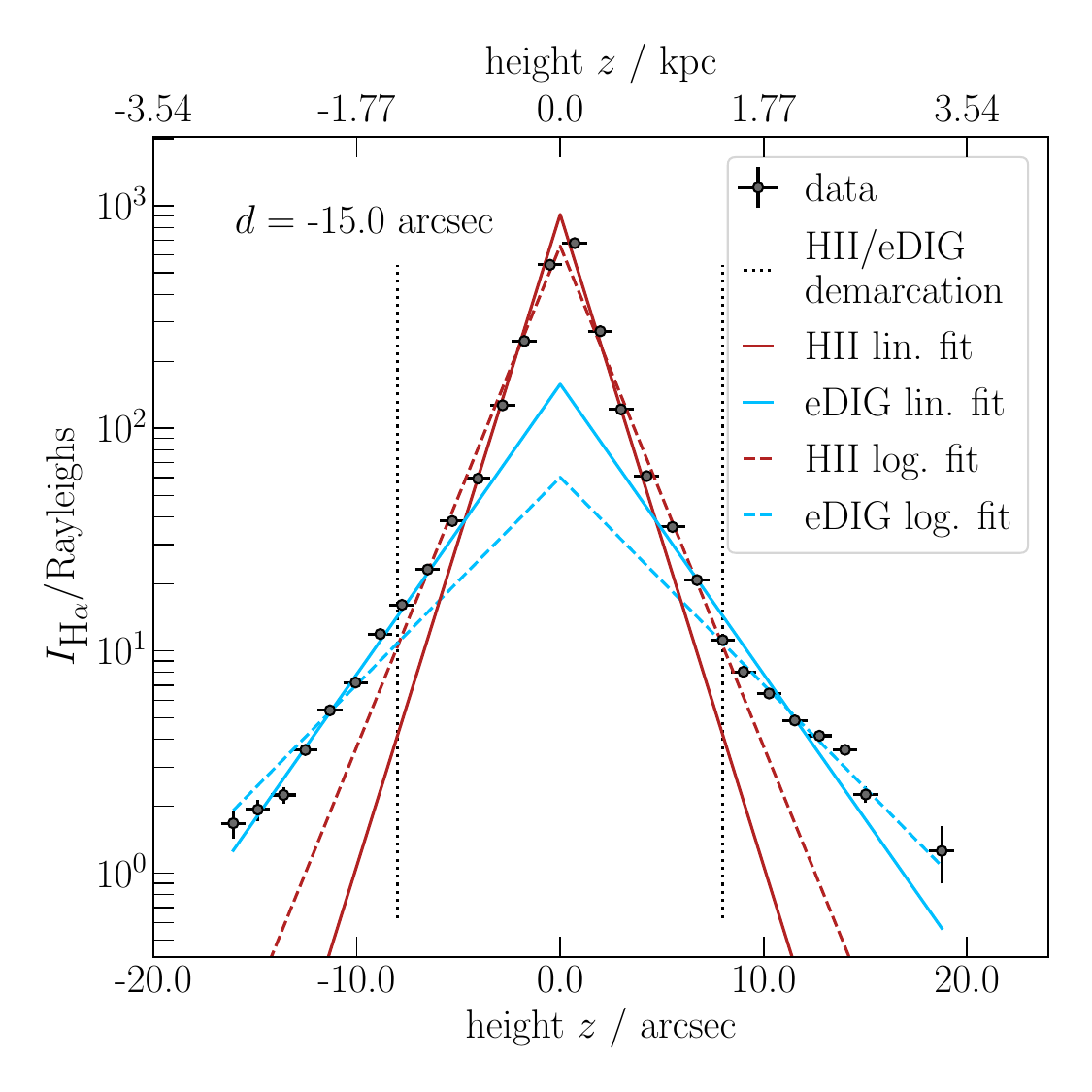}
    \includegraphics[width=0.5\textwidth]{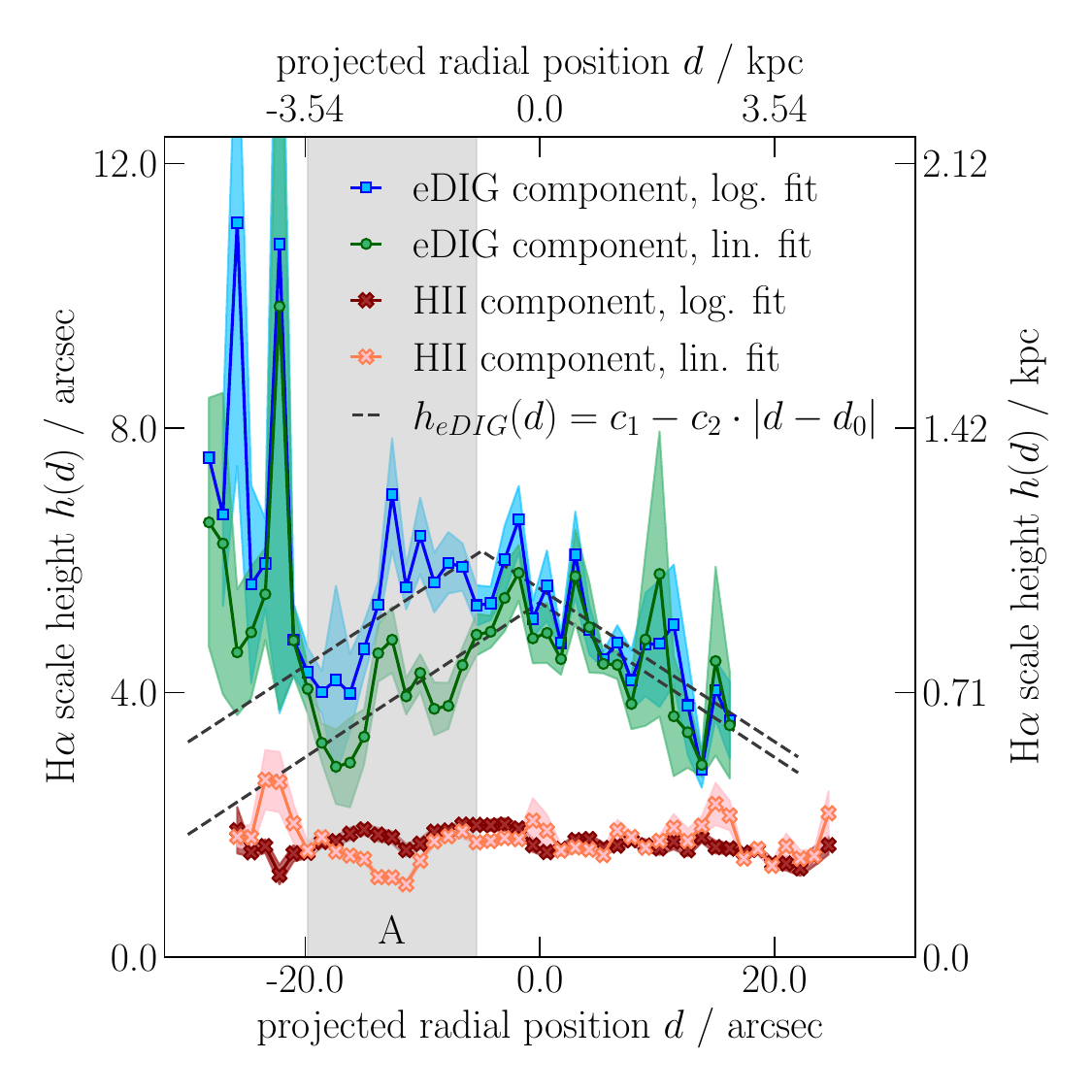}
    \caption{Visualization and results of the H$\alpha$ scale height fitting.
    \textit{Left}: Visualization of the $I_{\textrm{H}\alpha}$ fit at the example projected radial distance of $d=-15.0\arcsec$. The dotted lines mark the separation height, $z_{\textrm{sep}}=8\arcsec$, between bins used for fitting the disk \ion{H}{ii} (red curves) and eDIG (blue curves) components. The solid curves are derived with linear weighting of $I_{\textrm{H}\alpha}$ according to Eq. (\ref{eq:scaleheight_onecompfit}) and the dashed curves result from logarithmic weighting of $I_{\textrm{H}\alpha}$ during the fit.
    \textit{Right}: Radial distribution of scale heights for the eDIG (blue squares from logarithmic weighting, green circles from linear weighting) and disk \ion{H}{ii} (red crosses from logarithmic weighting, orange crosses from linear weighting) components obtained from the fitting. The shaded areas around the scale height curves give the $1\sigma$ errors obtained from the corresponding estimated parameter covariance. The dashed lines are best fits of Eq. (\ref{eq:radialshrelation}) to the eDIG component scale heights within $|d| \leq 20\arcsec$. The gray shaded area, \textbf{A}, marks the region where the logarithmic weighting produces significantly larger eDIG scale heights than the linear weighting.}
    \label{fig:scaleheights}
\end{figure*}

First, we attempted a two-component fit to the measured H$\alpha$ flux $I_{\textrm{H}{\alpha}}$ along the vertical axis, $z$, at every radial position, $d$, of the form
\begin{align}\label{eq:twocompfit}
    I_{\textrm{H}{\alpha},d}(z) = k_{\ion{H}{ii}}(d) \cdot \exp \left( \frac{-|z|}{h_{\ion{H}{ii}}(d)} \right) + k_{\textrm{eDIG}}(d) \cdot \exp \left( \frac{-|z|}{h_{\textrm{eDIG}}(d)} \right) ,
\end{align}
where the first component characterizes the \ion{H}{ii} gas contained in the disk and the second component characterizes the extraplanar diffuse ionized gas (eDIG), but we found that the data is not sufficient to constrain the free parameters in this approach. We therefore fit the two components individually by treating bins in the plane of the disk separately from bins in the halo:
\begin{align}\label{eq:scaleheight_onecompfit}
    I_{\textrm{H}{\alpha},d}(z) = 
    \begin{cases}
    k_{\ion{H}{ii}}(d) \cdot \exp \left( \frac{-|z|}{h_{\ion{H}{ii}}(d)} \right), & \textrm{if } |z| \leq z_{\textrm{sep}}\\
    k_{\textrm{eDIG}}(d) \cdot \exp \left( \frac{-|z|}{h_{\textrm{eDIG}}(d)} \right), & \textrm{if } |z| > z_{\textrm{sep}}
    \end{cases}
\end{align}
In this approach the height, $z_{\textrm{sep}}$, where the two components are separated should be large enough to suppress a bias in the eDIG parameters from the disk emission and small enough for the eDIG component to contain sufficient data points for the fit. In addition to fitting the linear H$\alpha$ flux according to Eq. (\ref{eq:scaleheight_onecompfit}) we have used logarithmic weighting to give more power to the low-brightness bins at high altitudes. Initially, we separated the disk and eDIG components at $z_{\textrm{sep}} = 5\arcsec$, but found that the logarithmic weighting resulted in systematically larger eDIG scale heights than those resulting from the linear weighting at all radial positions, which we interpreted as contamination from the disk component. We, therefore, increased $z_{\textrm{sep}}$ to $8\arcsec$ in this analysis, at which point both weighting schemes produce consistent eDIG scale heights throughout the northern half of IC\,1553 (for a discussion about the southern half, see below). The left panel of Fig. \ref{fig:scaleheights} shows an example of the results from this fitting process at $d=-15.0\arcsec$ for both components and both weighting schemes. Appendix \ref{appendixc} contains analogous plots for the fitting at every projected radial distance.

The right panel of Fig. \ref{fig:scaleheights} presents the radial behavior of all scale heights we derived in this way for IC\,1553. For the galaxy's northern half we obtain eDIG scale heights only out to $d=16.2\arcsec$ because the H$\alpha$ flux falls below the $5\sigma$ quality cut at $z_{\textrm{sep}}$ beyond that point. Within the sampled range in the north it can be seen that the H$\alpha$ scale height declines linearly with the radial distance from the center on average from $h_{\textrm{eDIG}} \approx 1\,\textrm{kpc}$ to 0.6\,kpc, and that both weighting schemes yield consistent results. For the linear weighting scheme this trend is mirrored in the southern half down to $d \lesssim -20\arcsec$, leading to a behavior of the form 
\begin{align}\label{eq:radialshrelation}
    h_{\textrm{eDIG}}(d) = c_1 - c_2 \cdot |d-d_0|
\end{align}
with the parameter values $c_{1}=0.949\,\textrm{kpc},$ $c_{2}=0.118,$ and $d_{0}=-0.002\,\textrm{kpc}$. The logarithmic weighting, however, produces significantly larger $h_{\textrm{eDIG}}$ at $-20\arcsec \leq d \leq -5\arcsec$ (the shaded gray area in Fig. \ref{fig:scaleheights} and boxed area in the last panel of Fig. \ref{fig:ha_map}, both labeled as \textbf{A}) than the linear weighting, leading to best-fitting parameters for Eq. (\ref{eq:radialshrelation}) of $c_{1}=1.087\,\textrm{kpc},$ $c_{2}=0.116,$ and $d_{0}=-0.880\,\textrm{kpc}$. 
At the southern edge of the disk ($d \leq 20\arcsec$) $h_\textrm{eDIG}$ appears to increase with the radial distance, but the fit is not entirely reliable here, because of several problematic bins in the H$\alpha$ map and the smaller sampled $z$-range resulting from the masking of the background galaxy (compare with the right panel of Fig. \ref{fig:ha_map}). We interpret and discuss this result in Sect. \ref{sect:shdiscussion} below.

\subsection{Ionization structure of the DIG}\label{sect:ionization}

\begin{figure*}
        \centering
        \begin{subfigure}{.33\linewidth}
        \centering
        \includegraphics[width=\hsize]{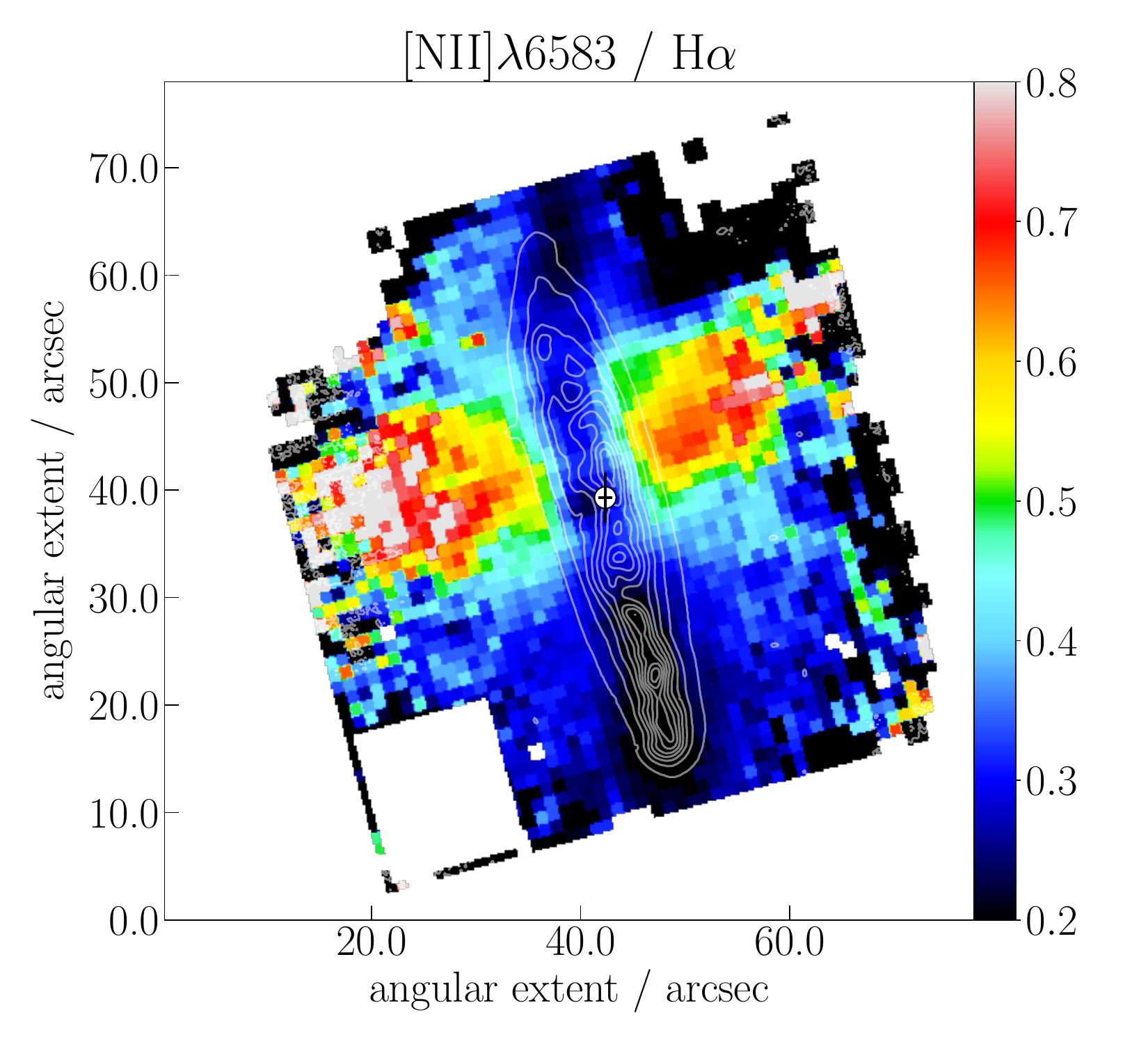}
    \end{subfigure}
        \begin{subfigure}{.33\linewidth}
        \centering
        \includegraphics[width=\hsize]{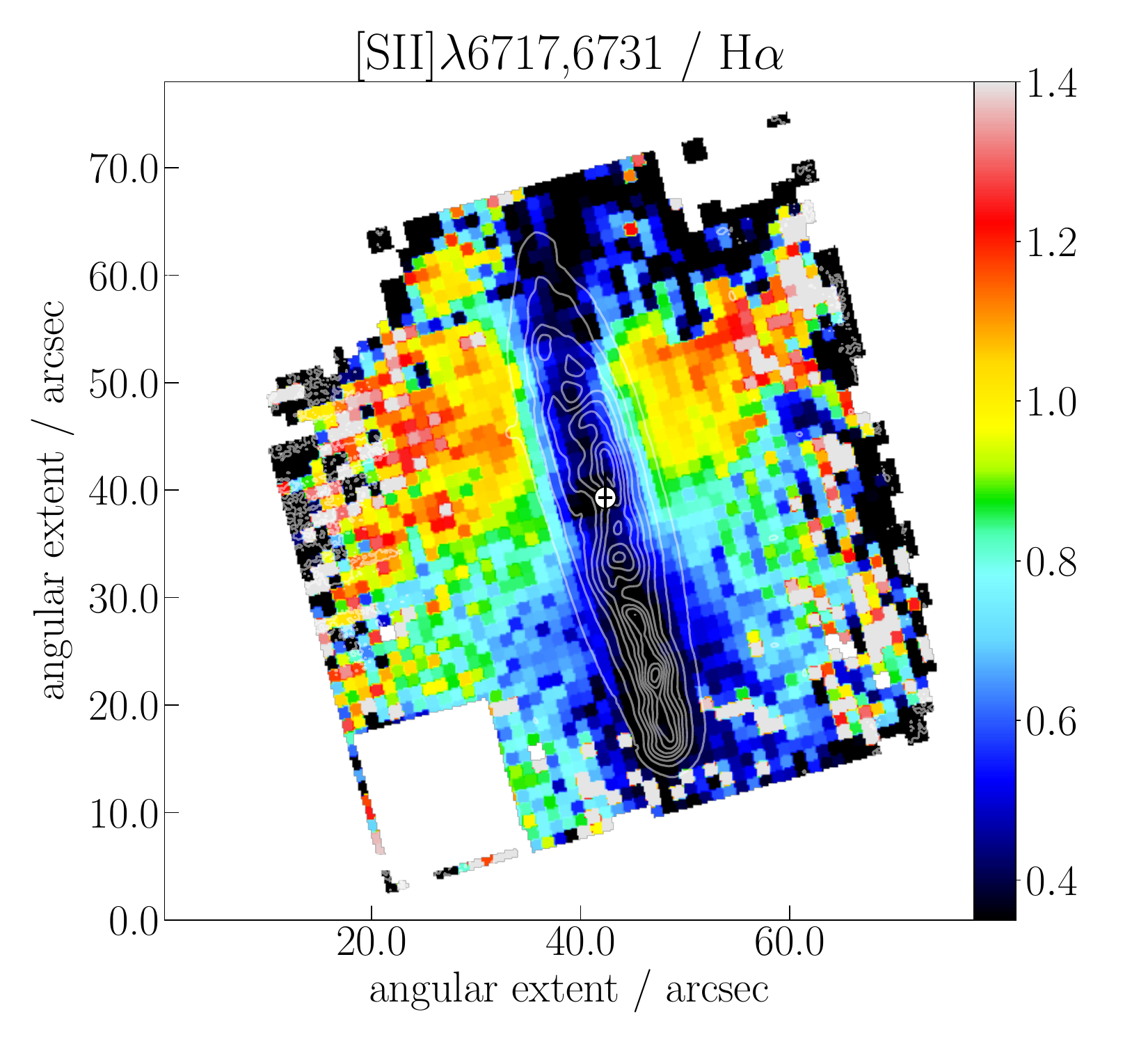}
        \end{subfigure}
        \begin{subfigure}{.33\linewidth}
        \centering
        \includegraphics[width=\hsize]{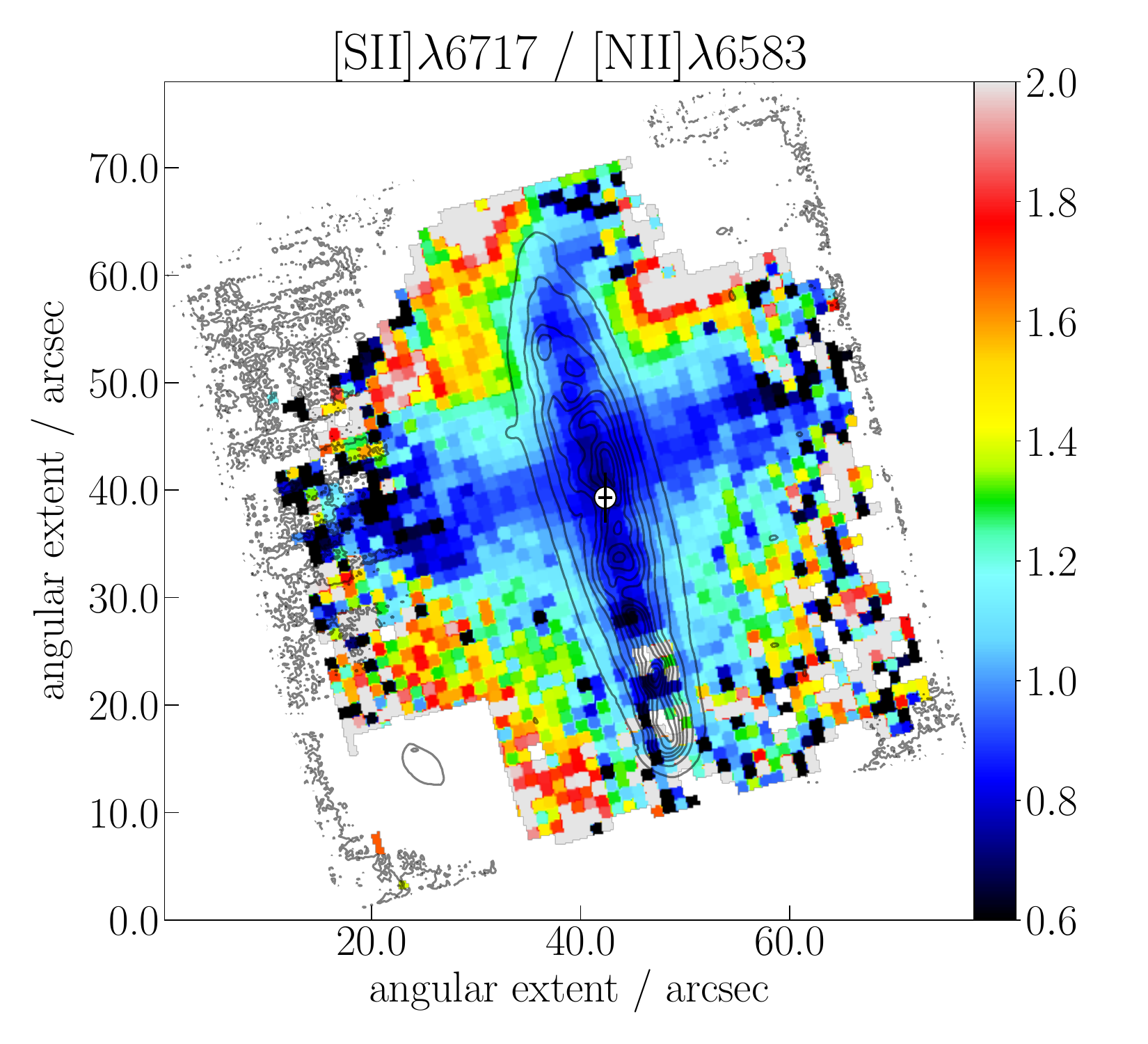}
        \end{subfigure} 
    \\
        \begin{subfigure}{.33\linewidth}
        \centering
        \includegraphics[width=\hsize]{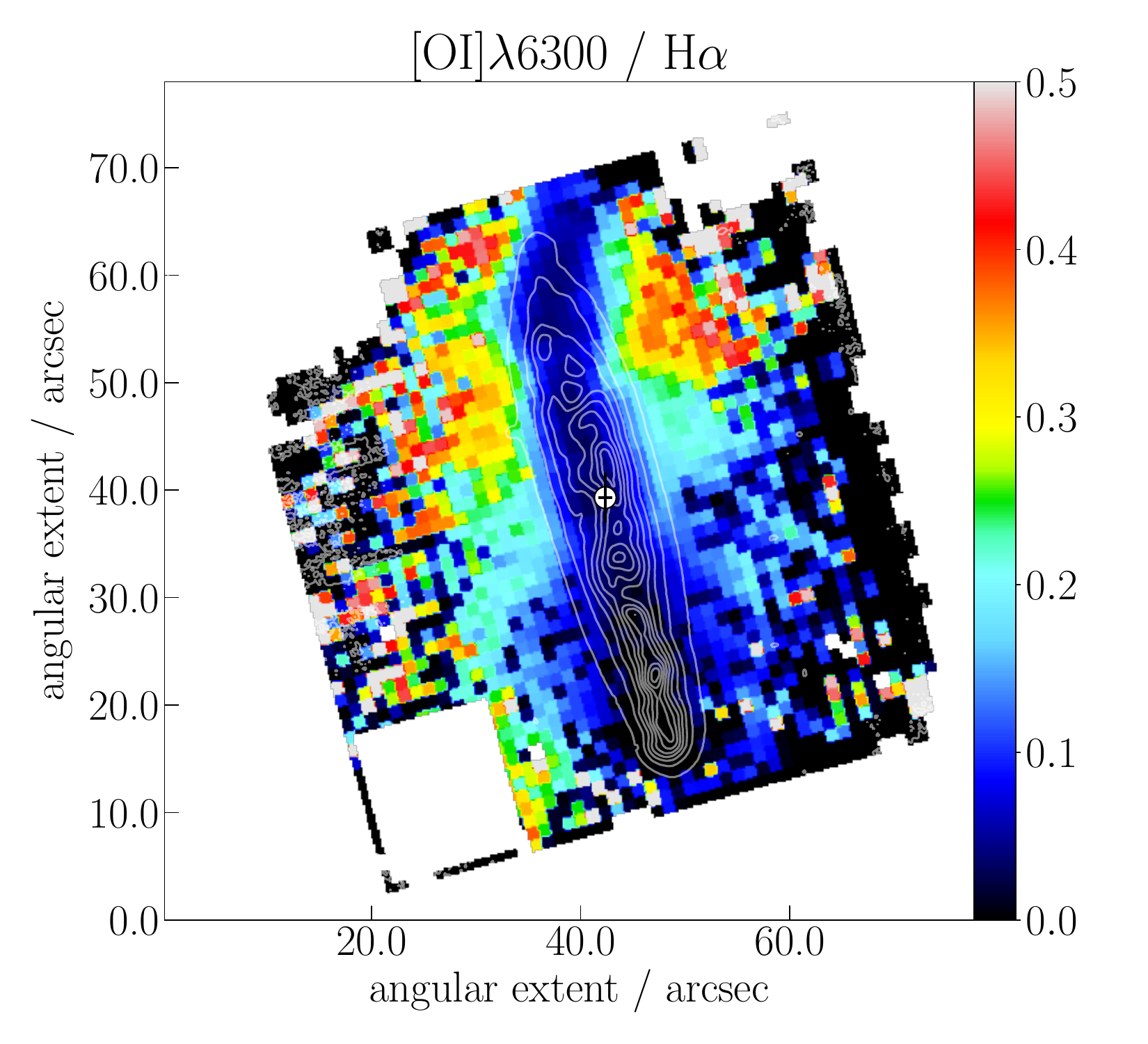}
        \end{subfigure}
        \begin{subfigure}{.33\linewidth}
        \centering
        \includegraphics[width=\hsize]{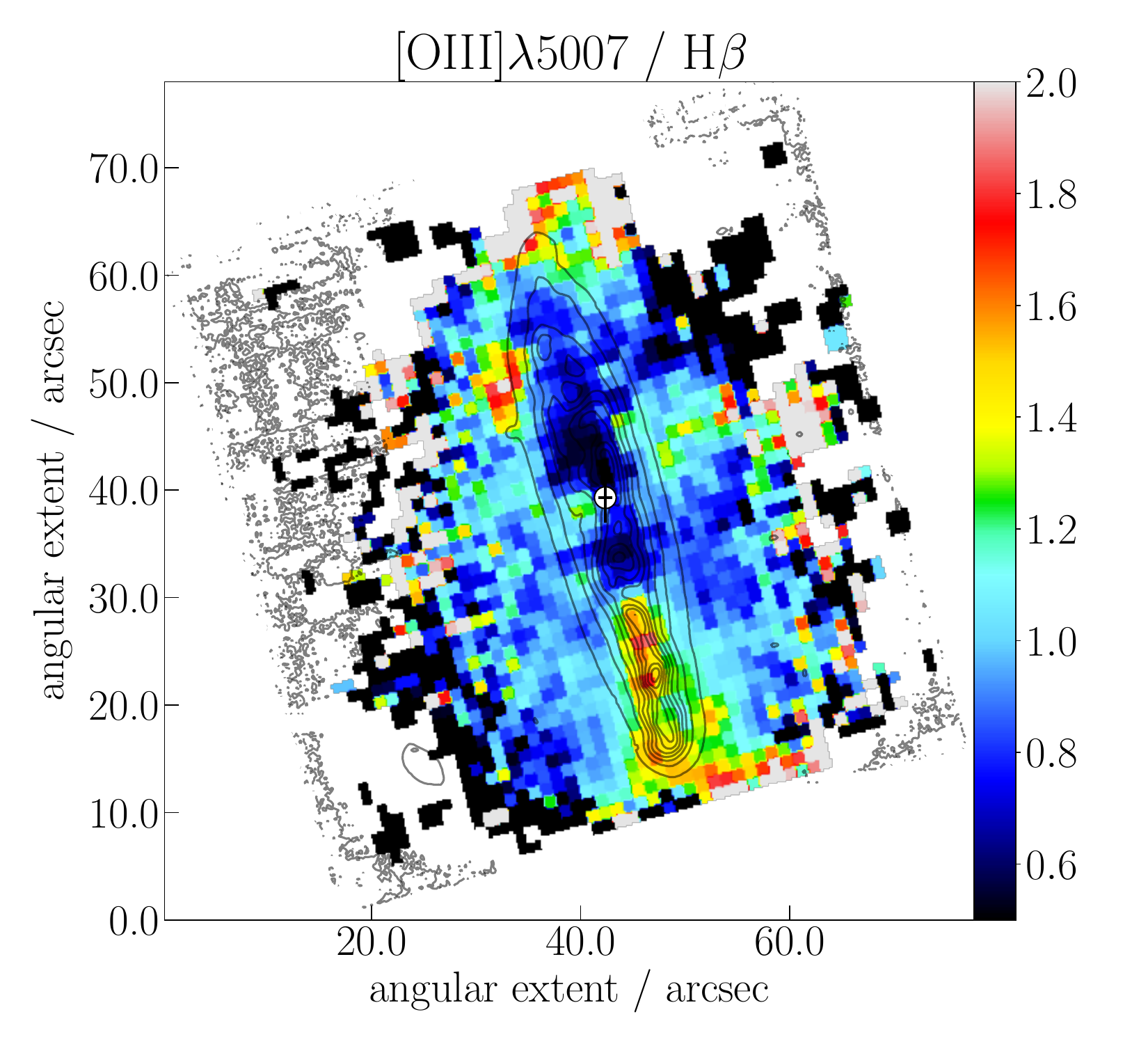}
        \end{subfigure}
        \begin{subfigure}{.33\linewidth}
        \centering
        \includegraphics[width=\hsize]{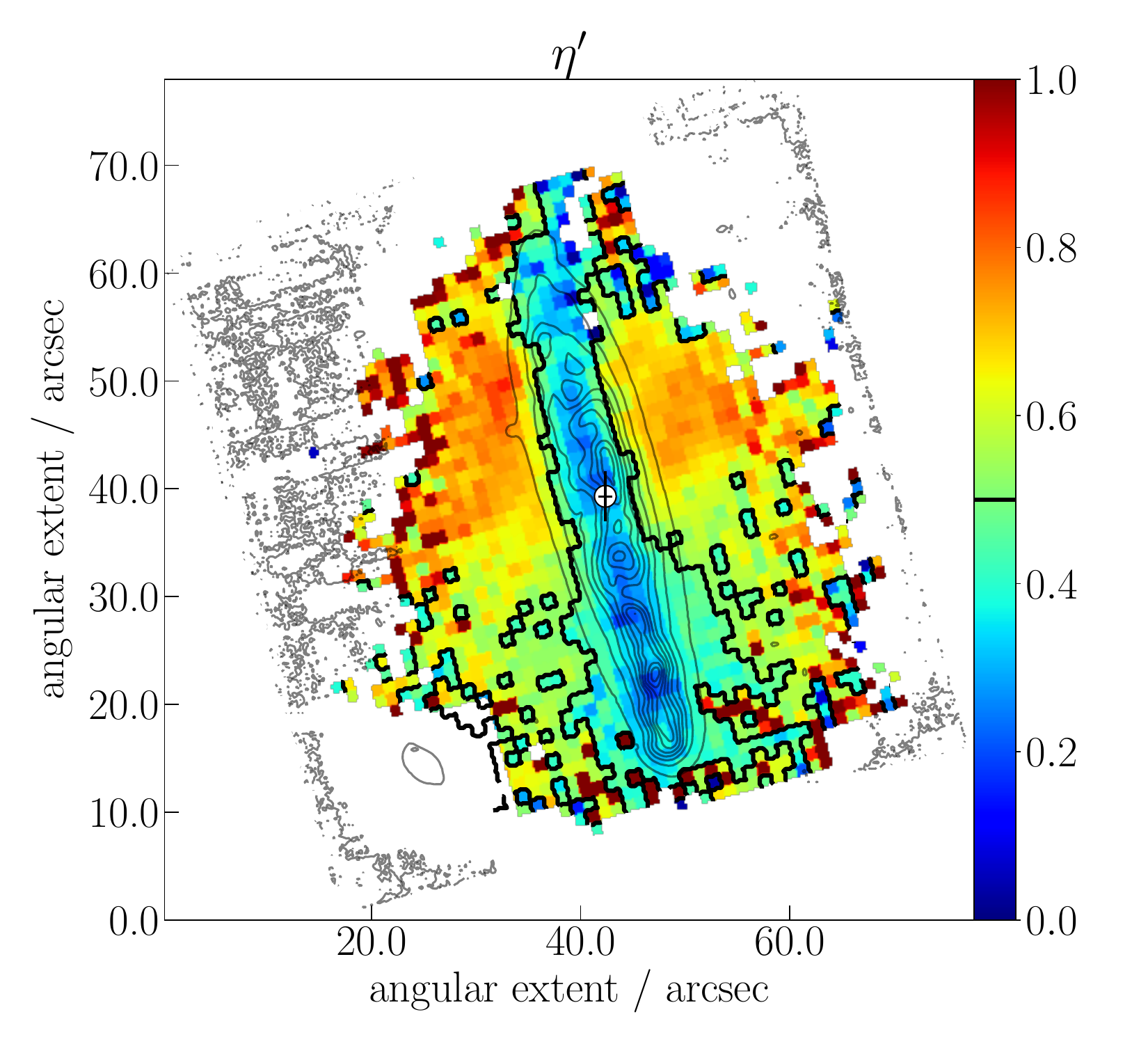}
        \end{subfigure}

\caption{Maps of the emission line ratios commonly used in the classical diagnostic diagrams. \textit{First row}:
$[\ion{N}{ii}]\lambda 6583$/H$\alpha$ \textit{(left)}, 
$[\ion{S}{ii}]\lambda $6717,6731/H$\alpha$ \textit{(middle)}, and 
$[\ion{S}{ii}]\lambda 6717$/$[\ion{N}{ii}]\lambda 6583$ \textit{(right)}, which is commonly analyzed in studies of the DIG. 
\mbox{\textit{Second row}:} 
$[\ion{O}{i}]\lambda 6300$/H$\alpha$ \textit{(left)} and
$[\ion{O}{iii}]\lambda 5007$/H$\beta$ \textit{(middle)}, as well as the $\eta^{\prime}$ parameter \textit{(right)} as described in Sect. \ref{sect:ionization} and defined in Appendix \ref{appendix:eta}, which quantifies the distance to the \cite{kewley01} starburst limit in the [\ion{O}{iii}]/H$\beta$ vs. [\ion{S}{ii}]/H$\alpha$ diagnostic diagram. The black demarcation lines in the $\eta^{\prime}$ map separate bins below the starburst limit ($\eta^{\prime} \leq 0.5$) from bins above that limit ($\eta^{\prime} > 0.5$). The gray and white contours are at the same equidistant arbitrary levels from the synthetic $V$-band image as in Fig. \ref{fig:ha_map}.}\label{fig:elineratios_maps} 
\end{figure*} 

Figure \ref{fig:elineratios_maps} shows maps of the emission line ratios commonly used to derive information about the ionization conditions of interstellar gas. These maps are consistent with some of the observed properties of the DIG in other galaxies and the Reynolds-layer of the Milky Way: 
1) The line ratios $[\ion{N}{ii}]\lambda 6583$/H$\alpha$ and $[\ion{S}{ii}]\lambda 6717,6731$/H$\alpha$ generally rise with increasing height above the disk \citep[e.g.,][]{golla96, haffner99, collins01, levy19} and 2) $[\ion{S}{ii}]\lambda 6717,6731$/H$\alpha$ and $[\ion{N}{ii}]\lambda 6583$/H$\alpha$ reach elevated levels compared to classical \ion{H}{ii} regions \citep[e.g.,][]{reynolds85, reynolds88, golla96, haffner99, levy19}. For IC\,1553, however, the magnitude of these effects varies drastically over the projected radial position. $[\ion{N}{ii}]\lambda 6583$/H$\alpha$ has a clear conical structure of increased values centered slightly north ($\sim 5\arcsec$) of the galactic center. The same can be seen for $[\ion{O}{i}]\lambda 6300$/H$\alpha$, though here the structure is even further off-centered ($\sim$$10\arcsec$) and less pronounced. $[\ion{S}{ii}]\lambda 6717,6731$/H$\alpha$ displays a similar structure of increased value, producing a clear asymmetry between the northern and southern halves of IC\,1553. $[\ion{O}{iii}]\lambda 5007$/H$\beta$ on the other hand peaks in the plane of the disk coincident with the prominent star-forming regions in the south (compare with Fig. \ref{fig:ha_map}), but everywhere else its value is larger in the eDIG than in the planar gas.

One unusual feature of the DIG in IC\,1553 is the map of the $[\ion{S}{ii}]\lambda 6717$/$[\ion{N}{ii}]\lambda 6583$ ratio. Multiple studies found this ratio to remain largely constant with $z$ \citep[e.g.,][]{golla96, otte99, haffner99, rand00}. Here, however, this is only observed in the volume coincident with the conus of increased $[\ion{N}{ii}]\lambda 6583$/H$\alpha$. Outside of this volume, IC\,1553 displays a rising trend of $[\ion{S}{ii}]\lambda 6583$/$[\ion{N}{ii}]\lambda 6583$ with increasing height.

\begin{figure*}
    \centering
    \includegraphics[width=\textwidth]{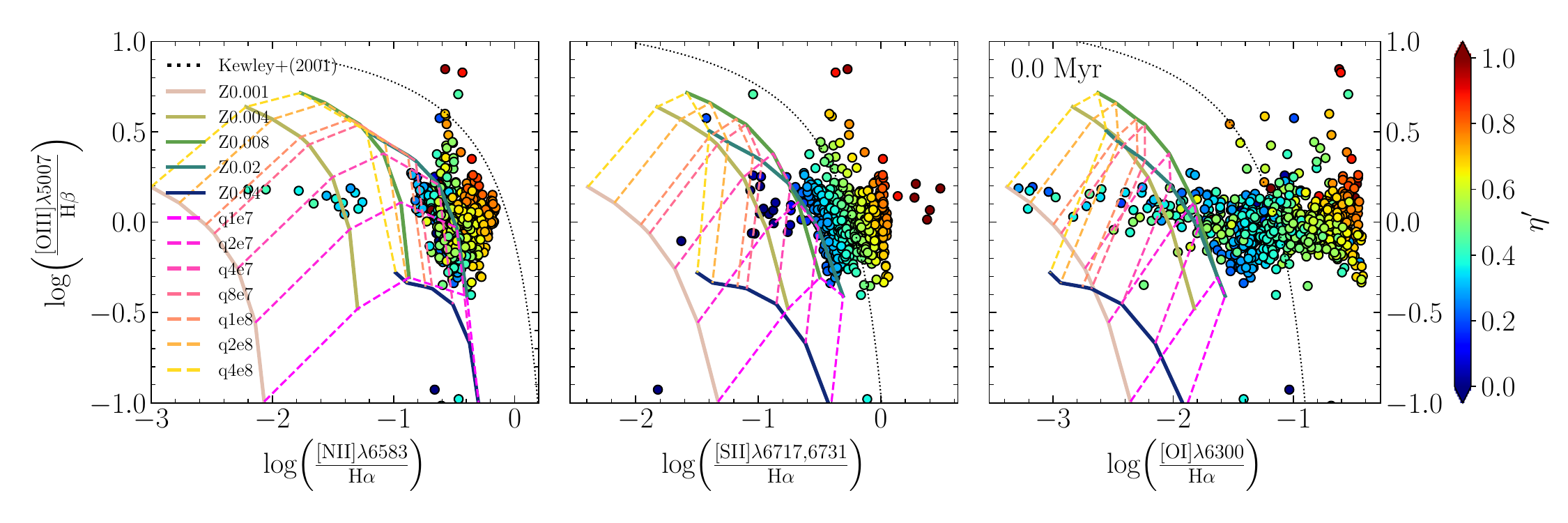}
    \caption{Typical emission line diagnostic diagrams for the analysis of ionization sources. The data points are the population of bins with H$\alpha$ flux $\geq 5\sigma$, colored by their $\eta^\prime$ value.
    The black dotted line marks the theoretical limit for starburst galaxies from \cite{kewley01}. The solid and dashed lines mark the metallicity ($Z$) and ionization parameter ($q$) grid of photoionization models described in \cite{levesque10} for the best-fitting star formation age of $0\,\textrm{Myrs}$ and electron density of $100\,\textrm{cm}^{-3}$.}\label{fig:bpt_photmodels}
\end{figure*}

\begin{figure*}
    \centering
    \includegraphics[width=\textwidth]{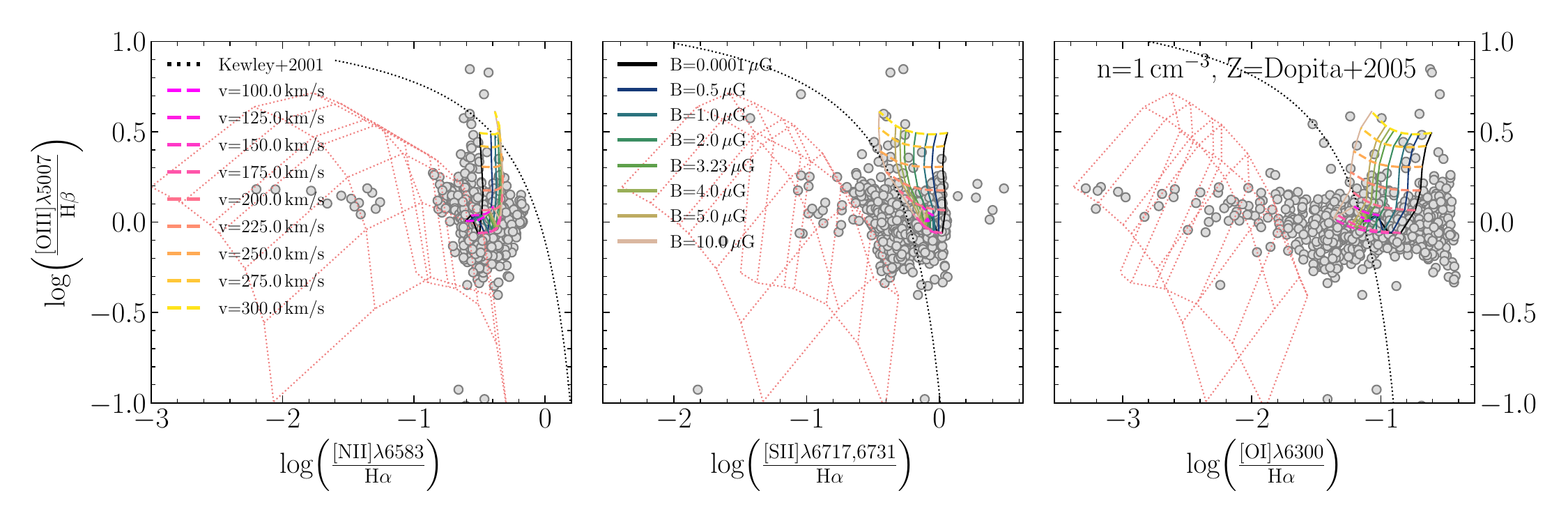}
    \caption{The same diagnostic diagrams as shown in Fig. \ref{fig:bpt_photmodels} (dotted black and red lines are the starburst limit and photoionization models described therein), now including the velocity ($v$) and magnetic field ($B$) grid (dashed and solid lines) for the best-fitting shock models from \cite{allen08}, in which the preshock density $n=1\,\textrm{cm}^{-3}$ and the chemical abundances described in \cite{dopita05} were used.}
    \label{fig:bpt_schockmodels}
\end{figure*}

\begin{figure*}
    \centering
    \includegraphics[width=\textwidth]{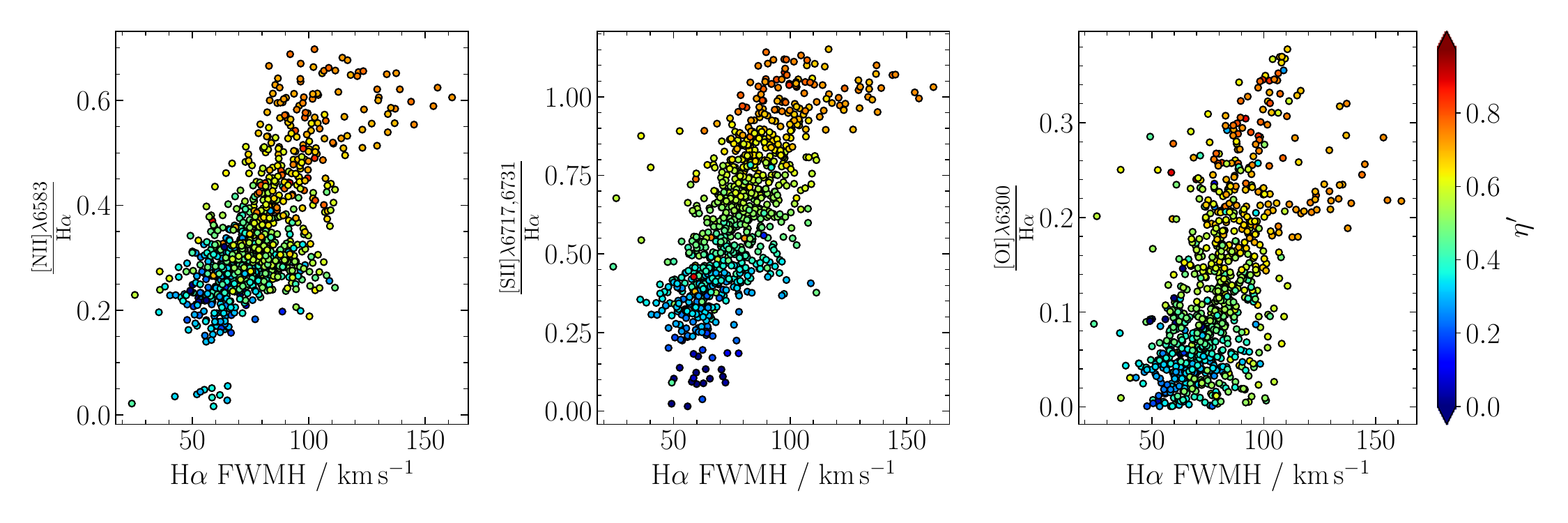}
    \caption{Plots of the shock-sensitive emission line ratios $[\ion{N}{ii}]\lambda 6583$/H$\alpha$,  [\ion{S}{ii}]$\lambda$6717,6731/H$\alpha$, and [\ion{O}{i}]$\lambda$6300/H$\alpha$ vs. the H$\alpha$ FWHM of all bins with H$\alpha$ flux $\geq 5\sigma$, colored by their $\eta^\prime$ value.}
    \label{fig:ratio_width_correlation}
\end{figure*}

In Fig. \ref{fig:bpt_photmodels} we show the classical ionization diagnostic diagrams \citep{baldwin81, veilleux87} with the limiting surface for starburst galaxies from \cite{kewley01} and the best-fitting photoionization model grid from \cite{levesque10} for comparison. Fig. \ref{fig:bpt_photmodels} (as well as Figs. \ref{fig:bpt_schockmodels} and \ref{fig:ratio_width_correlation}) contains only data points where H$\alpha$ was detected at a $5\sigma$-level ($1.06\,\textrm{R}$). The population of bins most closely resembles the models for a current instantaneous star formation burst of age $0\,\textrm{Myrs}$ with an electron density of $n_e = 100\,\textrm{cm}^{-3}$. The grid covers a range of metallicities, $Z$, between 0.001 and 0.04 where $Z_\odot \equiv 0.02$, and ionization parameters, $q$, between $1 \times 10^7\,\textrm{cm}\,\textrm{s}^{-1}$ and $4 \times 10^8\,\textrm{cm}\,\textrm{s}^{-1}$. The photoionization models can reproduce the observed population in the $\log$([\ion{O}{iii}]/H$\beta$) vs. $\log$([\ion{N}{ii}]/H$\alpha$) diagram quite nicely (left panel of Fig. \ref{fig:bpt_photmodels}) for roughly solar metallicity ($Z \approx 0.02$) and an ionization parameter around $2-4\times 10^7\,\textrm{cm}\,\textrm{s}^{-1}$. Roughly half of the bins fall slightly outside of the volume covered by the grid and closer to the \cite{kewley01} starburst limit. In principle this could be explained by spectral hardening of the photons during their propagation from the disk to the halo \citep{wood04}. However, the other two diagnostic diagrams display a clear excess of [\ion{S}{ii}] and [\ion{O}{i}] emission compared to the models, with a significant fraction of the population actually crossing the starburst limit. 

Since the behavior of the emission line ratios varies strongly over the projected radial position, we are interested in mapping the bins in the plane of the sky to their distance from the \cite{kewley01} starburst limit in the diagnostic diagrams to gauge the spatial distribution of the ionization conditions more clearly. \cite{rautio22} used the $\eta$-parameter from \cite{erroz-ferrer19} for this purpose, which quantifies the position in the [\ion{O}{iii}]/H$\beta$ vs. [\ion{N}{ii}]/H$\alpha$ diagram. For IC\,1553 there is merit in using the [\ion{S}{ii}]/H$\alpha$ or [\ion{O}{i}]/H$\alpha$ diagnostic instead, because there the data deviates much more strongly from the photoionization models. We therefore define the parameter $\eta^{\prime}$ analogously to $\eta$ so that $\eta^{\prime} < 0.5$ marks a position to the left of the \cite{kewley01} starburst limit in the [\ion{O}{iii}]/H$\beta$ vs. [\ion{S}{ii}]/H$\alpha$ diagram, and $\eta^{\prime} > 0.5$ to its right. The specific calculation of $\eta^{\prime}$ is described in Appendix \ref{appendix:eta}. The bottom-right panel of Fig. \ref{fig:elineratios_maps} shows the map of $\eta^{\prime}$ for IC\,1553 and visualizes that most of the volume occupied by the sampled eDIG produces emission line ratios beyond the starburst limit. Only the innermost halo above the star-forming cluster in the southern disk exhibits bins with $\eta^{\prime} < 0.5$.

The \cite{wood04} radiative transfer models can reproduce the observed [\ion{S}{ii}]/H$\alpha$ and [\ion{O}{i}]/H$\alpha$ values in principle, but not without also producing an elevated [\ion{N}{ii}]/H$\alpha$. They can therefore explain the ionization behavior in the eDIG above the center of the galaxy, but not above the outskirts of the disk, where [\ion{S}{ii}]/[\ion{N}{ii}] rises with increasing height. 

This deficit in [\ion{N}{ii}] emission compared to [\ion{S}{ii}] and [\ion{O}{i}] in conjunction with the moderate ratio values of [\ion{O}{iii}]/H$\beta$ observed in the data leads us to consider shocks as an additional source of ionization for the DIG of IC\,1553. To this end, we compare the observed emission line ratio to the library of shock models from \cite{allen08}. The best agreement with our data is achieved under the conditions of solar abundances with dust depletion, as described by \cite{dopita05}, and a preshock density of $n=1.0\,\textrm{cm}^{-3}$. These models are plotted for a range of shock velocities of $v = 100 - 300\,\textrm{km}\,\textrm{s}^{-1}$ and magnetic fields of $B = 10^{-4} - 10\,\mu \,\textrm{G}$ in Fig. \ref{fig:bpt_schockmodels}. 

The line ratios for the majority of bins fall somewhere between the parameter space covered by the \cite{levesque10} photoionization models, and the curve given by the \cite{allen08} shock ionization models for the shock velocity of $v = 225\,\mathrm{km\,s}^{-1}$. This distribution indicates that both photo- and shock ionization are relevant for the ionization state of the eDIG in IC\,1553, and that the emission lines in each spatial bin may contain components from both processes with differing relative weights. If we want to characterize properties of the shock, we should consider the edge of the distribution, which is furthest away from the photoionization model grid, because there the contribution from the shocked region to the total emission is maximal. By this reasoning, the data points in Fig. \ref{fig:bpt_schockmodels} are consistent with a mixture of emission produced by photoionization on the one hand and a shock with the approximate velocity of $225\,\mathrm{km\,s}^{-1}$ on the other hand.
The highest observed ratios for [\ion{N}{ii}]/H$\alpha$ and [\ion{O}{i}]/H$\alpha$ cannot be reproduced in detail by the models, but it is conceivable that a more fine-tuned model, specifically in terms of a lower gas density or slightly adapted abundances, might be able to.

If shocks make a significant contribution to the ionization of the DIG we expect to see a positive correlation between observed line widths and shock-sensitive line ratios \citep{kewley19}. We therefore plot the corresponding ratio values over the H$\alpha$ FWHM in Fig. \ref{fig:ratio_width_correlation} and indeed such a correlation is present for all ratios with a Spearman correlation coefficient $r_\mathrm{s}$ of $\approx 0.64$ and $p$-values $\ll 0.1$. For [\ion{N}{ii}]/H$\alpha$ the correlation is only clear for bins with $\eta^\prime \gtrsim 0.5$, in other words for lines of sight in the halo sufficiently far away from the southern star-forming cluster (compare with Fig. \ref{fig:elineratios_maps}). Figure \ref{fig:ratio_width_correlation} also shows that the H$\alpha$ linewidth correlates with $\eta^\prime$, which indicates that shocks play a significant role in shaping the ionization structure of the disk-halo interface of IC\,1553.


\subsection{DIG kinematics}\label{sect:kinematics}
\begin{figure*}[t]
    \begin{subfigure}{0.5\linewidth}
    \includegraphics[width=\hsize]{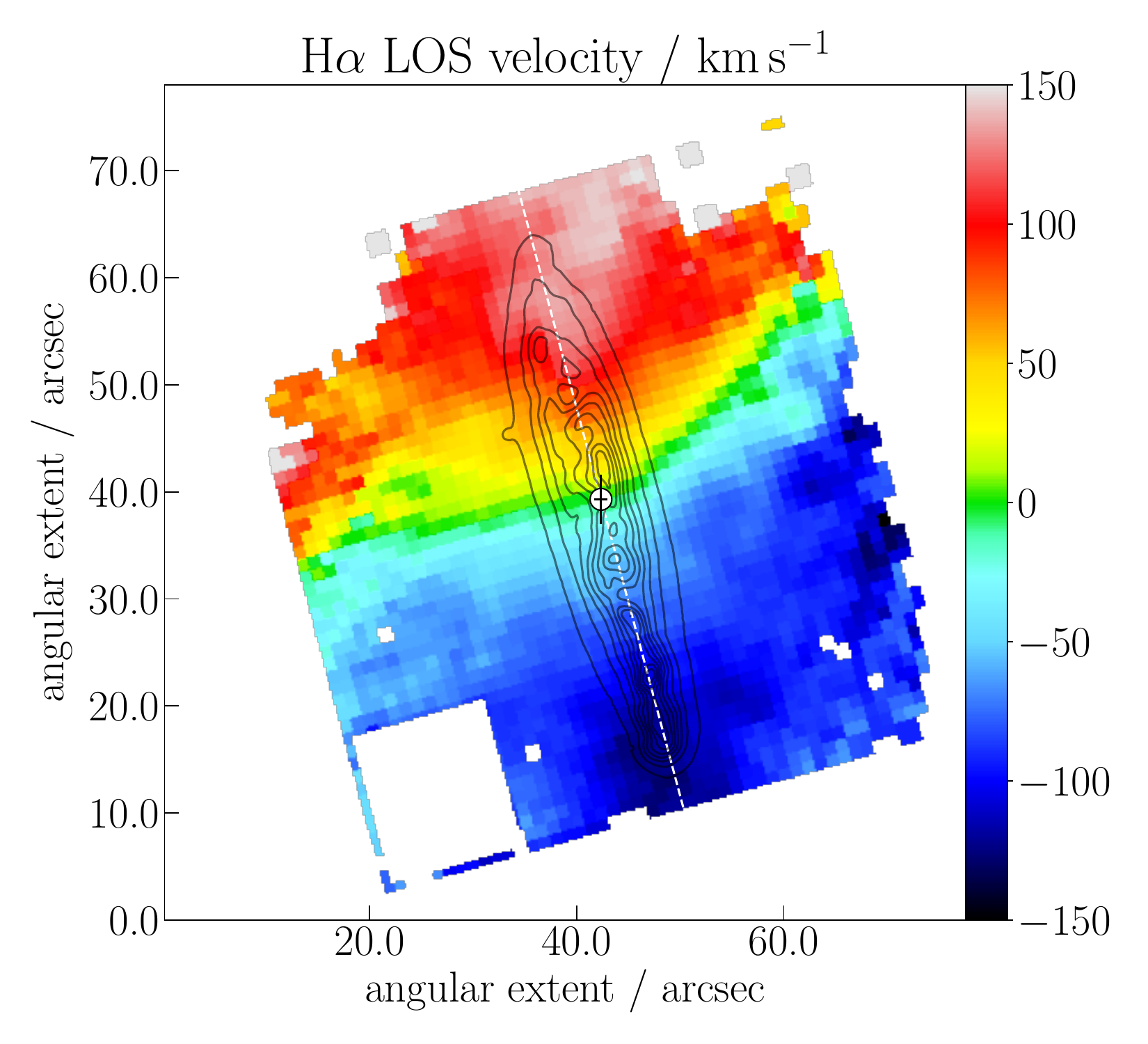}
    \end{subfigure}
    \begin{subfigure}{0.5\linewidth}
    \includegraphics[width=\hsize]{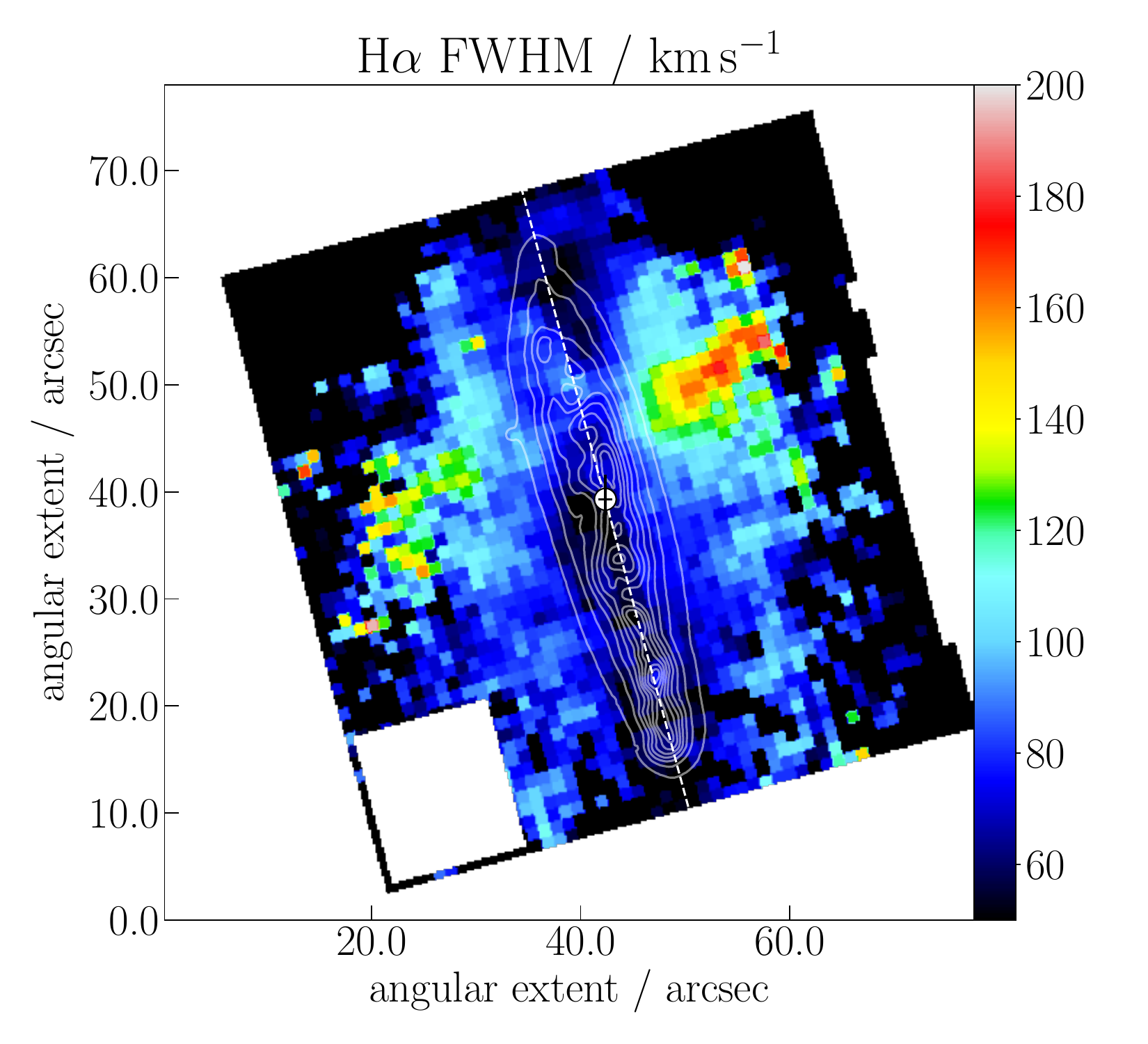}
    \end{subfigure}\\
    \begin{subfigure}{0.5\linewidth}
    \includegraphics[width=\hsize]{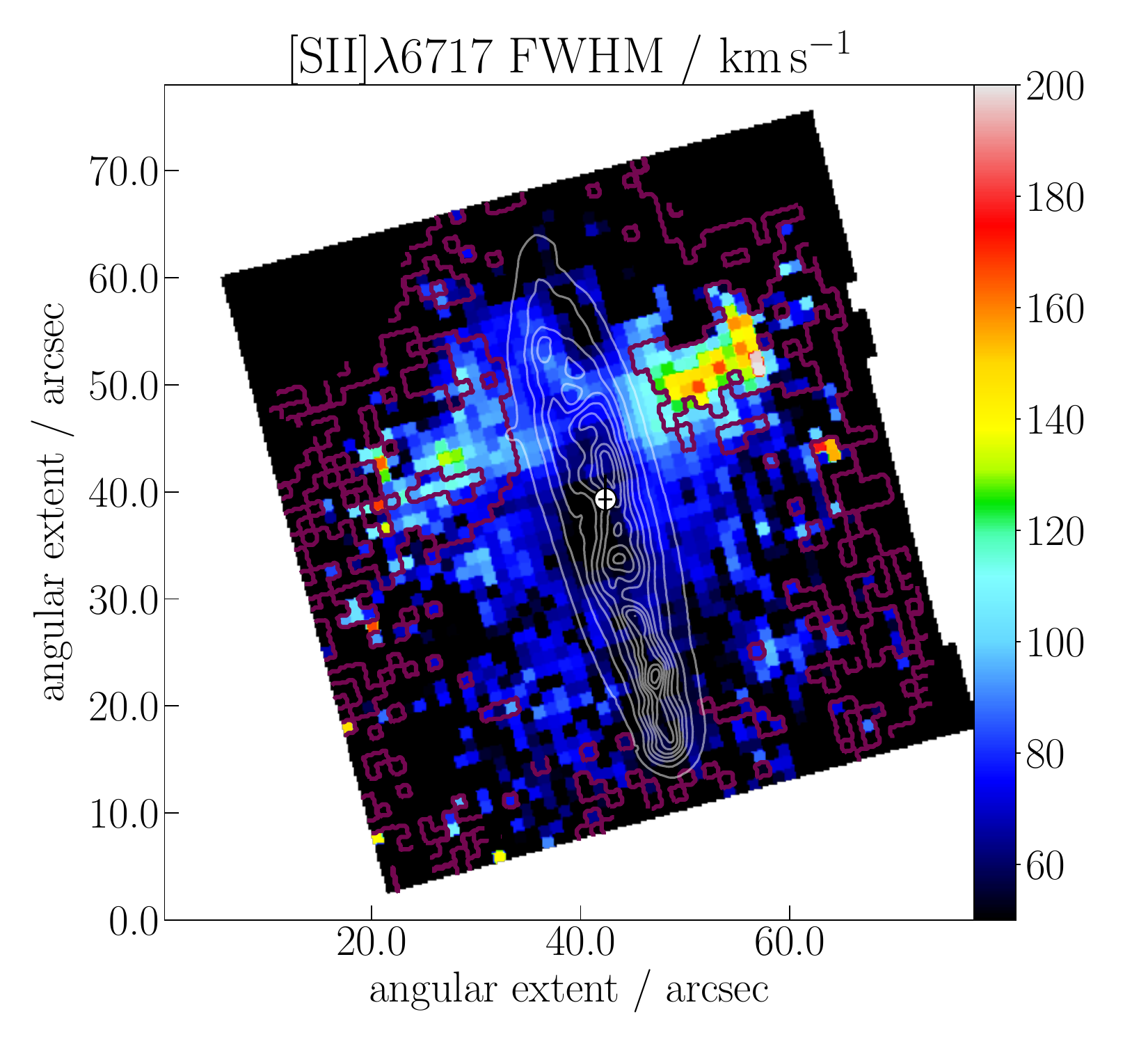}
    \end{subfigure}
    \begin{subfigure}{0.5\linewidth}
    \includegraphics[width=\hsize]{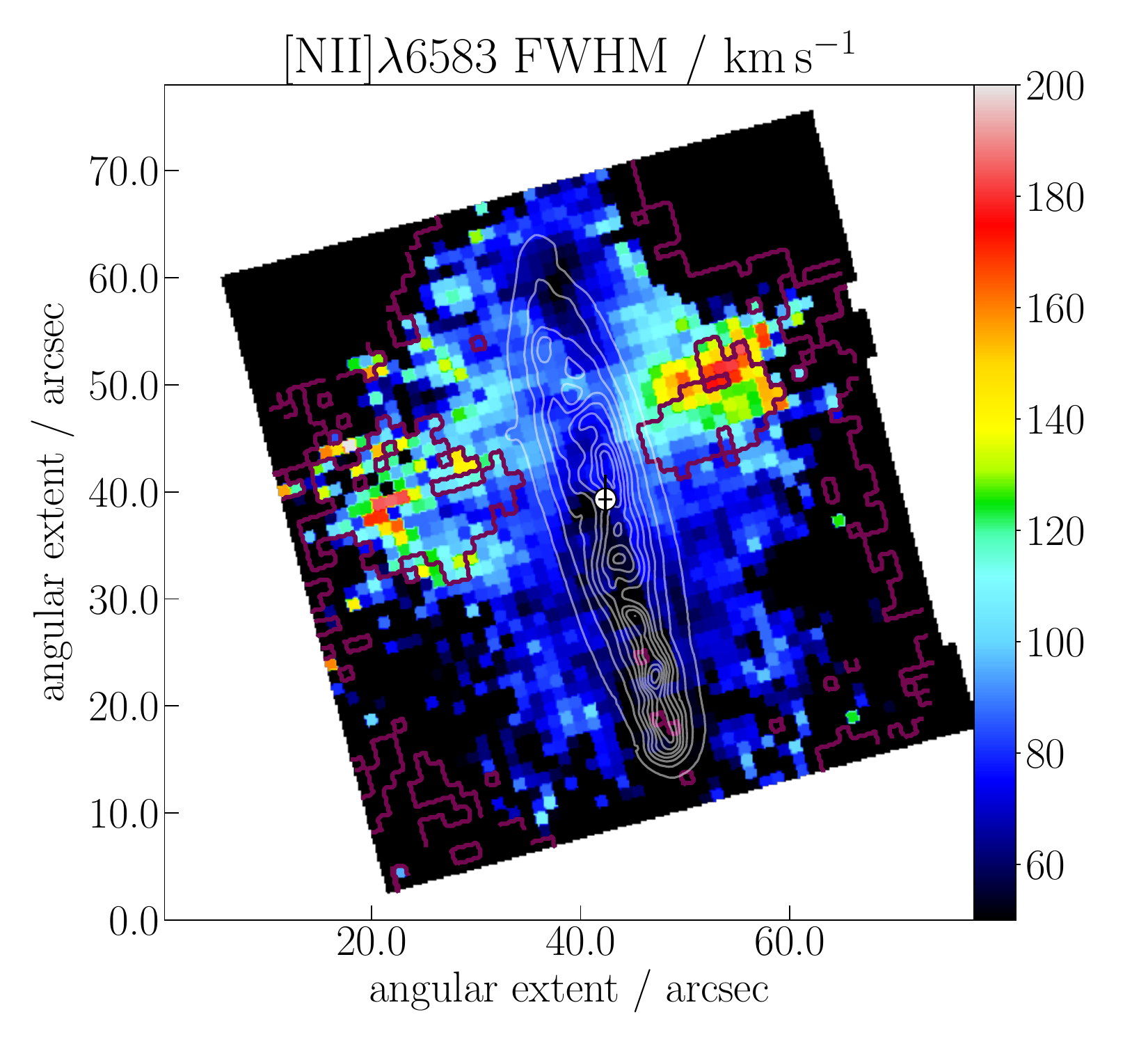}
    \end{subfigure}
    \caption{Maps of the emission line kinematics of the DIG in IC\,1553. \textit{Top left}: Map of the H$\alpha$ line-of-sight velocity in $\textrm{km}\,\textrm{s}^{-1}$ with contours from the continuum map. \textit{Top right}: Map of the H$\alpha$ FWHM in $\textrm{km}\,\textrm{s}^{-1}$ after correcting for LSF broadening with Eq. (\ref{eq:lsf_correct}). \textit{Bottom row:} FWHM maps measured on [\ion{S}{ii}]$\lambda 6717$ \textit{(left)} and [\ion{N}{ii}]$\lambda 6583$ \textit{(right),} LSF-corrected in the same way. The white contours are from the continuum map as before. The violet contours visualize the alignment with the respective emission line flux by separating bins based on the following flux thresholds relative to H$\alpha$: [\ion{S}{ii}]$\lambda$6717,6731/H$\alpha \gtrless 1.0$ \textit{(left)} and [\ion{N}{ii}]$\lambda 6583$/H$\alpha \gtrless 0.60$ \textit{(right)}.}
    \label{fig:ha_dynamics}
\end{figure*}
The top panels of Fig. \ref{fig:ha_dynamics} show maps of the line-of-sight component of the velocity, and the FWHM as a proxy for the velocity dispersion, measured on the H$\alpha$ emission line. The line-of-sight velocity map is consistent with the ones published by \citet{comeron19} and \cite{rautio22}, displaying observed rotation velocities up to $v_\mathrm{c} \sim 140\,\textrm{km\,s}^{-1}$ as well as a northward bend in lines of constant $v_\mathrm{c}$. In the velocity map as well as the H$\alpha$ flux map (right-hand panel in Fig. \ref{fig:ha_map}) we also see a slight warp of the star-forming disk at the northernmost edge of the FOV, but not in the continuum contours. The NASA/IPAC Extragalactic Database lists no potential companion in a 1\,deg radius, making recent interaction an unlikely explanation for this warp.

The FWHM map shows a fairly homogenous velocity dispersion across the disk of $\textrm{FWHM} = 60 - 70\,\textrm{km\,s}^{-1}$, with one exception: in the northern half at $d = 8\arcsec$ there appears to be a channel with $\textrm{FWHM} \approx 90\,\textrm{km\,s}^{-1}$ in H$\alpha$ connecting roughly conical volumes in the halo with significantly increased dispersion of $\textrm{FWHM} \geq 120\,\textrm{km\,s}^{-1}$. This picture strongly suggests the presence of an outflow, though the spectral resolution of $\gtrsim55\textrm{km\,s}^{-1}$ is not sufficient for a separation of kinematic components.

Outside of the volume of this suspected outflow, the velocity dispersion in the eDIG reaches elevated levels of $\textrm{FWHM} \sim 100-110\,\textrm{km\,s}^{-1}$ if it is measured on H$\alpha$, but only up to $\sim 80\,\textrm{km\,s}^{-1}$ if measured on [\ion{S}{ii}]$\lambda$6717 and [\ion{N}{ii}]$\lambda$6583, which are shown in the bottom panels of Fig. \ref{fig:ha_dynamics}. Aside from this difference, the linewidths of [\ion{S}{ii}]$\lambda$6717 and [\ion{N}{ii}]$\lambda$6583 are generally morphologically consistent with H$\alpha$. The most noticeable difference is that [\ion{S}{ii}]$\lambda$6717 exhibits a smaller FWHM in the region of the suspected outflow. Among the bins with $I_{\mathrm{H}\alpha} \geq 5\sigma$ the FWHM peaks at the position $(d,z) = (9.0\arcsec, -13.9\arcsec)$ with the maximum $\mathrm{FWHM}=165\,\textrm{km\,s}^{-1}$ in [\ion{S}{ii}] and $\mathrm{FWHM}=178\,\textrm{km\,s}^{-1}$ in [\ion{N}{ii}] and H$\alpha$.
The maximum FWHM approaches the expected shock velocity of $\sim 225$\,km\,s$^{-1}$ inferred from the ionization diagnostic diagrams (Fig. \ref{fig:bpt_schockmodels}). 

The bottom panels in Fig. \ref{fig:ha_dynamics} also show a violet contour line that encompasses bins of high [\ion{S}{ii}]/H$\alpha$ and [\ion{N}{ii}]/H$\alpha$ values, respectively, to compare the alignment of the line width- and flux-producing mechanisms. The corresponding thresholds for these contour lines are [\ion{S}{ii}]$\lambda$6717,6731 / H$\alpha \gtrless 1.0$ and [\ion{N}{ii}]$\lambda 6583$ / H$\alpha \gtrless 0.6$. While for [\ion{S}{ii}] the volume of highest relative flux coincides with the volume of highest FWHM, there appears to be a misalignment for [\ion{N}{ii}], where the volume of highest relative flux is centered a few arcseconds further south. This misalignment indicates that the [\ion{N}{ii}] emission traces a composite of photo- and shock ionization -- both mechanisms produce similar quantities of [\ion{N}{ii}] / H$\alpha$ in the models shown in Fig. \ref{fig:bpt_schockmodels} -- while [\ion{S}{ii}] primarily traces shock ionization in the DIG of IC\,1553.

Assuming that an off-centered outflow is responsible for some of the discussed kinematic and ionization properties of IC\,1553, we are interested in whether or not the escape velocity is reached; depending on the answer, the flow of matter can enrich primarily either the intergalactic medium or the eDIG. We therefore modeled the galactic gravitational potential with \texttt{galpy} \citep{bovy15} under the condition that it reproduces the rotation curve published by \cite{comeron19}. We have confirmed that the data processing in this study yields a consistent rotation curve (see Fig. \ref{fig:appendix_rotcurve} in Appendix \ref{appendixc}). The galactic potential is modeled with three components:
\begin{itemize}
    \item A Miyamoto-Nagai $(MN)$ potential component \citep{miyamoto75} for the disk with a minor-to-major axis ratio of $b/a=0.1$. For comparison, \cite{comeron18} measured radial scale lengths for the disk of IC\,1553 between $12.8\arcsec$ and $14.6\arcsec$ and vertical scale heights of $0.9\arcsec$ and $2.7\arcsec$ for the thin and thick disk, respectively.
    \item A Hernquist $(H)$ potential component \citep{hernquist90} for the bulge.
    \item A spherically symmetric Navarro-Frenk-White $(NFW)$ potential component \citep{navarro96} for the halo.
\end{itemize}
Under these conditions, each component requires one parameter describing the scale radius of the potential $(a_{disk},a_{bulge}, a_{halo})$ and one parameter for the amplitude given by the mass of the component. We describe the individual amplitudes in terms of relative weights $(\alpha ,\beta ,\gamma)$, so that the superposed potential is given as 
\begin{equation}\label{eq:potfitting}
    \Phi = A \left( \alpha *MN(a_{\mathrm{disk}}) + \beta *H(a_{\mathrm{bulge}}) + \gamma *NFW(a_{\mathrm{halo}})\right),
\end{equation}
where A gives the absolute mass scale of the amplitude.

\renewcommand{\arraystretch}{1.5}
\begin{table*}
    \caption{Parameter grid used for the grid search during the gravitational potential fitting according to Eq. (\ref{eq:potfitting}).}
    \label{tab:gridsearch}
    \centering
    \begin{tabular}{c c c c c c c c c}
    \hline \hline 
         & $A$ / $10^9$\,M$_\odot$ & $a_{disk}$ / 30\arcsec & $a_{bulge}$ / 30\arcsec & $a_{halo}$ / 30\arcsec & $\alpha$ & $\beta$ & $\gamma$ \\
         \hline
         minimum value & 15 & 0.05 & 0.05 & 0.05 & 0 & 0 & 0 \\
         maximum value & 151 & 1 & 1 & 1 & 1 & 0.2 & 1 \\
         step& 2 & 0.05 & 0.05 & 0.05 & 0.05 & 0.01 & 0.05 \\
         best fit& 67 & 0.7 & 1 & 1 & 0.85 & 0.15 & 0.1 \\
         \hline
    \end{tabular}
\end{table*}
We constrain the seven free parameters with a grid search over the parameter grid given in Table \ref{tab:gridsearch}. The scanned ranges of $A$ and $\beta$ are informed by \cite{comeron18} giving a total mass of the luminous matter in IC\,1553 of around $15-16 \times 10^9\,\textrm{M}_\odot$, and a mass ratio between the bulge (or central matter concentration) and disk of around $\sim 0.05$. The other parameter ranges and increments were chosen to facilitate reasonable computation times.
The grid search was conducted in two steps. First, we produced potentials for all combinations of the six shape-determining parameters (all except the amplitude $A$) and normalized the resulting rotation curve so that it reaches a value of $v=1$ at the end of the sampled radius interval. The similarity to the rotation curve model of \cite{comeron19} sampled to the observed extent and normalized in the same way was then quantified with the $R^2$ score function for regression. For the second step, all potential models with scores of $R^2 < 0.99$ were discarded. For each remaining model, we obtained the best-fitting amplitude from the physical velocity scale of the rotation curve, again using the $R^2$ score and discarding parameter combinations with $R^2 < 0.99$, leaving $612499$ combinations for further evaluation. While these models reproduce the observed rotation curve quite well, they suggest an unusually high mass fraction of the disk component of more than 50\%. This property and its influence on the escape velocity estimate is further discussed in Sect. \ref{sect:potfit_discussion}.

\begin{figure}
    \centering
    \includegraphics[width=0.5\textwidth]{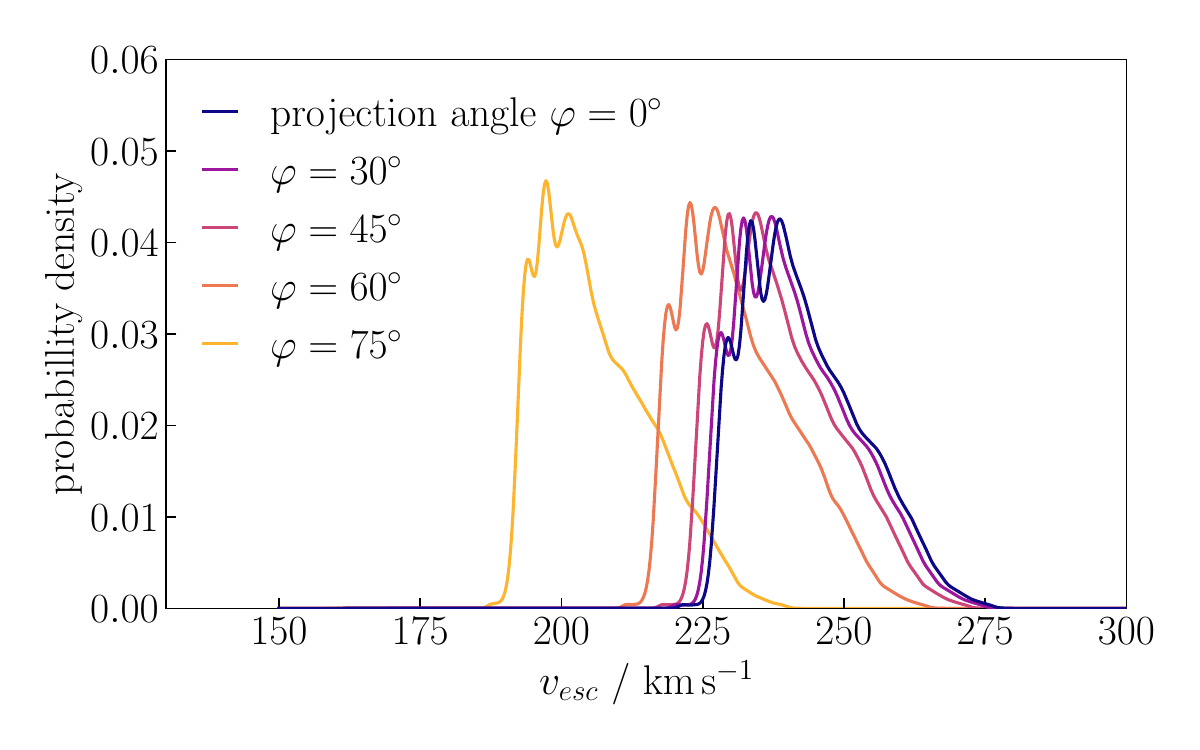}
    \caption{Gaussian kernel density estimates of the escape velocity, $v_{\mathrm{esc}}$, at the position of the spaxel with the highest H$\alpha$ FWHM derived from fitting gravitational potentials to the rotation curve of IC\,1553. $v_{\mathrm{esc}}$ was modeled for multiple projection angles, $\varphi$, between the plane of the sky and the deprojected positional vector from the galactic center to this spaxel.}
    \label{fig:v_esc_pdfs}
\end{figure}

Since \texttt{galpy} supports the evaluation of escape velocities only in the midplane in its current version (v1.7), we obtained the escape velocity at the position $\Vec{x}_2 = (9.0\arcsec, -13.9\arcsec)$, where the highest FWHM is observed in the data, from the relation
\begin{equation}
    v_{\mathrm{esc}}(\Vec{x}_2) = \sqrt{v_{\mathrm{esc}}^2(\Vec{x}_1)+2\left( \Phi (\Vec{x}_1) - \Phi (\Vec{x}_2) \right)}
\end{equation}
using the known quantities at the corresponding position $\Vec{x}_1 = (9.0\arcsec,0\arcsec)$ within the midplane. Assuming the projected radial distance of these positions is subject to projection effects, the actual radial distance will be $r=9.0\arcsec / \cos{\varphi}$ for a projection angle, $\varphi$, between the radial vector and the plane of the sky. $v_{\mathrm{esc}}(\Vec{x}_2)$ was calculated for multiple projection angles up to $\varphi = 75^\circ$, at which point the deprojected radial distance approaches the radial extent of the disk of IC\,1553. Finally, the resulting $v_{\mathrm{esc}}$ populations were translated into probability density distributions using a Gaussian kernel density estimate; these are shown in Fig. \ref{fig:v_esc_pdfs}.

The models yield lower limits on the escape velocity in the range $190 \leq v_{\mathrm{esc}} / \textrm{km\,s}^{-1} \leq 225$ depending on the projection angle, $\varphi$, which is exactly the velocity range between the maximum measured FWHM in the region of the outflow and the shock velocity inferred from the best-fitting shock models in the ionization diagnostic diagrams (Fig. \ref{fig:bpt_schockmodels}). The probability density distribution of $v_{\mathrm{esc}}$ only peaks below $225\,\textrm{km\,s}^{-1}$ for high projection angles $\varphi > 45^\circ$ and extends up to $275\,\textrm{km\,s}^{-1}$. Assuming that the highest velocity is reached within the sampled volume and neglecting drag forces from the high-altitude halo gas, it is therefore unclear whether the bulk of the warm ionized phase of the supposed outflow reaches the escape velocity or not, according to our gravitational potential models. 
This result is generally consistent with other studies concerning kinematics of ionized gas produced by stellar feedback \citep[e.g.,][]{Martin1998, Bomans2014}.


\section{Discussion}
\subsection{eDIG scale heights}\label{sect:shdiscussion}
In Sect. \ref{sect:Ha_dist} we derived H$\alpha$ scale heights as a function of the projected radial position for two single-component exponential models representing the eDIG and the ionized ISM confined to the disk. We fit the model parameters to both the $I_{\ion{H}{ii}}$ and $\log \left( I_{\ion{H}{ii}} \right)$ profiles to give more weight to the emission from low- and high-altitude gas respectively. 

The resulting eDIG scale heights range from $2.9\arcsec \equiv 0.51\,\textrm{kpc}$ to $7.0\arcsec \equiv 1.24\,\textrm{kpc}$ within the central $20\arcsec$ of projected radial distance to the dynamical center of IC\,1553, which is generally comparable to eDIG scale heights obtained in other studies \citep[e.g.,][]{levy19}. The southern tip of the disk at $d \leq -20\arcsec$ might host larger scale heights but the fit in this region is not reliable. \cite{rautio22} report the averaged scale height $h_{\textrm{eDIG}} = 1.39\,\textrm{kpc}$ for IC\,1553, which falls slightly above the highest values we find. \cite{rautio22} obtained this value from narrowband H$\alpha$ imaging extending to an altitude of 6\,kpc, while we sampled the MUSE data out to $\sim$ 3.5\,kpc. The biggest methodological difference is that we fit single-component exponential functions to the eDIG H$\alpha$ flux at $|z| \geq 8\arcsec \equiv 1.4\,\textrm{kpc}$ whereas \cite{rautio22} use a two-component model for data at $|z| \geq 0.25\,\textrm{kpc}$. It is therefore possible that this discrepancy is caused by a bias in our results from the tail of the disk component, though we see no evidence for this within the sampled $z$-range (see Fig. \ref{fig:shfitting_full} in Appendix \ref{appendixc}).

The two weighting schemes produce consistent eDIG scale heights in the northern half of IC\,1553 but diverge above the prominent star-forming clusters in the south, where the logarithmic weighting results in scale heights up to $\sim 50\%$ larger than when using the linear weighting. This indicates that the eDIG above the active star formation produces more emission at high altitudes than expected from a single-component exponential behavior. Since \cite{rautio22} sample almost twice the $z$-range, it is possible that their results are dominated by this more extended component that we only detect above the star-forming clusters, which could also cause the discrepancy. This would indicate that the profile of the eDIG halo at larger heights may be dominated by the same process that shapes the eDIG profile above the star-forming region on the southern side of the galaxy at lower heights.

If the emitting ionized gas is optically thin, then the H$\alpha$ flux in each bin is given in Rayleighs by the line-of-sight integral \citep[e.g.,][]{reynolds91two, dettmar92one}
\begin{equation}\label{eq:fluxequation}
    I_{\mathrm{H}\alpha}(d,z) = \frac{0.36\,\textrm{R}}{\textrm{pc}}   \int^{\infty}_{0} \left(\frac{n_{\mathrm{e}}(s)}{\textrm{cm}^{-3}}\right)^2~ \left(\frac{T_{\mathrm{e}}(s)}{10^{4}\,K}\right)^{-0.9}~ \diff s.
\end{equation}
The increased emission at high-altitude therefore corresponds to regions along the line-of-sight that are denser and/or cooler compared to expectations from the innermost disk-halo interface. Since this extended emission coincides with the actively star-forming regions in IC\,1553, gas transport from feedback processes may play a role here. Additionally, \cite{rautio22} suggest that the galaxy is currently accreting gas from a diffuse source based on the kinematics of the ionized gas, which could also cause overdensities at large $z$.

Effects from a relatively low gas temperature at higher altitudes are also interesting to consider, given that other studies have found that the observations instead require additional heating mechanisms beyond ionizing radiation from planar OB stars for the eDIG \citep[e.g.,][]{reynolds92, reynolds99, collins01, flores11}. This additional heating is usually invoked to explain the near constancy of [\ion{S}{ii}]/[\ion{N}{ii}] and the rising trend of [\ion{O}{iii}]/H$\beta$ (or equivalently [\ion{O}{iii}]/H$\alpha$ assuming an approximately constant Balmer decrement) with increasing altitude. However, neither of these trends is observed in the southern eDIG of IC\,1553 (compare with Fig. \ref{fig:elineratios_maps}). If gas accretion is indeed relevant here, lower temperatures at larger heights may be indicative of significant mixing of the eDIG with a colder phase of the accreted material.

\subsection{Ionization structure}\label{sect:discussion_ionization}
The most intriguing property of IC\,1553 related to the ionization state of its eDIG is certainly the [\ion{N}{ii}]$\lambda$6583 emission. The bins with the highest flux relative to H$\alpha$ are not directly spatially related with the potential sources of ionization and instead form a slightly off-centered conical structure. The usually observed DIG property that [\ion{S}{ii}]$\lambda$6717/[\ion{N}{ii}]$\lambda$6583 remains roughly constant with increasing altitude only holds for this conical volume. A rising trend of this ratio with increasing altitude -- like we see outside of this cone -- has been observed before \citep{golla96}, but never before in tandem with the usual constancy within the same galaxy. Outside of this conical region the gaseous halo of IC\,1553 exhibits a deficit of [\ion{N}{ii}]$\lambda$6583 emission with respect to [\ion{S}{ii}]$\lambda$6717 compared to the radiation transfer models of \cite{wood04}. We note that most studies of the DIG were conducted with long-slit spectroscopy, so that in a case like IC\,1553 the analysis will yield dramatically different interpretations depending on where exactly the slit was placed. It might therefore be necessary to revisit the target galaxies of these previous studies using IFU data, to extract the full picture of the relevant physical processes.

This behavior can be explained in principle, if the primary ionization mechanism for the eDIG of IC\,1553 is actually shock ionization from stellar feedback. Shocks have been invoked before as a significant contributor to the maintenance of the ionization state in the DIG of dwarf galaxies \citep{martin97} and in the warm component of galactic winds \citep[e.g.,][]{sharp10, Rich2010}.
Following the \cite{allen08} models, slow shocks naturally produce a lower [\ion{N}{ii}]$\lambda$6583 flux relative to [\ion{S}{ii}]$\lambda$6717 when compared to photoionization (compare with Fig. \ref{fig:bpt_schockmodels}), assuming \cite{dopita05} abundances, resulting in the observed deficit. The observed rising trend of this emission line ratio with altitude in the outskirts of the halo can then be expected from a mixing sequence between the predominant shock ionization in the halo and photoionization in the disk. The evident correlations between the diagnostic line ratios and the H$\alpha$ FWHM (see Fig. \ref{fig:ratio_width_correlation}), which is valid for the majority of observed bins, certainly supports this idea, and the ionization diagnostic diagrams are consistent with this scenario as well. It is unclear, however, how the observed morphology of the [\ion{N}{ii}]$\lambda$6583 emission would be produced, since the highest fluxes relative to H$\alpha$ occur closer to the galactic center than the origin of the outflow. 

\begin{figure*}
        \begin{subfigure}{.5\linewidth}
        \includegraphics[width=\hsize]{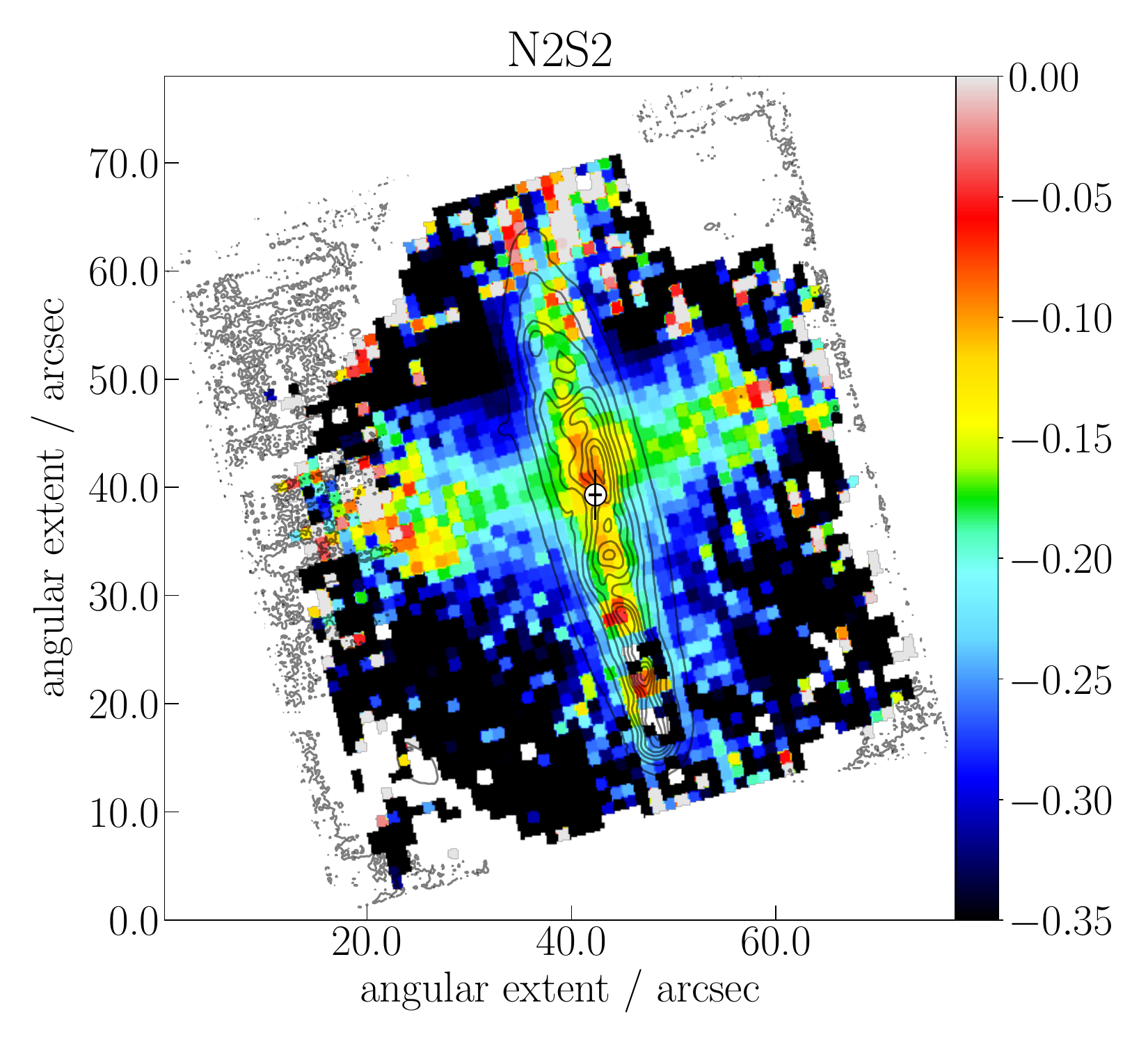}
        \end{subfigure}
        \begin{subfigure}{.5\linewidth}
        \includegraphics[width=\hsize]{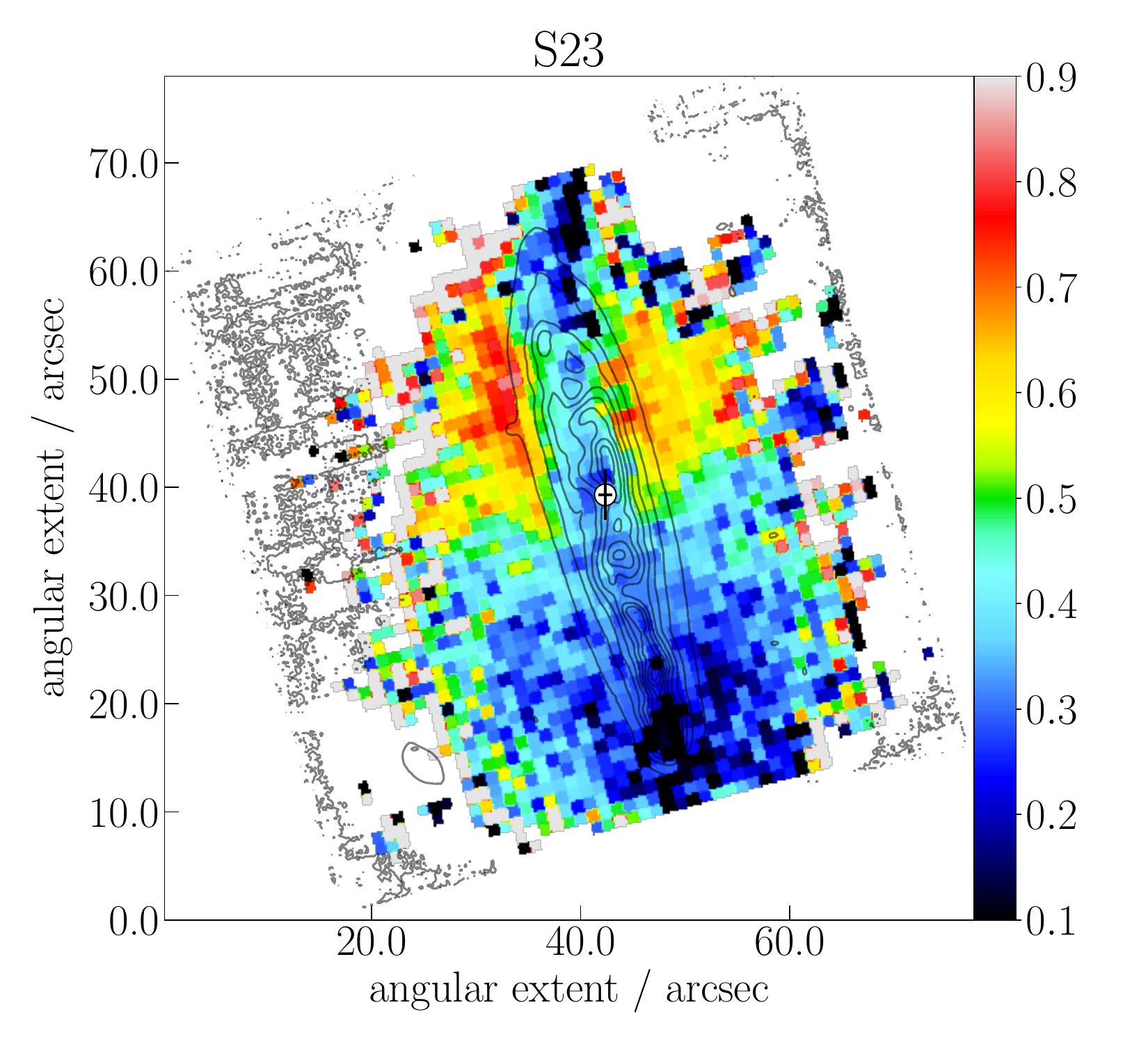}
        \end{subfigure}
    \caption{Maps of the metallicity diagnostic ratios 
$\log ([\ion{N}{ii}]\lambda 6584/\{[\ion{S}{ii}]\lambda$6717,6731\}) \textit{(N2S2, left)} and 
$\log$(\{[\ion{S}{ii}]$\lambda$6717,6731 + 3.44$\times$[\ion{S}{iii}]$\lambda$9069\} / H$\beta$) \textit{($S_{23}$, right)}.}
    \label{fig:metallicity_maps}
\end{figure*}

Alternatively, if photoionization from planar OB stars is responsible for the maintenance of the eDIG ionization state, one needs to explain the [\ion{N}{ii}] emission deficit outside of the conical structure and the [\ion{N}{ii}] morphology. The low [\ion{N}{ii}] flux can be the result of deficient nitrogen abundances in the eDIG or the ionization conditions favoring other ionization states for nitrogen. Fig. \ref{fig:metallicity_maps} shows maps of the abundance diagnostic ratios 
\begin{align}
    N2S2 &= \log \left( \frac{[\ion{N}{ii}]\lambda 6583}{[\ion{S}{ii}]\lambda6717,6731}\right) \hspace{5mm} \textrm{and}\\\label{eq:S_23}
    S_{23} &= \log \left( \frac{[\ion{S}{ii}]\lambda6717,6731 + [\ion{S}{iii}]\lambda9069,9531}{\mathrm{H}\beta}\right),
\end{align}
though as [\ion{S}{iii}]$\lambda$9531 lies outside the MUSE spectral range we assume its flux is approximately equal to $2.44\times$[\ion{S}{iii}]$\lambda$9069 from atomic physics \citep{perezmontero17}. 

The central conus has an $N2S2$ value similar to that of the thin disk, leading to a slim, cross-shaped structure with values in the range $N2S2 \approx -0.1$ to $-0.18$, with a few localized knot-like regions reaching $-0.05$. Using the calibrations of \cite{kewley02} with an oxygen abundance of $\log \mathrm{(O/H)}\approx 8.45$ (see Table \ref{tab:general_props}) this gives a lower limit on the ionization parameter in these regions of $q \geq 2 \times 10^7 \mathrm{cm\,s}^{-1}$. This cross-shape appears encased in a thicker layer of gas with $N2S2 \approx -0.2$ to $-0.3$, which would indicate a lower ionization parameter around $1 - 2 \times 10^7 \mathrm{cm\,s}^{-1}$. In this map IC\,1553 appears fairly symmetrical; the biggest difference between the northern and southern disk is that this thicker layer is slightly less extended in the north. Outside of this layer, $N2S2$ decreases further to values between $-0.3$ and $-0.4$, corresponding to a smooth further decrease of $q$ down to  $0.5 - 1 \times 10^7 \mathrm{cm\,s}^{-1}$.

In $S_{23}$, however, the eDIG of IC\,1553 exhibits a pronounced asymmetry. Outside of the thin disk $S_{23}$ appears to have an almost regular gradient along the $d$-axis from $S_{23} \approx 0.2$ in the south up to $ 0.75$ in the region of the northern outflow. Again using the \cite{kewley02} calibration, the ionization parameters inferred from the $N2S2$ map at the same oxygen abundance can only produce $S_{23} \leq 0.35$, with higher ratio values requiring much smaller ionization parameters. This is contradicting the inference from the $N2S2$ map for the central eDIG, where a large $q$ is required to explain the central conus, and produces a discrepancy in the required ionization parameter in the northern eDIG spanning orders of magnitude. The gas within the thin disk exhibits values of $S_{23} = 0.1$ to $0.4$, which translates to $q \leq 1.5 \times 10^{8}\,\mathrm{cm\,s}^{-1}$ and is consistent with the $N2S2$ map.

In summary, the $N2S2$ and $S_{23}$ maps together indicate that the state of the ionized gas in the plane of the disk and in the southern eDIG above the prominent star formation is consistent with the \cite{kewley02} photoionization models, while the diagnostics are contradictory in the central and northern eDIG. Consequently, photoionization from OB stars is unlikely to be the primary ionization mechanism outside of the southern eDIG.

\cite{rautio22} discuss HOLMES as a possible additional ionizing source. The corresponding model grids published by \cite{flores11} fit slightly better to the observed emission line fluxes in the [\ion{O}{iii}]/H$\beta$ vs. [\ion{N}{ii}]/H$\alpha$ diagram than the \cite{allen08} shock models, since they can reproduce the highest observed [\ion{N}{ii}]/H$\alpha$ values. The only other model grid presented by \cite{flores11} uses the [\ion{O}{ii}]$\lambda$3726,3729 lines, which are outside of the spectral range of MUSE for local galaxies.
While the HOLMES models do provide the best fit to the observed line ratio, it is particularly hard to differentiate sources of ionization in the [\ion{O}{iii}]/H$\beta$ vs. [\ion{N}{ii}]/H$\alpha$ diagram alone because the relevant model grids all fall relatively close to each other and partially overlap. Additionally, photoionization from HOLMES is not expected to produce the observed correlation between emission line ratios and the H$\alpha$ FWHM and the [\ion{N}{ii}]/H$\alpha$ and [\ion{S}{ii}]/[\ion{N}{ii}] morphologies are even more puzzling than in the shock ionization scenario, since HOLMES are expected to be fairly smoothly distributed throughout the thick disk and halo. Shocks, therefore, appear to be the more promising ionization mechanism in the northern and central eDIG.

\subsection{Models of the gravitational potential of IC\,1553}\label{sect:potfit_discussion}

\begin{figure*}
        \begin{subfigure}{.5\linewidth}
        \includegraphics[width=\hsize]{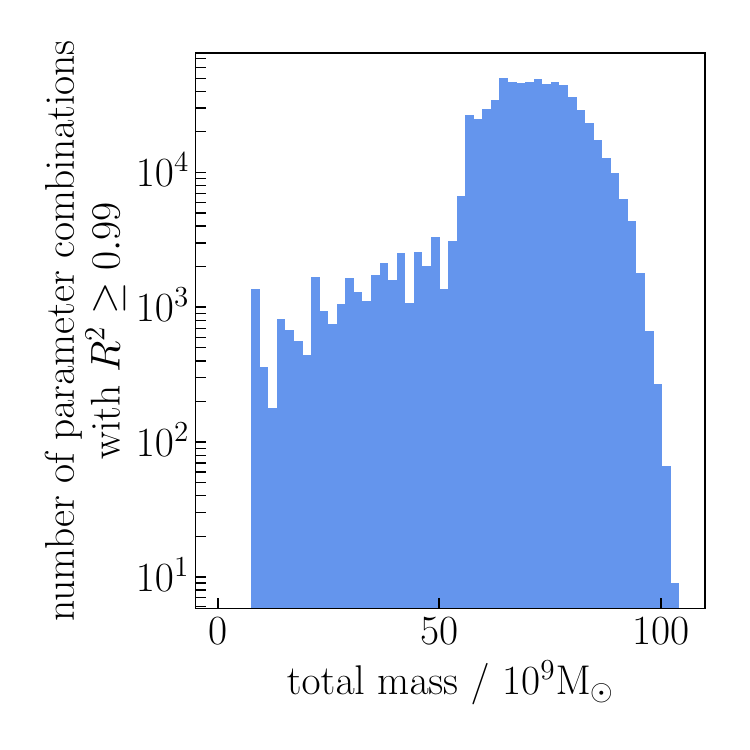}
        \end{subfigure}
        \begin{subfigure}{.5\linewidth}
        \includegraphics[width=\hsize]{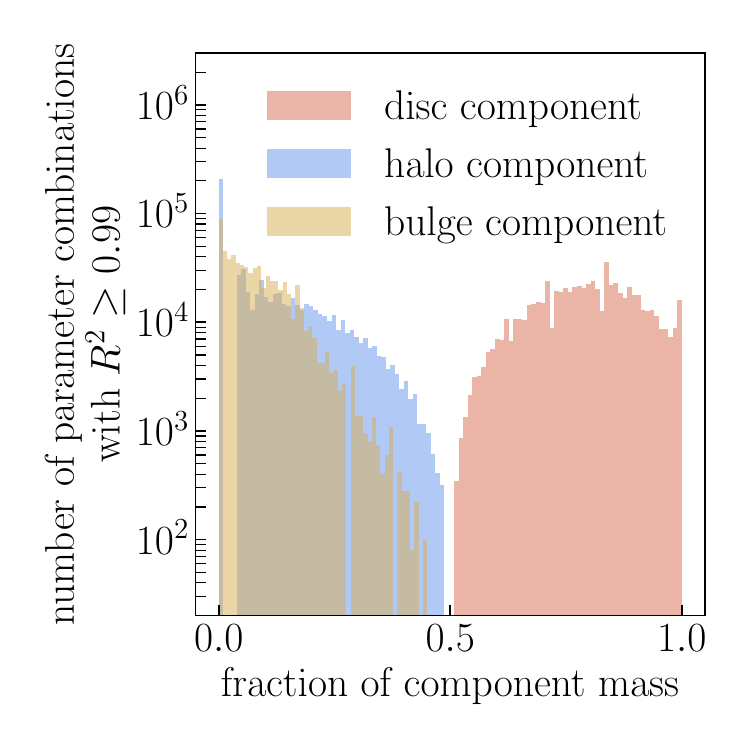}
        \end{subfigure}
    \caption{Histograms of the mass distribution properties in the model gravitational potentials. \textit{Left}: Distribution of the total mass in the potential models that reproduce the rotation curve from \cite{comeron19} for IC\,1553, as described in Sect. \ref{sect:kinematics} ($R^2 \geq 0.99$). \textit{Right}: Distribution of the component masses for the same set of gravitational potential models.}
    \label{fig:potfit_masses}
\end{figure*}

As described in Sect. \ref{sect:kinematics}, we assembled a set of model gravitational potentials, which reproduce the rotation curve of IC\,1553 measured by \cite{comeron19}. The first panel of Fig. \ref{fig:potfit_masses} shows a histogram of the resulting total masses in these model potentials. On average the models predict a total mass for IC\,1553 of $M_\mathrm{tot} = (70 \pm 12) \times 10^9\,\mathrm{M}_\odot$. If we assume its luminous mass is equal to $16.36 \times 10^9\,\mathrm{M}_\odot$ \citep[the sum of all component masses reported by][]{comeron18}, then this corresponds to a dark matter fraction of ($81 \pm 3$){\%}.

However, the potential models move the majority of the mass into the disk component, as can be seen in the right panel of Fig. \ref{fig:potfit_masses}. On average the halo component contains only $\sim 10\%$ and the disk component $\sim 80\%$ of the total mass. It is therefore evident that the total mass distribution in our model potentials is significantly more concentrated in the plane of the disk than would be expected. The average mass of the disk component in this set of models is $M_{\mathrm{disk}} = (55 \pm 8) \times 10^{9}\,\mathrm{M}_\odot$, which is more than three times as much mass as the $16.36 \times 10^9\,\mathrm{M}_\odot$ measured by \cite{comeron18}.
If the total mass estimated from these models can be trusted then the models may overestimate the escape velocities we present in Sect. \ref{sect:kinematics} as a result.

It is possible that this underestimation of the halo component is caused by the fact that the radial range covered by the MUSE data does not extend sufficiently to where the rotation curve plateaus, making it hard to constrain the properties of the halo component. 

\subsection{Evidence for an off-centered outflow}
The off-centered galactic outflow we find in IC\,1553 is certainly unusual. The energetic source of galactic outflows and winds discussed in the literature mostly falls into one of two categories: nuclear starbursts, typically in more massive disk galaxies, and global starbursts, often in dwarf galaxies and merging galaxies. Additionally, starburst regions at the end of an elongated stellar main body or bar-like structure appear in Magellanic dwarf galaxies (type IBm) like NGC\,2366 \citep{vanEymeren2007, vanEymeren2009a}, NGC\,4861 \citep{vanEymeren2007, vanEymeren2009b}, or Kiso\,5639 \citep{Elmegreen2016}.
Global starburst-like activity over the whole disk has also been observed in massive galaxies like NGC\,3079 \citep[e.g.,][]{Cecil2001, Li2013}, where the outflow is not confined to the famous starburst bubble in the core \citep{Cecil2001}. Another special case is the dwarf galaxy NGC\,1569, where a wind seems to be driven by the entire stellar disk \citep{martin02, veilleux05}. In NGC\,1569 a clear age gradient is apparent and the two older star clusters carved a large hole into the gas disk and drive a large scale outflow \citep{Heckman1995}, while the youngest
cluster is still embedded in its dusty, dense \ion{H}{ii} region \citep{clark13}. In more massive galaxies, localized star-forming regions driving large ($\sim 1$ kpc size) bubbles out of the disks are not uncommon, for example NGC\,55 \citep{Graham1982, Bomans1994} and other edge-on galaxies \citep[e.g.,][]{Rossa2003a}. The case of IC\,1553, however, may be unique in that it appears to have a large-scale outflow, with its base located far from the dynamical center of the host galaxy but not at the end of the disk. Depending on projection effects, though, the base of the outflow may indeed be located near the end of the disk of IC\,1553, similar to what has been observed in some dwarf galaxies.
 
While the spectral resolution of the MUSE data is not sufficient for us to identify multiple kinematic components in the nebular emission lines, the following indirect evidence for the presence of an outflow was found:
1) The H$\alpha$ FWHM, which is sensitive to the velocity of gas flows in the presence of a shock, shows a roughly biconical structure centered on a region of the disk approximately $1.4\,\textrm{kpc}$ away from the galactic center in projection. 2) The emission line ratios [\ion{S}{ii}]/H$\alpha$ and [\ion{O}{i}]/H$\alpha$ outside of the disk are consistent with shock ionization models, rather than photoionization models. [\ion{N}{ii}]/H$\alpha$ is also consistent with this scenario, but cannot distinguish between shock- and photoionization. 3) The shock-sensitive emission line ratios [\ion{N}{ii}]/H$\alpha$, [\ion{S}{ii}]/H$\alpha$ and [\ion{O}{i}]/H$\alpha$ are correlated with the H$\alpha$ FWHM.

\cite{rautio22} originally suspected that IC\,1553 hosts a large biconical outflow related to superbubbles. The observations listed above together make a strong case that this suspicion is indeed correct. This would make IC\,1553 the first known host of a localized, off-centered galactic outflow or wind.

\subsection{The disk-halo interface}\label{sect:dischalo_interface}

\begin{figure}[t]
    \includegraphics[width=0.5\textwidth]{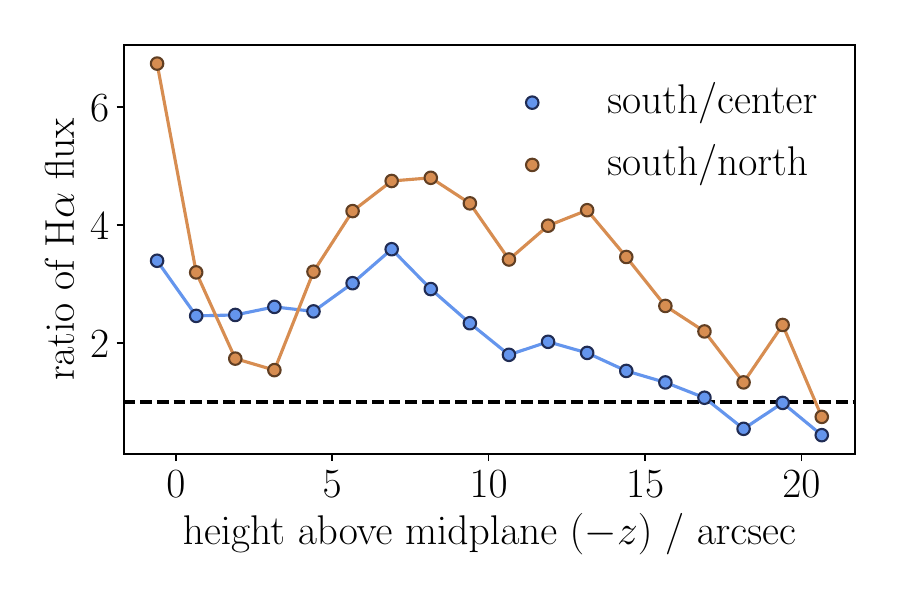}
        \caption{Ratios of the median H$\alpha$ fluxes between the southern, central, and northern fiducial regions of IC\,1553, as defined in Sect. \ref{sect:dischalo_interface}, as a function of the height above the midplane in the northwestern direction (where $z < 0$). The blue data points give the south-to-center flux ratio and the orange data points give the south-to-north ratio. The dashed line marks the ordinate value of $1$.}
        \label{fig:Ha_fluxratios}
\end{figure}

The brightest spot of the eDIG in H$\alpha$ at any given altitude in the inner halo is the region above the prominent star formation region in the south. As a rough quantification of this observation we selected fiducial intervals in the radial direction representing the southern ($-23\arcsec \leq d \leq -13\arcsec$), central ($-5\arcsec \leq d \leq 5\arcsec$), and northern ($10\arcsec \leq d \leq 20\arcsec$) regions of IC\,1553, and compared the median H$\alpha$ fluxes in these regions at different heights in Fig. \ref{fig:Ha_fluxratios}. We only considered bins in the northwestern direction above the midplane (where $z < 0$) because of the contamination from the background galaxy in the field of view.

In terms of this estimate, within the inner $10\arcsec$ of altitude the southern halo is around $\sim 2.5$ times as bright in H$\alpha$ as the central halo. The north-to-south comparison at low altitude ($-z \lesssim 5\arcsec$) is modulated by the warp in the H$\alpha$-disk of IC\,1553 at its northern edge and therefore does not represent the general trend. At higher altitudes the ratio has a maximum around the value $\sim 4$. In general, though, both ratios tend to steadily decrease to a value of $\lesssim 1$ at altitudes above $\sim 17\arcsec$. We note that at this height the south-to-center ratio is contaminated by the extraplanar \ion{H}{ii} region mentioned by \cite{rautio22}.

According to Eq. (\ref{eq:fluxequation}) the higher H$\alpha$ flux in the southern eDIG corresponds to a decreased electron temperature and/or increased electron density along the line of sight (assuming that scattered light from the disk is negligible). \cite{wood04} argued that the (electron) temperature of DIG is expected to increase with distance to the star formation because the radiation field hardens along its path. For this effect to be the sole cause of the asymmetric H$\alpha$ morphology, the average temperature in the northern eDIG (at $|z| \leq 10\arcsec$) has to be around $\sqrt[\leftroot{-2}\uproot{2}0.9]{4} \approx 4.7$ times as high as in the southern halo, and the temperature in the central halo has to be around $\sqrt[\leftroot{-2}\uproot{2}0.9]{2.5} \approx 2.8$ as high as in the south. Temperatures of the eDIG are not well constrained in the literature, but it is a common assumption that the bulk of the gas is in the temperature range 6000 -- 10000\,K \citep[e.g.,][]{haffner09}, which would not allow for the required temperature ratios. However, IC\,1553 is certainly unusual in the distribution of its star formation, so it is conceivable that it produces an unusual temperature structure in its eDIG.

Alternatively, the H$\alpha$ morphology can be caused by increased densities in the southern eDIG relative to the other regions and, since $I_{\mathrm{H}\alpha}$ scales with an average $n_\mathrm{e}^2$, the required difference in the density is significantly lower than it is for the temperature. Such differences in the density could be caused by star formation feedback. In this case, since the southern eDIG produces approximately four times as much H$\alpha$ flux as the northern eDIG up to the height of $z \approx 13\arcsec$ (see Fig. \ref{fig:Ha_fluxratios}), this would correspond to twice the average eDIG density. Interestingly, if all ionized gas in the eDIG originates in the disk, then this would imply that the gas transport process into the disk-halo interface must be greater in magnitude for the active star formation in the south than it is for the outflow in the north. Given that the emission line ratios indicate that the shock velocity is on the same order as the escape velocity in our models, it is possible that the bulk of the gas in the outflow is instead transported well into the halo or beyond. Additionally, \cite{rautio22} concluded that IC\,1553 is currently undergoing accretion of gas from a diffuse source, which could contribute to the density of the extraplanar ionized gas.

In practice, it is likely that both the density- and temperature structure produce the relatively enhanced H$\alpha$ emission in the southern halo synergetically. It is interesting to note that, while the highest fluxes are therefore linked to the current star formation in the plane of the disk, the ionization of the eDIG appears to be driven in large part by the off-centered outflow on the other side of the galaxy. In this sense, we do not find evidence that the disk-halo interface is dominated by a single process. Rather, it appears to be shaped by star formation on different timescales in different aspects. 


\section{Summary and Conclusions}
We analyzed the optical emission line spectrum of the extraplanar DIG in the edge-on star-forming galaxy IC\,1553 using archival IFU data from the MUSE spectrograph. We derived the H$\alpha$ scale height $h_{\mathrm{eDIG}}(d)$ as a function of the projected radial distance from its center and find that it decreases roughly linearly with $|d|$. The scale height values that we measure for IC\,1553 are generally lower than those found by \cite{rautio22}, which sampled the eDIG to higher altitudes. This discrepancy might indicate the presence of a more extended layer of warm ionized gas, which would appear to be connected to the active star formation in the southern half of IC\,1553. Alternatively, it may be caused by peculiarities in the density or temperature structure of the eDIG.

\cite{rautio22} proposed that IC\,1553 hosts a biconical outflow originating from a superbubble in the disk, which we can confirm. Our analysis suggests that the shock produced by this outflow is a significant, if not dominant, source of ionization for the eDIG. The bulk of the gas transport into the disk-halo interface, however, is most likely coupled to the active star formation in the southern disk rather than the outflow in the north.
The spatial separation between the base of this outflow and the current star formation leads to very different ionization conditions across different regions in the eDIG. 

We compiled a population of gravitational potential models that reproduce the observed rotation curve for IC\,1553 from \cite{comeron19} and predict an escape velocity in the range 190 -- 275\,km\,s$^{-1}$, which encompasses the best-fitting shock velocity of the outflow of 225\,km\,s$^{-1}$. However, while these models have reasonable total masses for IC\,1553, they predict a very large mass fraction of the disk component, which may cause the escape velocity to be overestimated.

Based on the results of this study we argue that, in a case like IC\,1553, in which the physical conditions of the eDIG change significantly along the galaxy's major axis, IFU spectroscopy is a necessity for an accurate analysis of the physical processes involved. The majority of studies concerned with the DIG in other galaxies, however, only had access to long-slit spectra, so it might be necessary to revisit the corresponding target galaxies with the current generation of IFU spectrographs if we wish to understand the role of the DIG in the context of galaxy evolution. 

\begin{acknowledgements}
We thank S\'{e}bastien Comer\'{o}n for his helpful comments during the review process, which improved this work significantly. RJD and DJB acknowledge funding from the German Science Foundation DFG, via the Collaborative Research Center SFB1491 "Cosmic Interacting Matters - From Source to Signal". PK acknowledges financial support by the German Federal Ministry of Education and Research (BMBF)  Verbundforschung grant 05A20PC4  (Verbundprojekt D-MeerKAT-II).
\end{acknowledgements}

\bibliographystyle{aa}
\bibliography{literatur}

\begin{appendix}
\FloatBarrier

\section{Contamination from a background galaxy}\label{appendix}
The extended source at the southeastern edge of the field of view visible in the $V$-band continuum and H$\alpha$ kinematic maps (Figs. \ref{fig:continuumimage} and \ref{fig:ha_dynamics}) was determined to be a background galaxy based on its spectroscopic redshift of $z=0.052$ corresponding to a distance of $\gtrsim 200\,\textrm{Mpc}$, whereas IC\,1553 has a distance of only $36.5\,\textrm{Mpc}$ \citep{tully16}. The [OI]$\lambda 6300$ emission line of this background galaxy is redshifted to the same wavelength as H$\alpha$ at the systemic velocity of IC\,1553, which leads to false detections in Fig. \ref{fig:ha_dynamics} appearing like a satellite or coherent cloud of ionized gas. We have masked the corresponding area in figures showing any diagnostic that uses H$\alpha$ throughout this paper as a result.

\begin{figure}[t]
    \includegraphics[width=0.5\textwidth]{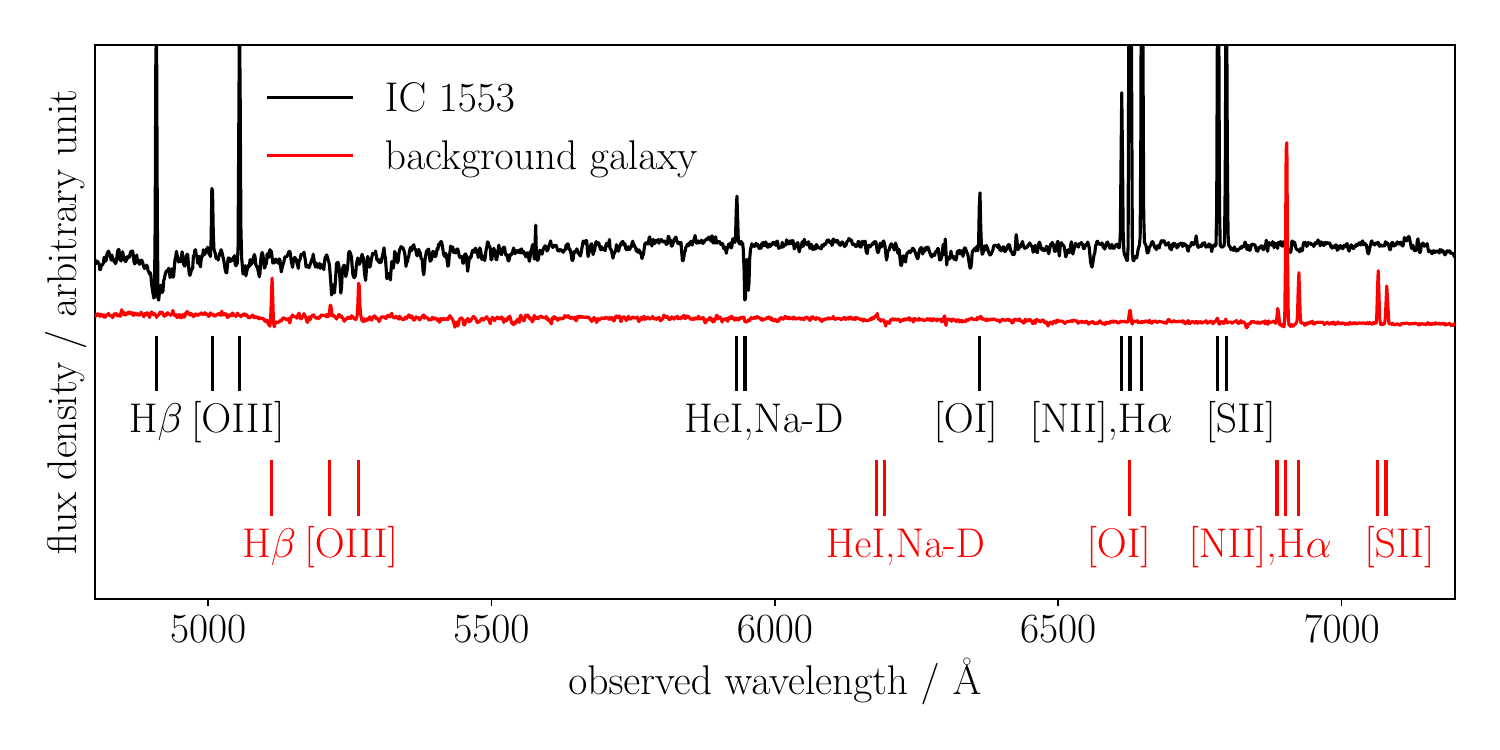}
        \caption{Comparison of example spectra of the central regions of IC\,1553 (black) and the background galaxy (red) extracted from the spectral cube.}
        \label{fig:appendix_spectrum}
\end{figure}

The background galaxy's spectrum contains typical nebular emission lines produced by star formation and has an estimated maximum rotational velocity of around 90 km/s. A comparison of example spectra from the central regions of IC\,1553 and this background galaxy is shown in Fig. \ref{fig:appendix_spectrum}.

\section{Calculation of the $\eta^{\prime}$-parameter}\label{appendix:eta}
The calculation of $\eta^{\prime}$ as described in Sect. \ref{sect:ionization} is largely analogous to the determination of $\eta$ in \cite{erroz-ferrer19}. The largest difference is that \cite{erroz-ferrer19} consider two demarcation lines (the empirical and theoretical curves from \cite{kauffmann03} and \cite{kewley01}) resulting in three separate regions in the [\ion{O}{iii}]/H$\beta$ vs. [\ion{N}{ii}]/H$\alpha$ diagram, while we only have the equivalent theoretical limit from \cite{kewley01} in the [\ion{O}{iii}]/H$\beta$ vs. [\ion{S}{ii}]/H$\alpha$ diagram.

Following the nomenclature in \cite{erroz-ferrer19}, the distance, $d_1$, between a bin's position in the ionization diagnostic diagram $\textrm{P}_1 = (x_1,y_1)$ and any point $(x,f(x))$ on the curve from the \cite{kewley01} starburst limit $f(x) = 0.72/(x-0.32)+1.30$ is
\begin{align}
    d_1(x) = \sqrt{(x_1 - x)^2 + \left[y_1 - \left(\frac{0.72}{x-0.32}+1.3\right)\right]^2}.
\end{align}
The minimal distance therefore corresponds to the solution of ${\partial}d_1(x)/{\partial}x = 0$, or equivalently to the solution of
\begin{multline}
    0 = 2x^4 -2x_1 x^3 -1.92 x^3 +1.92 x_1 x^2 +0.614 x^2 -0.614 x_1 x \\+ 1.44 y_1 x -1.942 x +0.066 x_1 -0.461 y_1 -0.436.
\end{multline}
Out of the two real roots of this polynomial, we selected the one that produces the smaller distance, $d_1(x_0)$. Following the definition of the $\eta$-parameter, we required that $\eta^{\prime} = 0.5$ if $P_1 \in f(x)$, so that
\begin{align}
    \eta^\prime = 
    \begin{cases}\label{eq:etaprime_def}
    0.5 - d_1(x_0), & \textrm{if } y_1 \leq f(x_1) \textrm{ and } x_1 < 0.32\\
    0.5 + d_1(x_0), & \textrm{else.}
    \end{cases}
\end{align}

\section{Additional figures}\label{appendixc}
\begin{figure}[t]
    \includegraphics[width=0.5\textwidth]{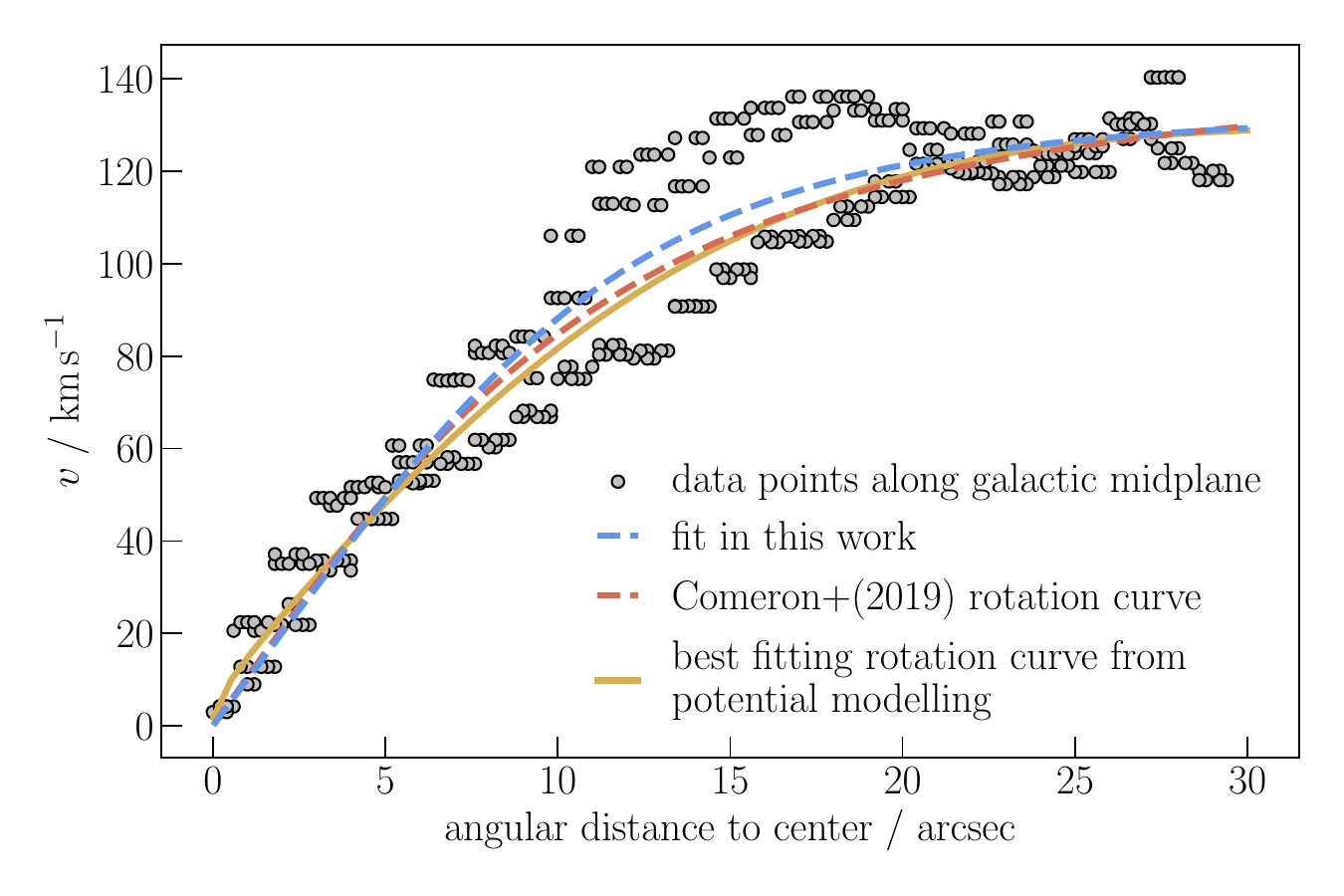}
        \caption{Line-of-sight velocity profile for all the bins intersecting the midplane of IC\,1553, measured on the H$\alpha$ emission line (\textit{gray circles}). The dashed lines mark fits of Eq. (\ref{eq:rotcurve}) to the data in this work (\textit{dashed blue line}) and from \cite{comeron19} (\textit{dashed red line}). The solid yellow curve marks the rotation curve from the best-fitting model of the gravitational potential obtained, as described in Sect. \ref{sect:kinematics}.}
        \label{fig:appendix_rotcurve}
\end{figure}

Figure \ref{fig:appendix_rotcurve} shows the line-of-sight velocity component for all bins in the midplane of IC\,1553 as a function of the projected angular distance to the dynamic center, measured on H$\alpha$. Following \cite{comeron19}, we show rotation curve fits to the expression \citep{courteau97}
\begin{align}\label{eq:rotcurve}
    v(\chi) = v_\mathrm{c} \frac{1}{(1+{\chi}^\gamma)^{1/\gamma}},
\end{align}
where $v_\mathrm{c}$ is the circular velocity, $\chi$ is the normalized axial coordinate, and $\gamma$ determines the sharpness of the transition from the inner increase in the rotation curve to its flattening further out. The curves shown in Fig. \ref{fig:appendix_rotcurve} include the best fit to the measured velocity profile in this work ($v_\mathrm{c} = 134\,\mathrm{km\,s}^{-1}, \chi = 13.3\arcsec, \gamma = 2.81$), the rotation curve obtained by \cite{comeron19} ($v_\mathrm{c} = 142\,\mathrm{km\,s}^{-1}, \chi = 13.5\arcsec, \gamma = 2.02$), and the rotation curve of the best-fitting gravitational potential model (see Sects. \ref{sect:kinematics} and \ref{sect:potfit_discussion}).

\begin{figure*}
        \begin{subfigure}{.24\linewidth}
        \includegraphics[width=\hsize]{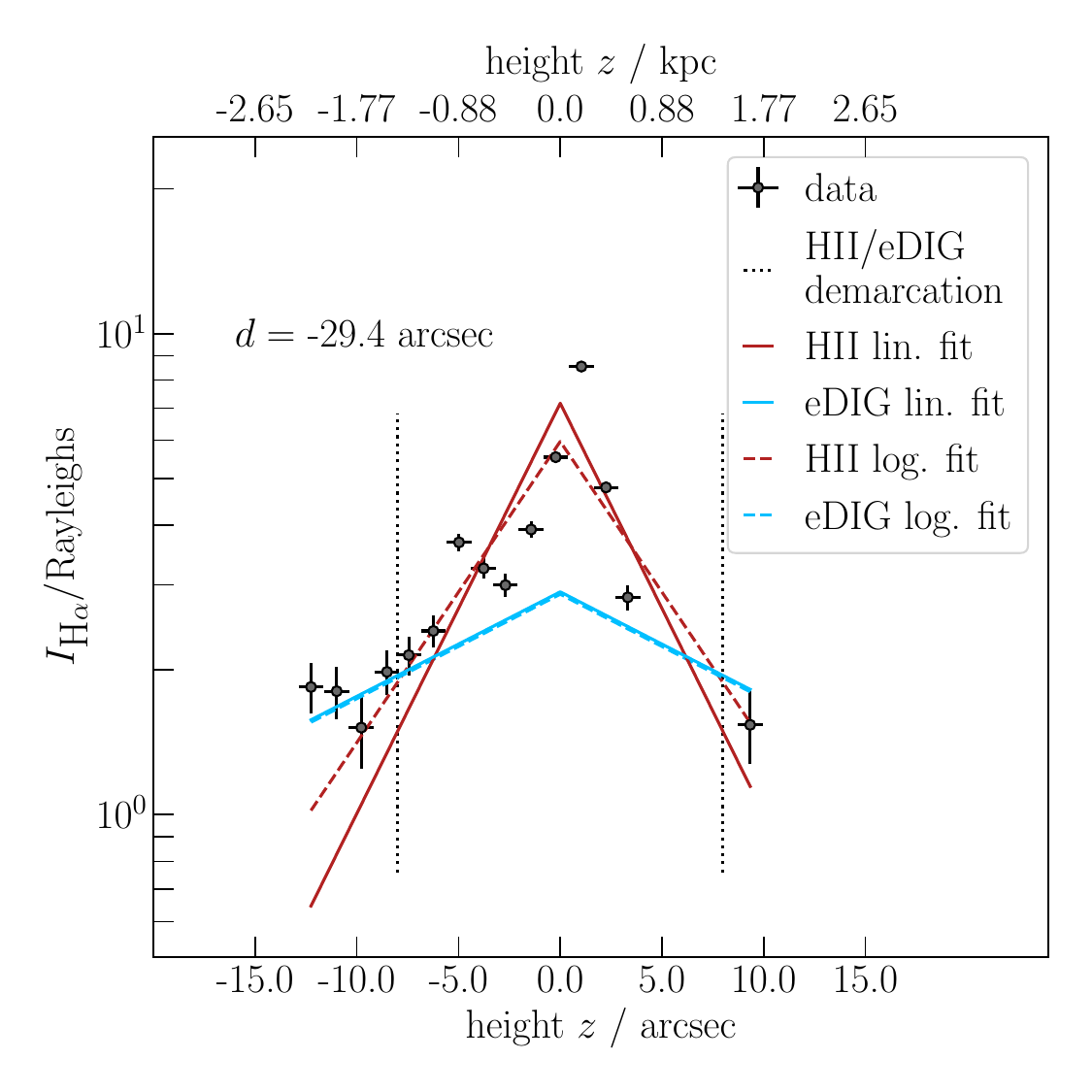}
    \end{subfigure}
        \begin{subfigure}{.24\linewidth}
        \includegraphics[width=\hsize]{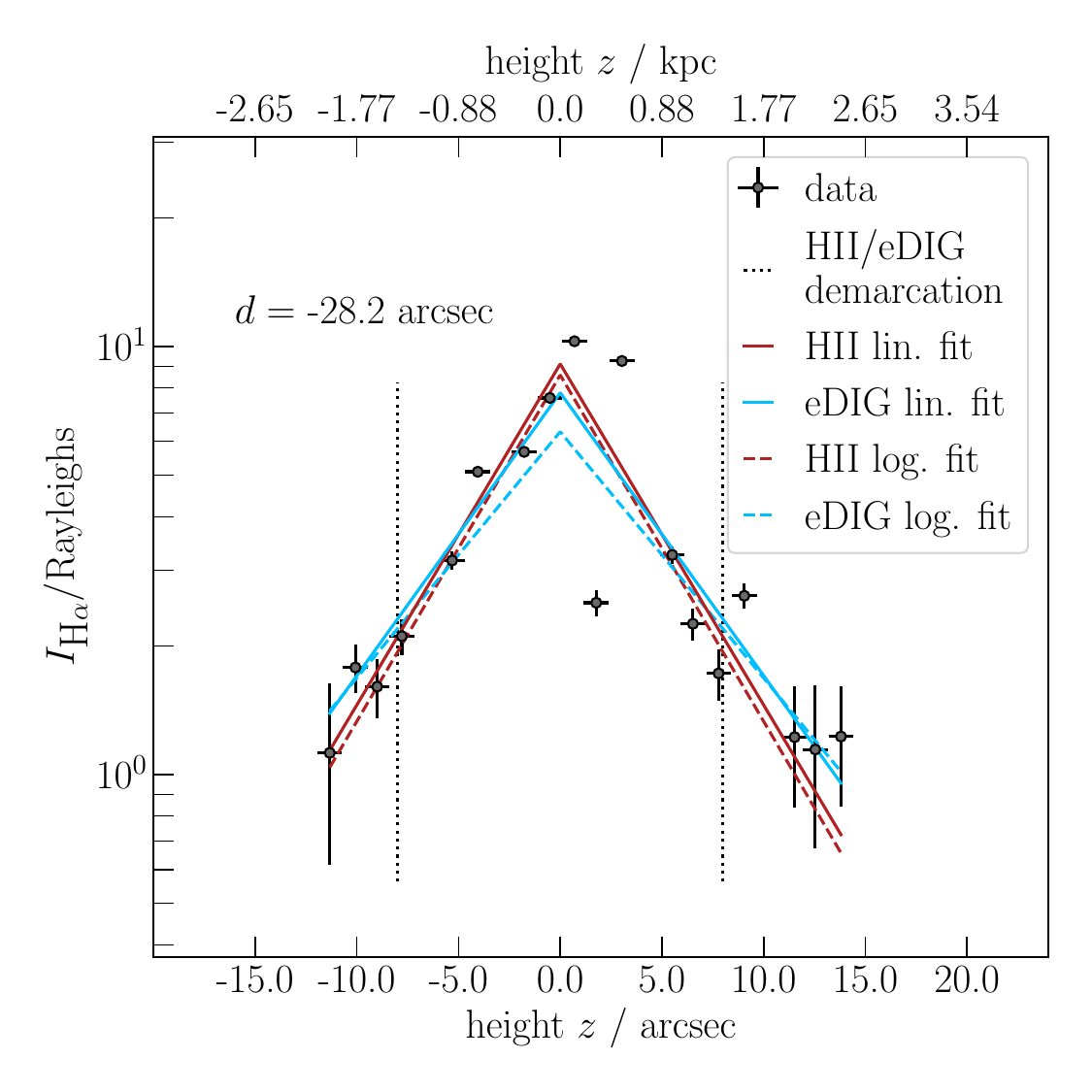}
    \end{subfigure}
        \begin{subfigure}{.24\linewidth}
        \includegraphics[width=\hsize]{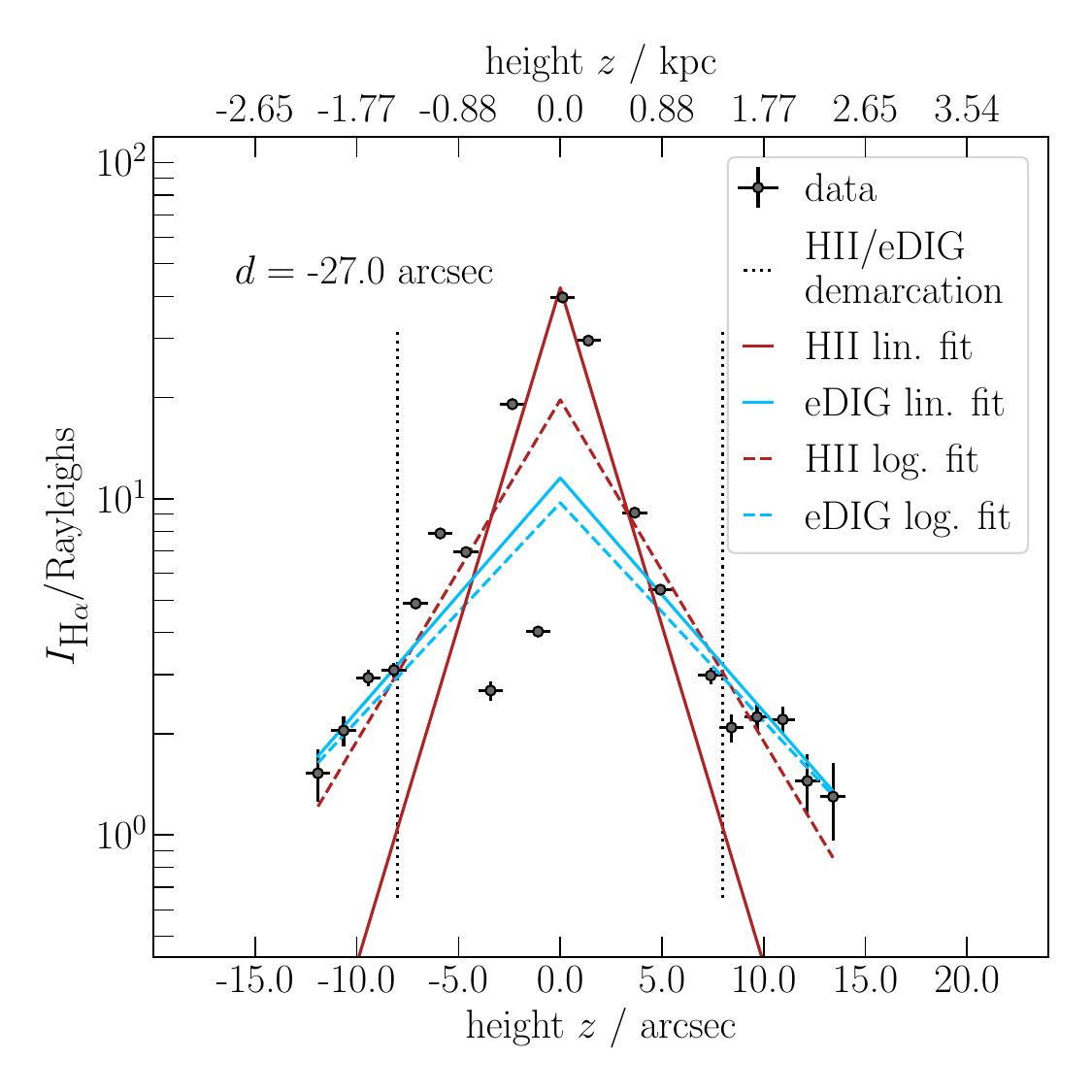}
    \end{subfigure}
        \begin{subfigure}{.24\linewidth}
        \includegraphics[width=\hsize]{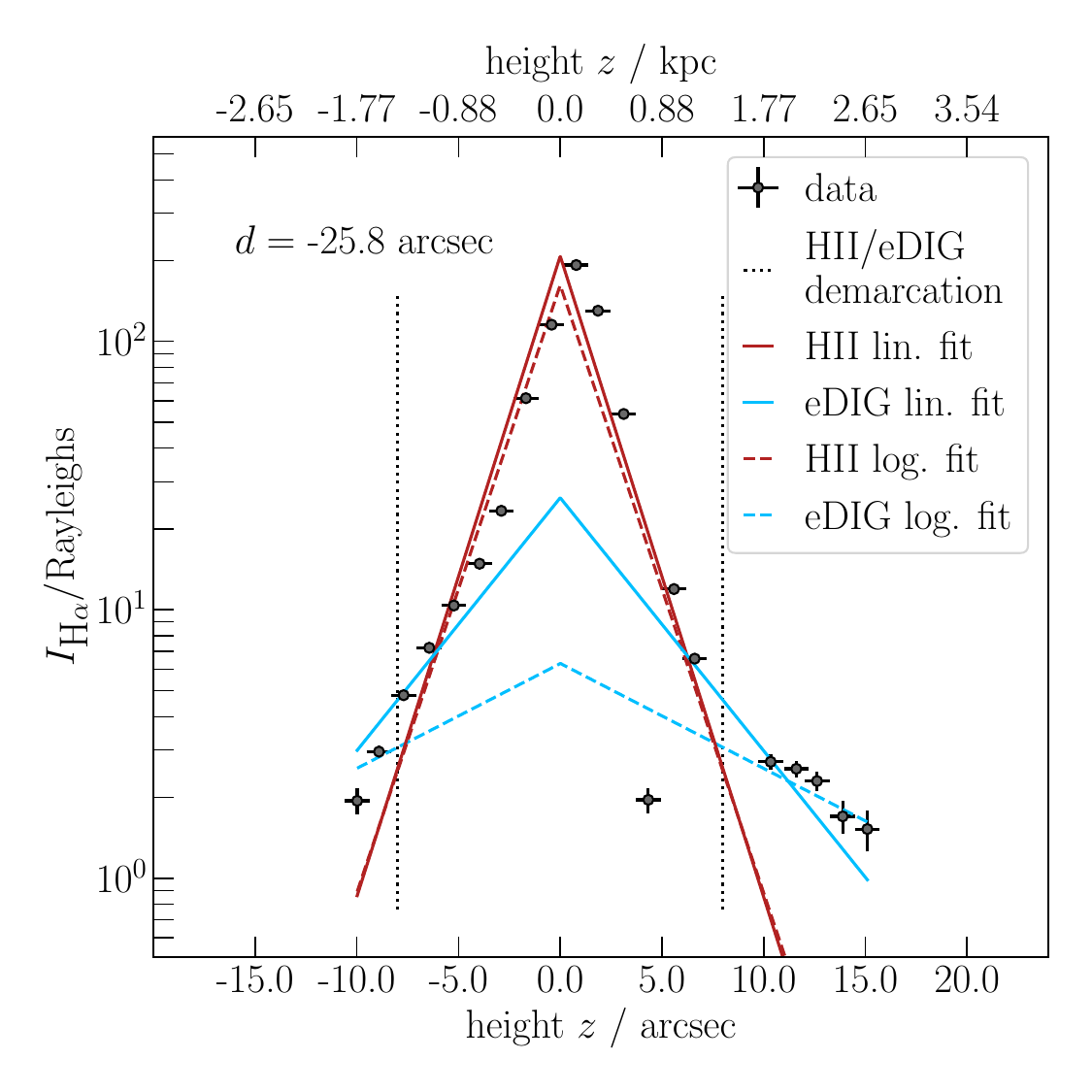}
    \end{subfigure}\\
        \begin{subfigure}{.24\linewidth}
        \includegraphics[width=\hsize]{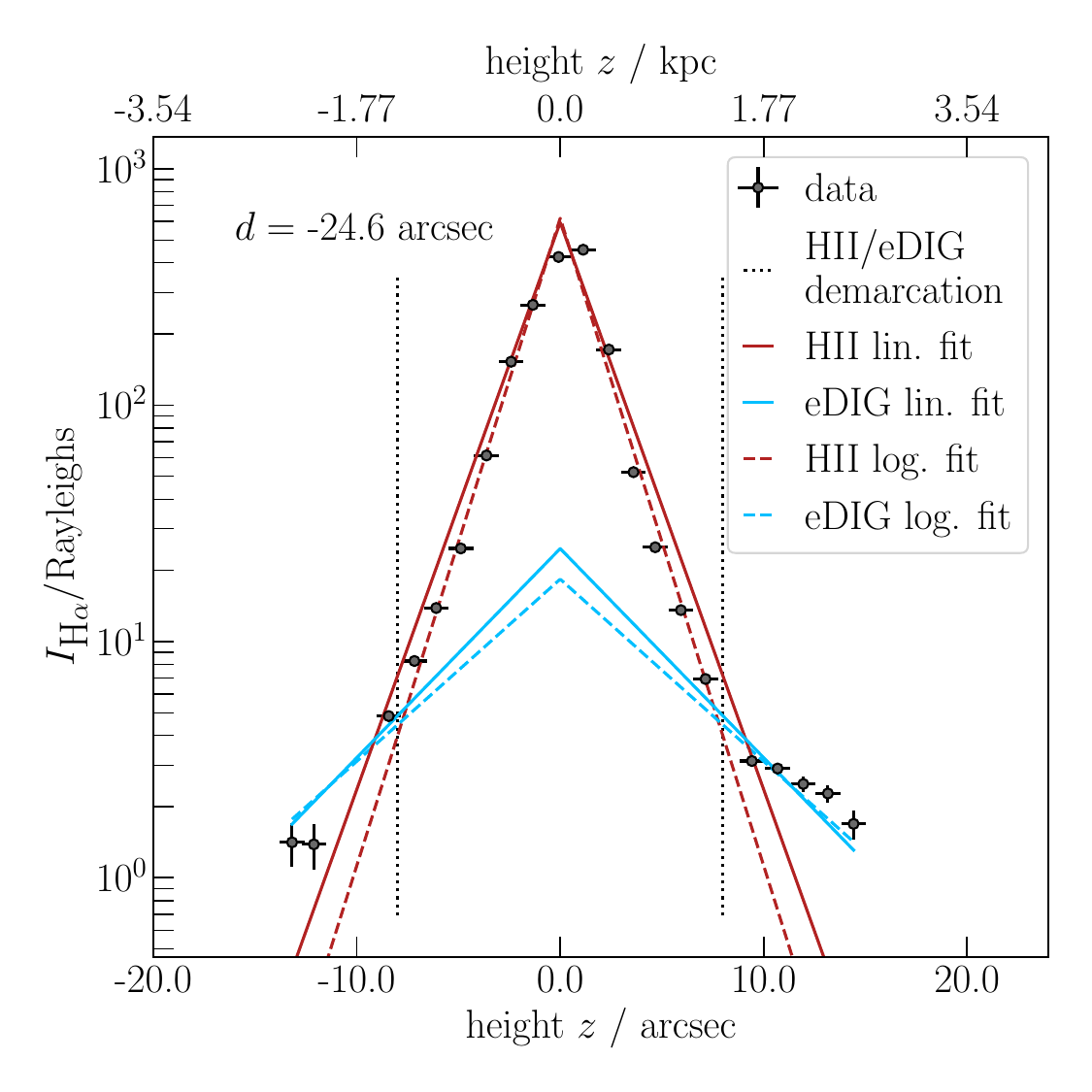}
    \end{subfigure}
        \begin{subfigure}{.24\linewidth}
        \includegraphics[width=\hsize]{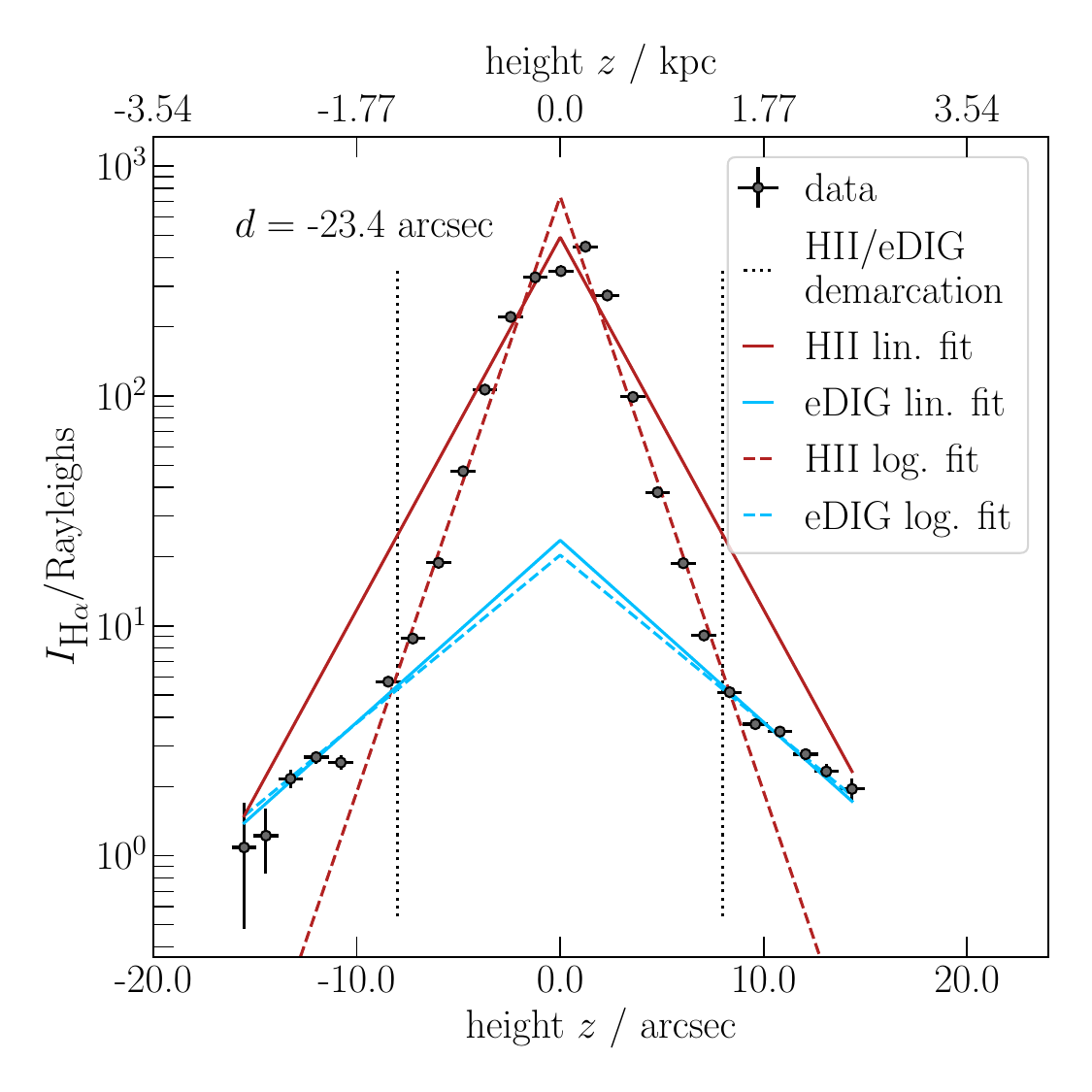}
    \end{subfigure}
        \begin{subfigure}{.24\linewidth}
        \includegraphics[width=\hsize]{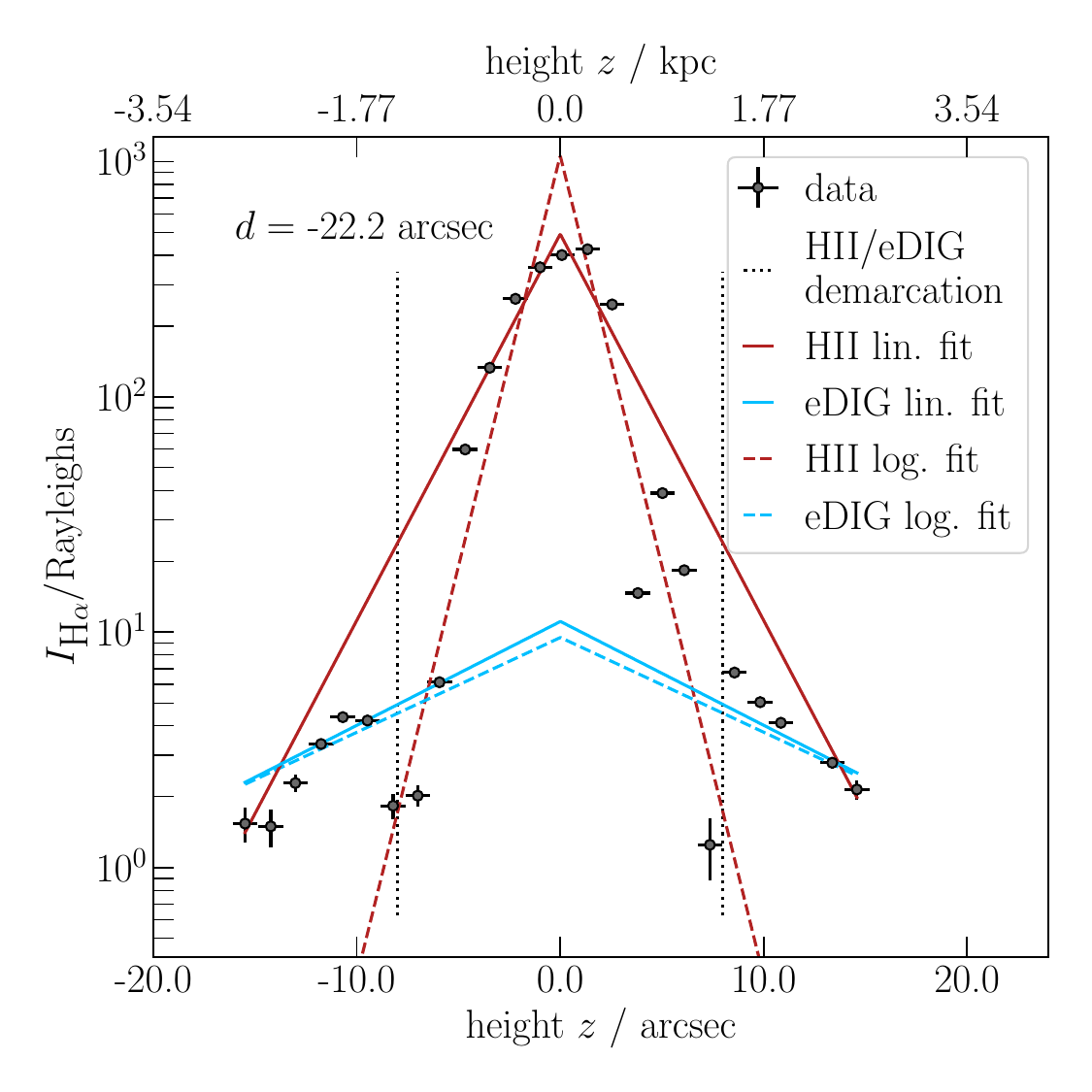}
    \end{subfigure}
        \begin{subfigure}{.24\linewidth}
        \includegraphics[width=\hsize]{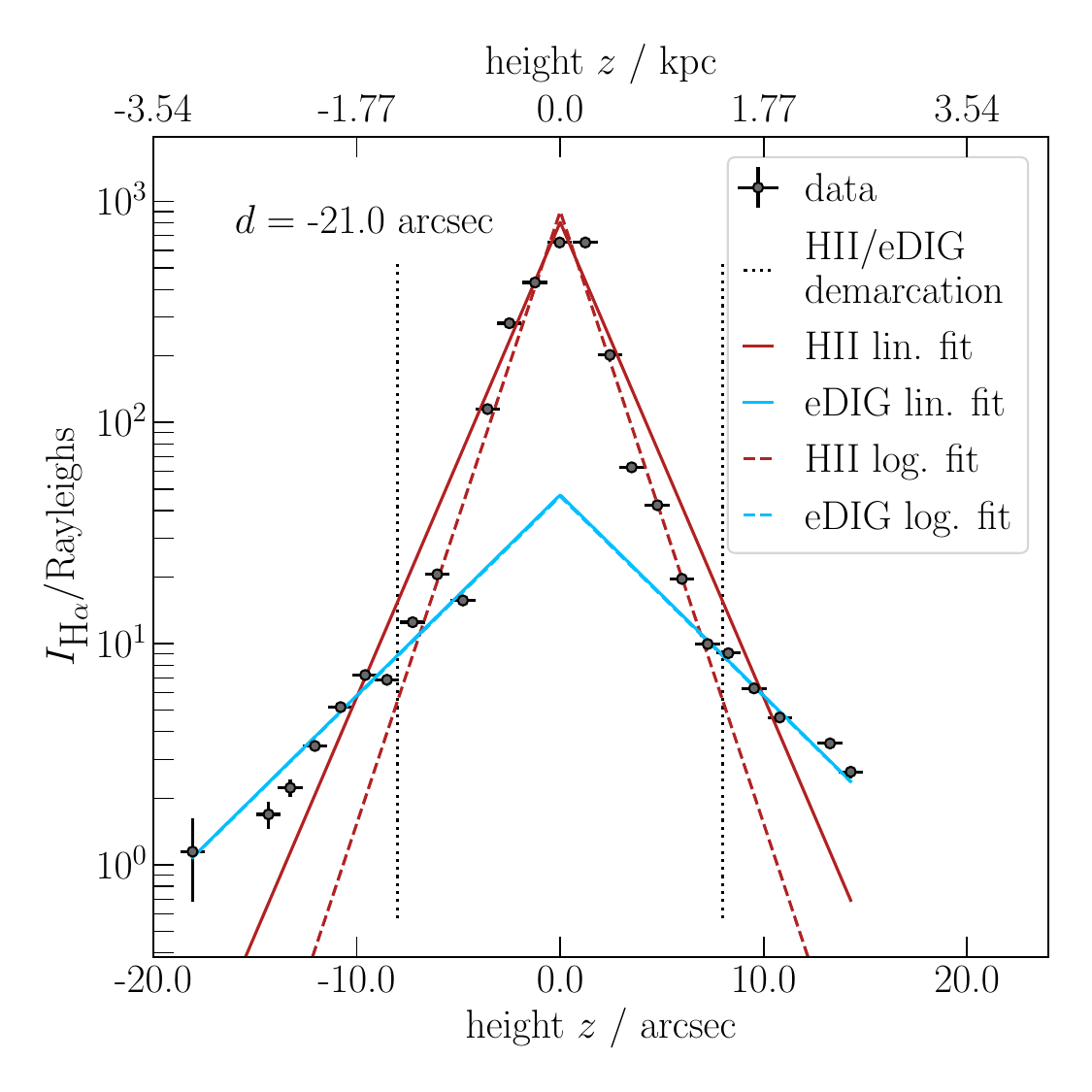}
    \end{subfigure}\\
        \begin{subfigure}{.24\linewidth}
        \includegraphics[width=\hsize]{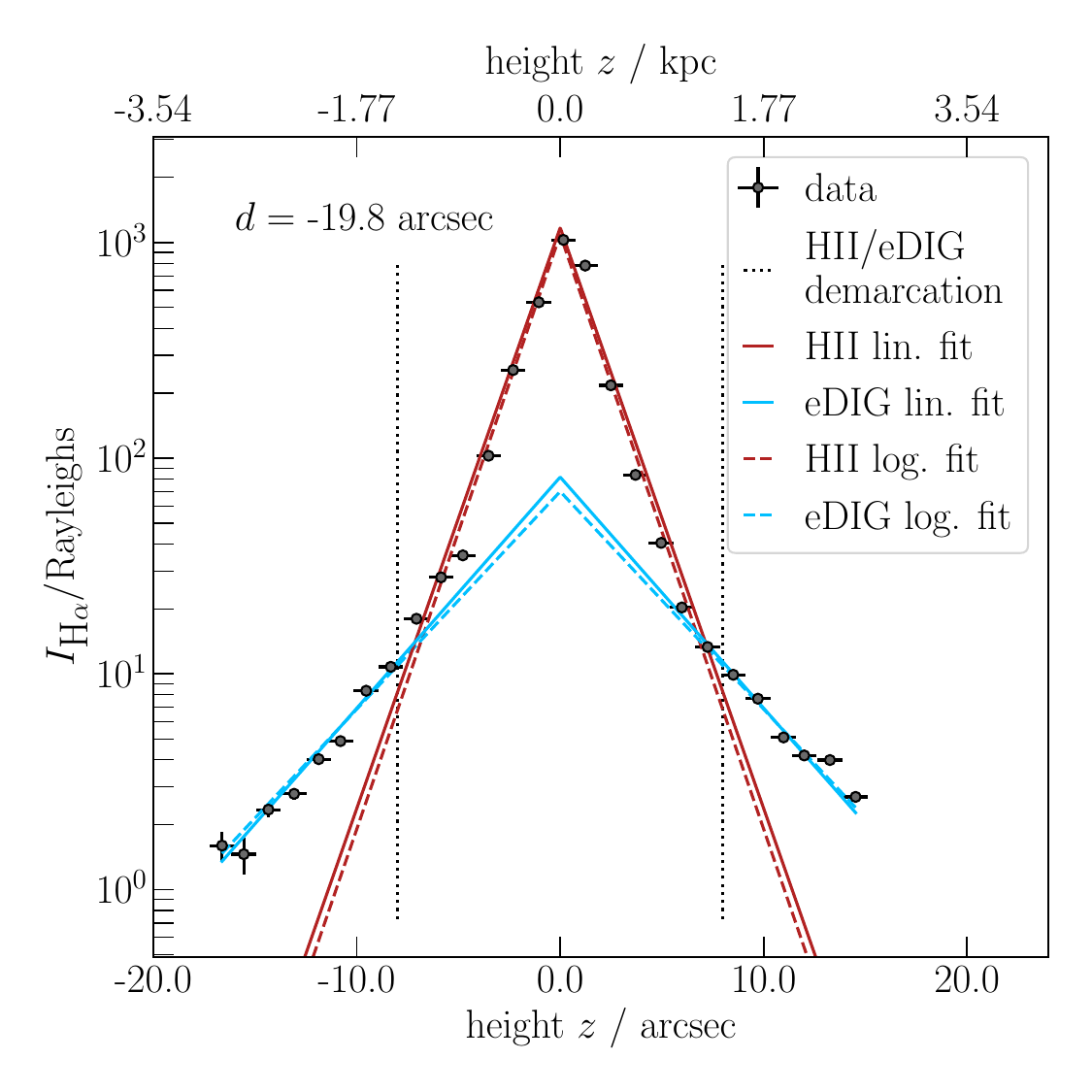}
    \end{subfigure}
        \begin{subfigure}{.24\linewidth}
        \includegraphics[width=\hsize]{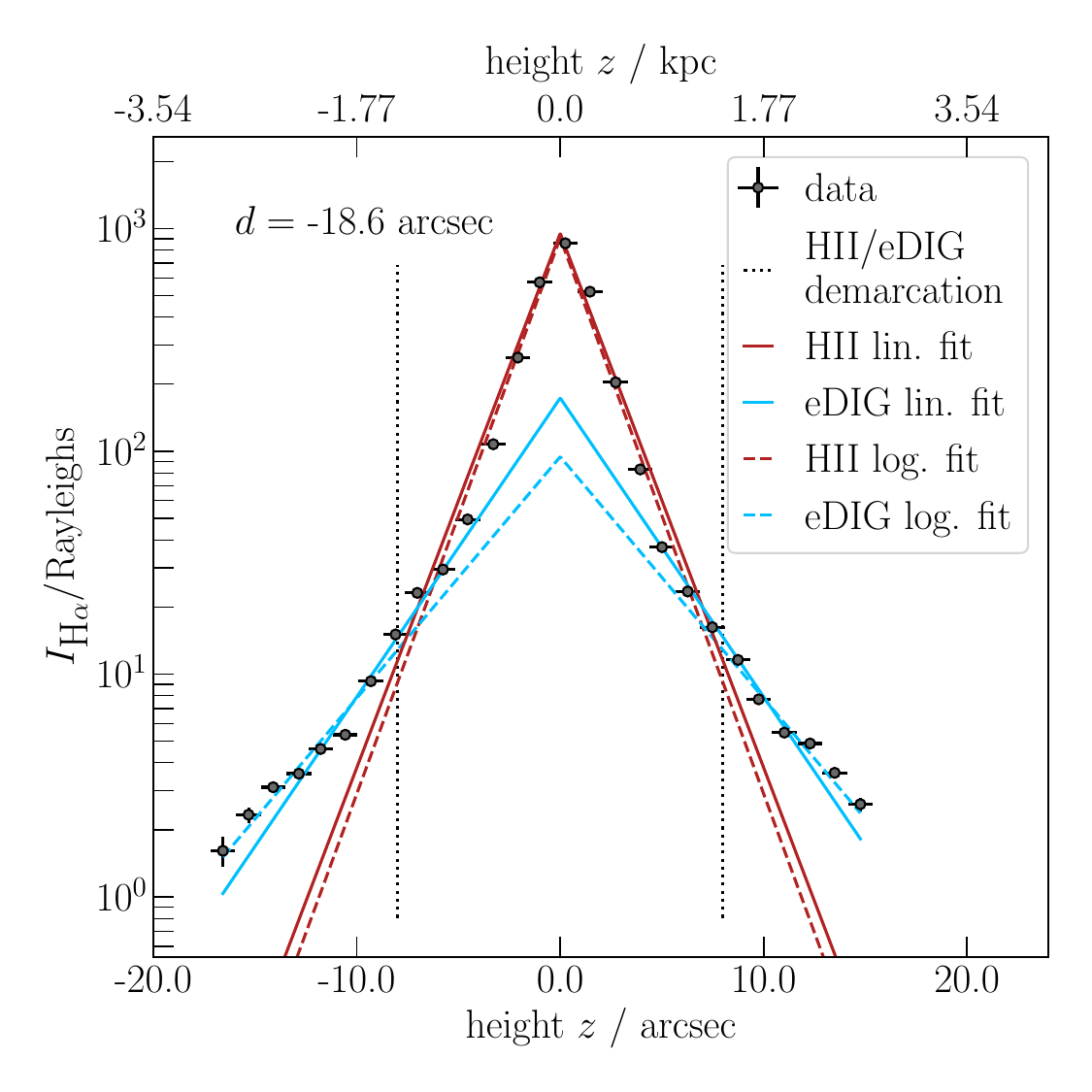}
    \end{subfigure}
        \begin{subfigure}{.24\linewidth}
        \includegraphics[width=\hsize]{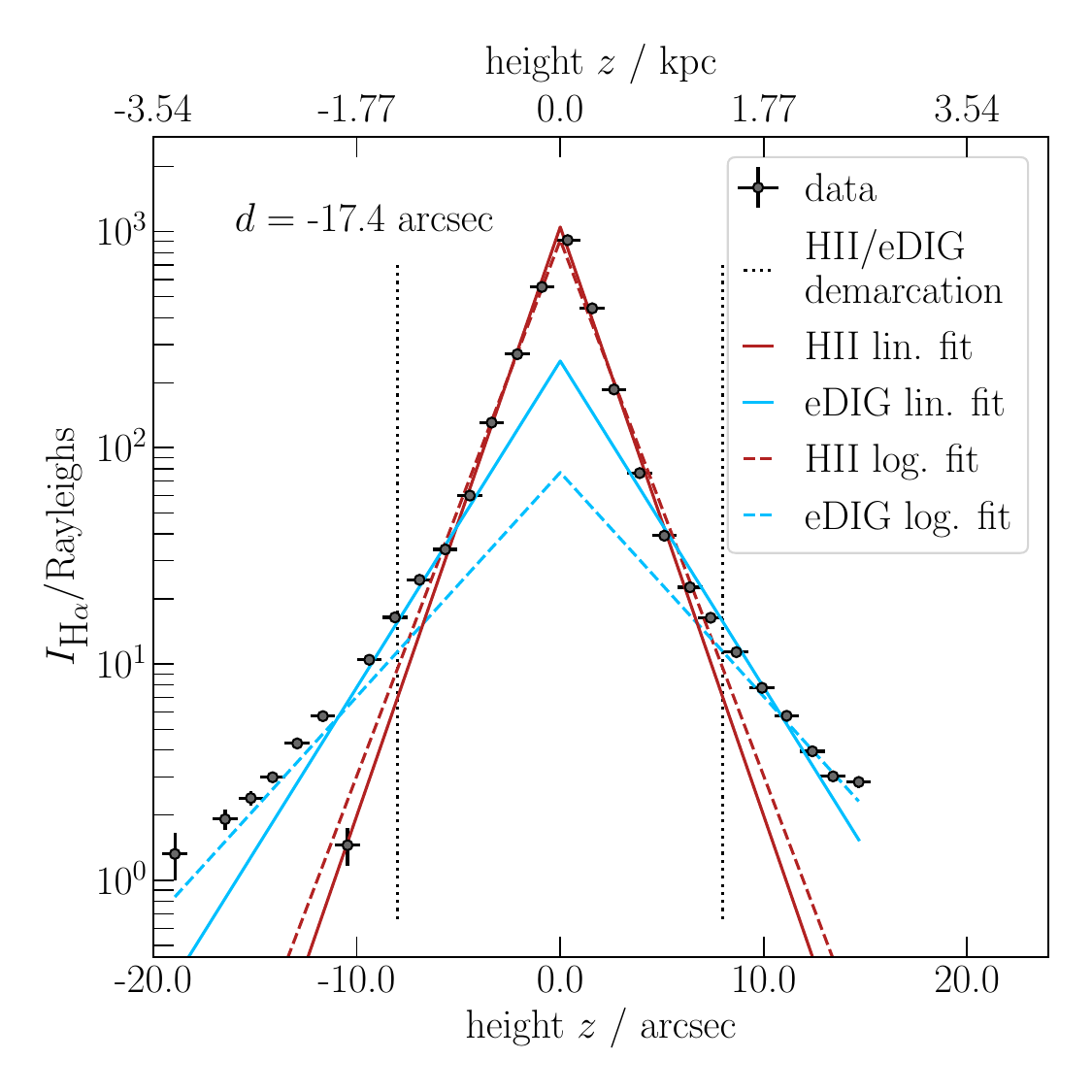}
    \end{subfigure}
        \begin{subfigure}{.24\linewidth}
        \includegraphics[width=\hsize]{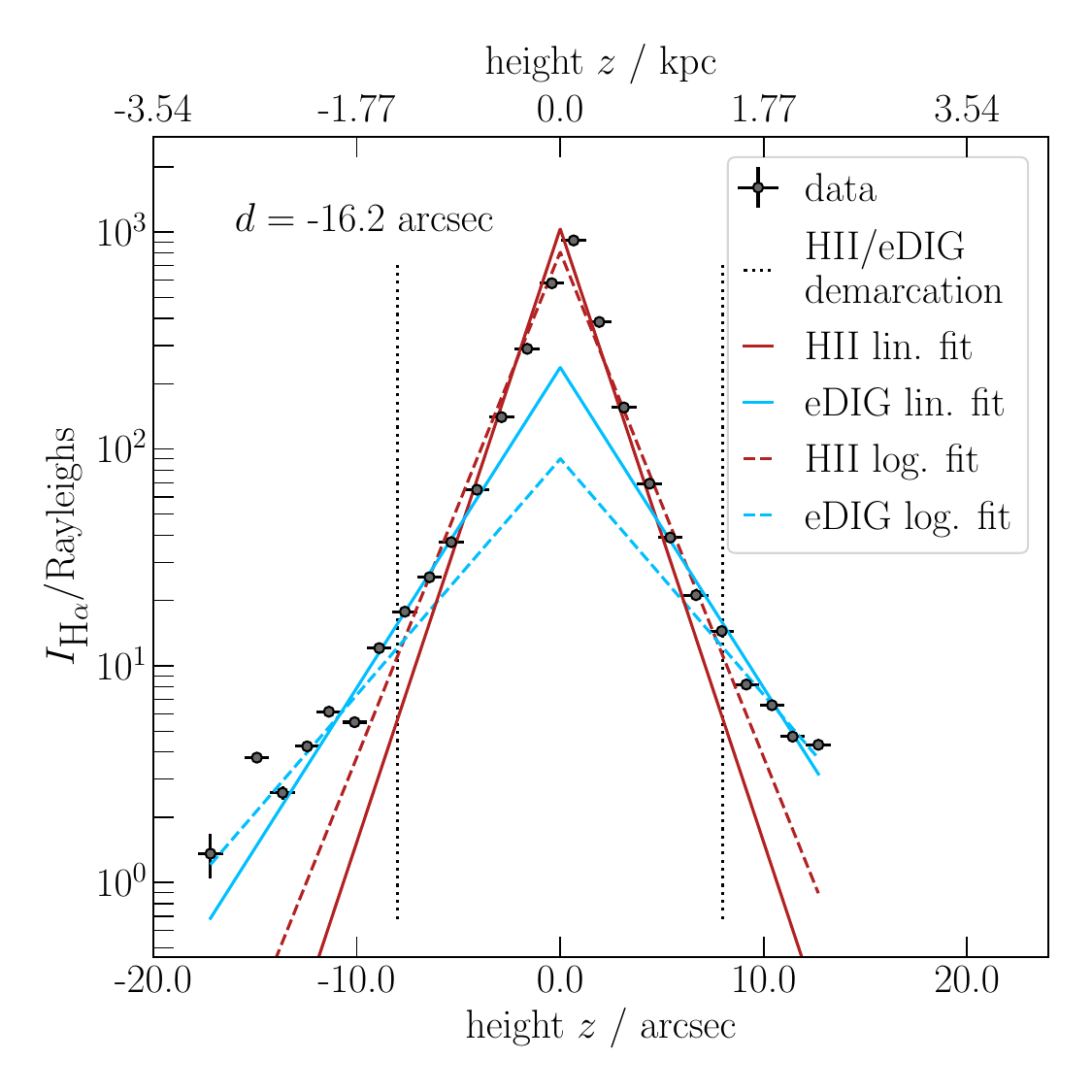}
    \end{subfigure}\\
        \begin{subfigure}{.24\linewidth}
        \includegraphics[width=\hsize]{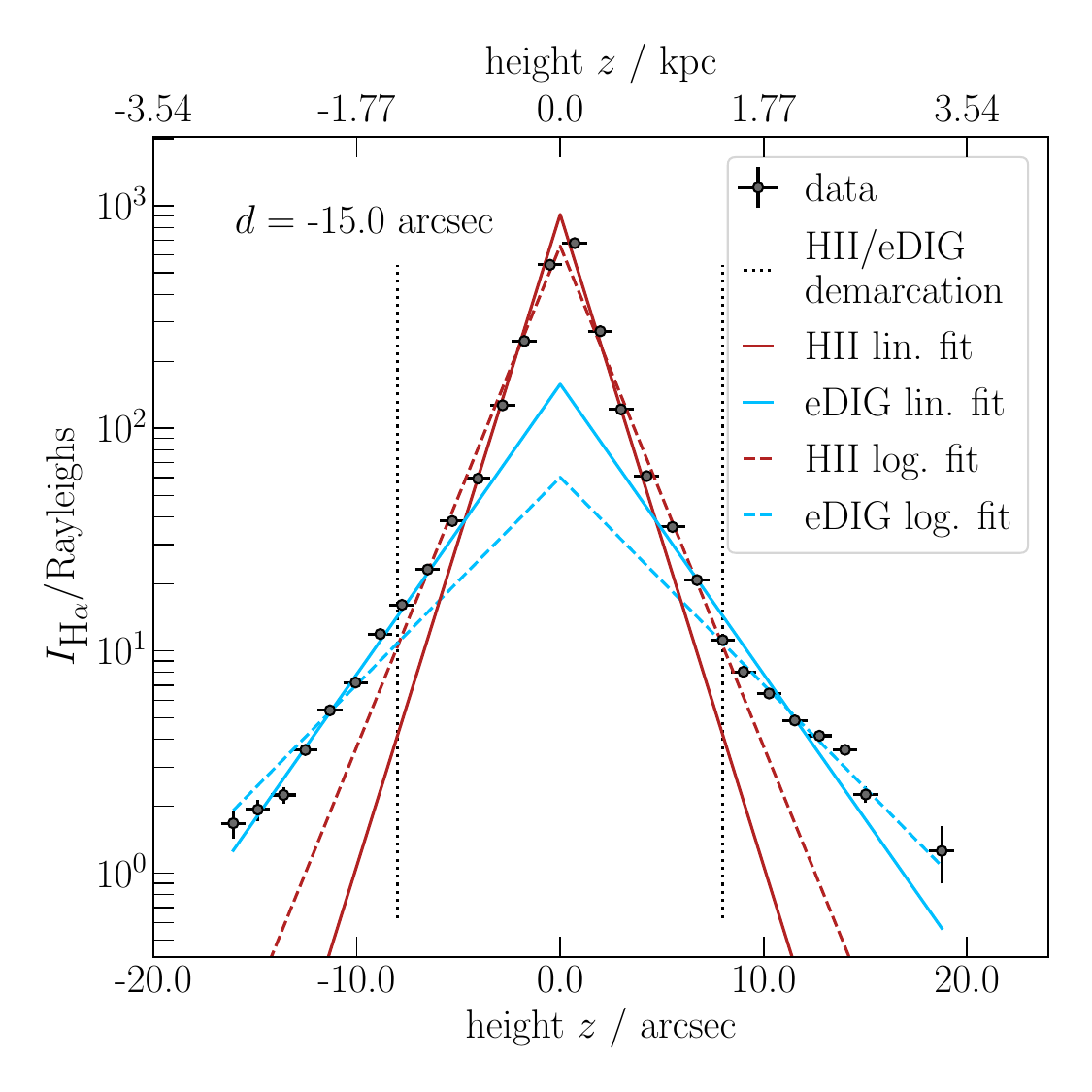}
    \end{subfigure}
        \begin{subfigure}{.24\linewidth}
        \includegraphics[width=\hsize]{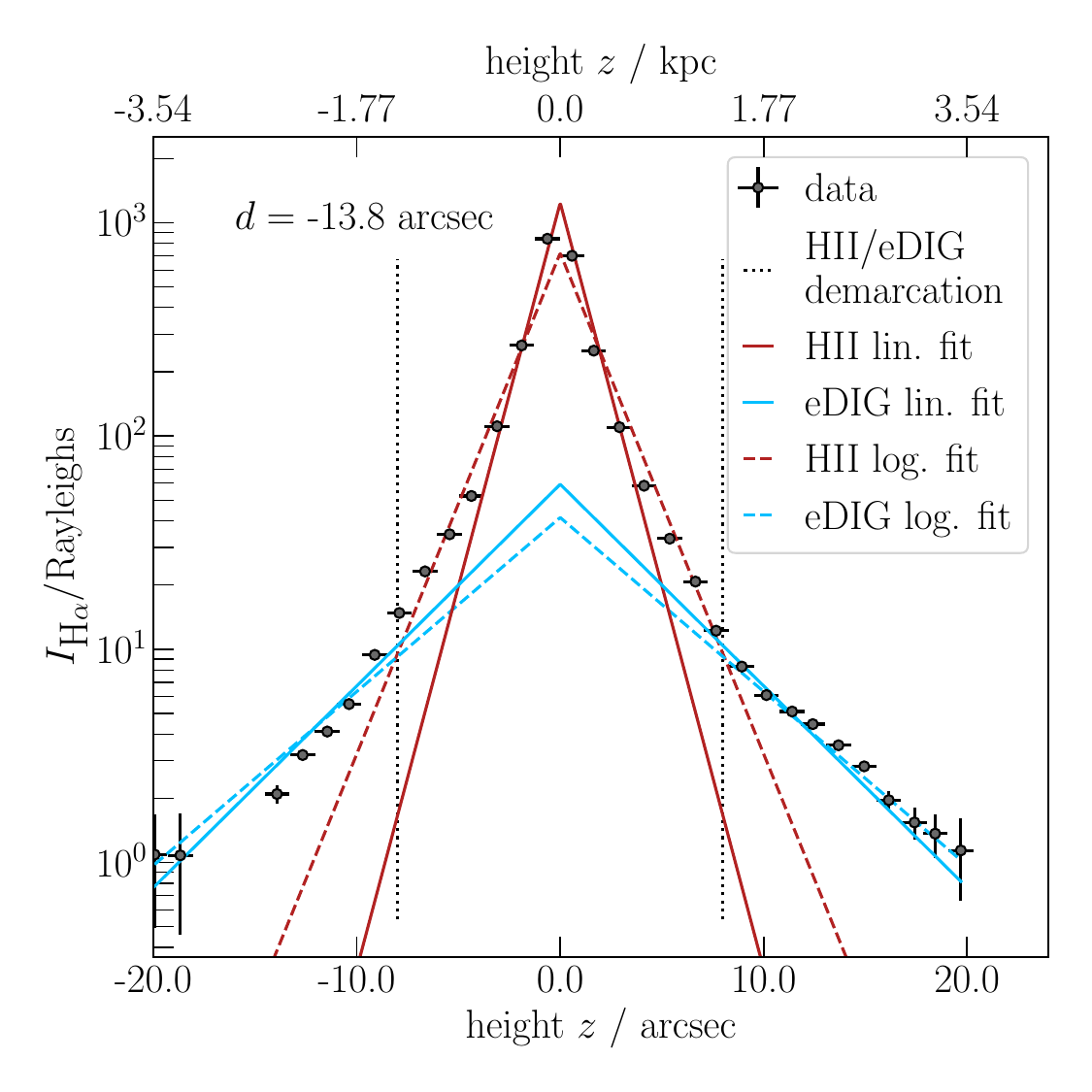}
    \end{subfigure}
        \begin{subfigure}{.24\linewidth}
        \includegraphics[width=\hsize]{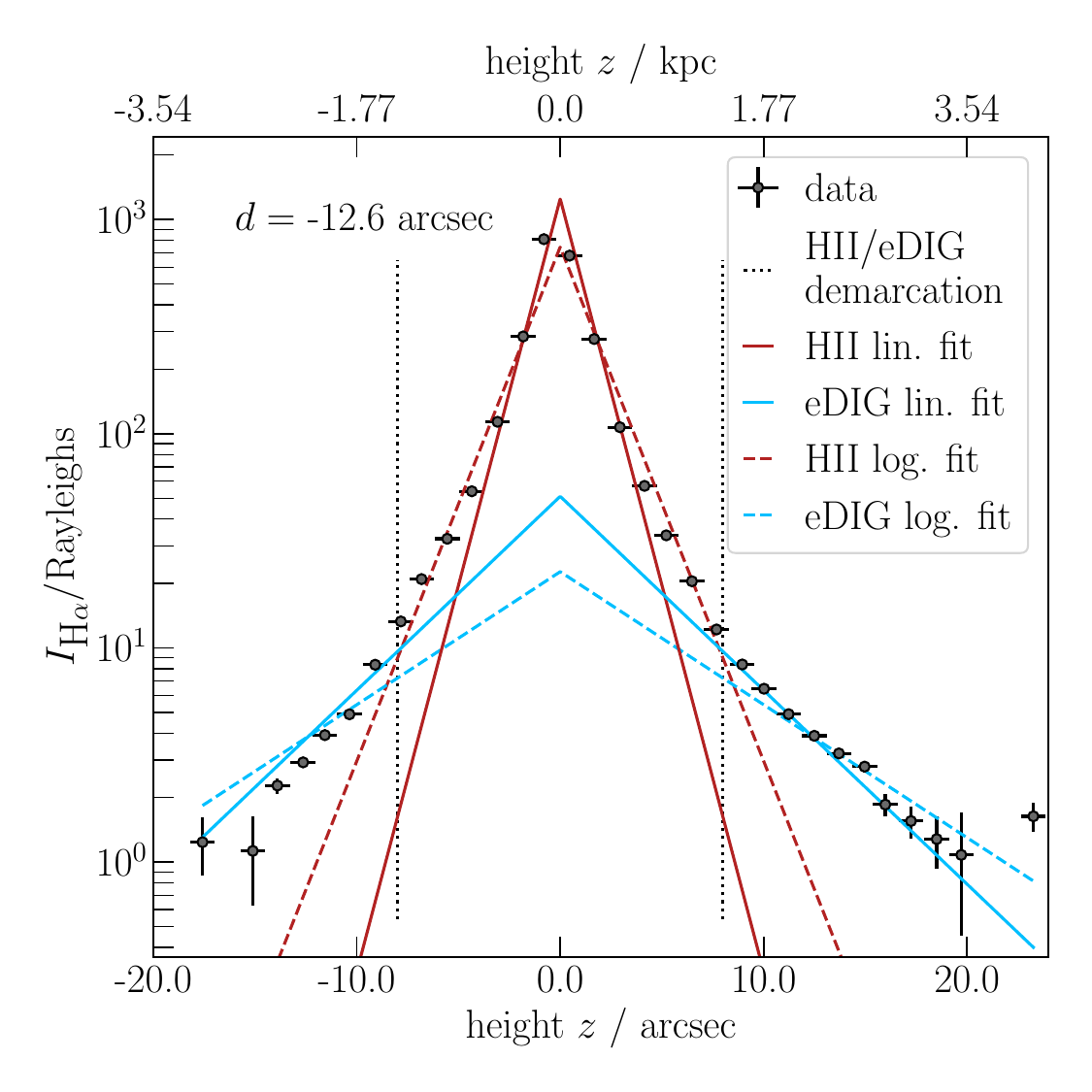}
    \end{subfigure}
        \begin{subfigure}{.24\linewidth}
        \includegraphics[width=\hsize]{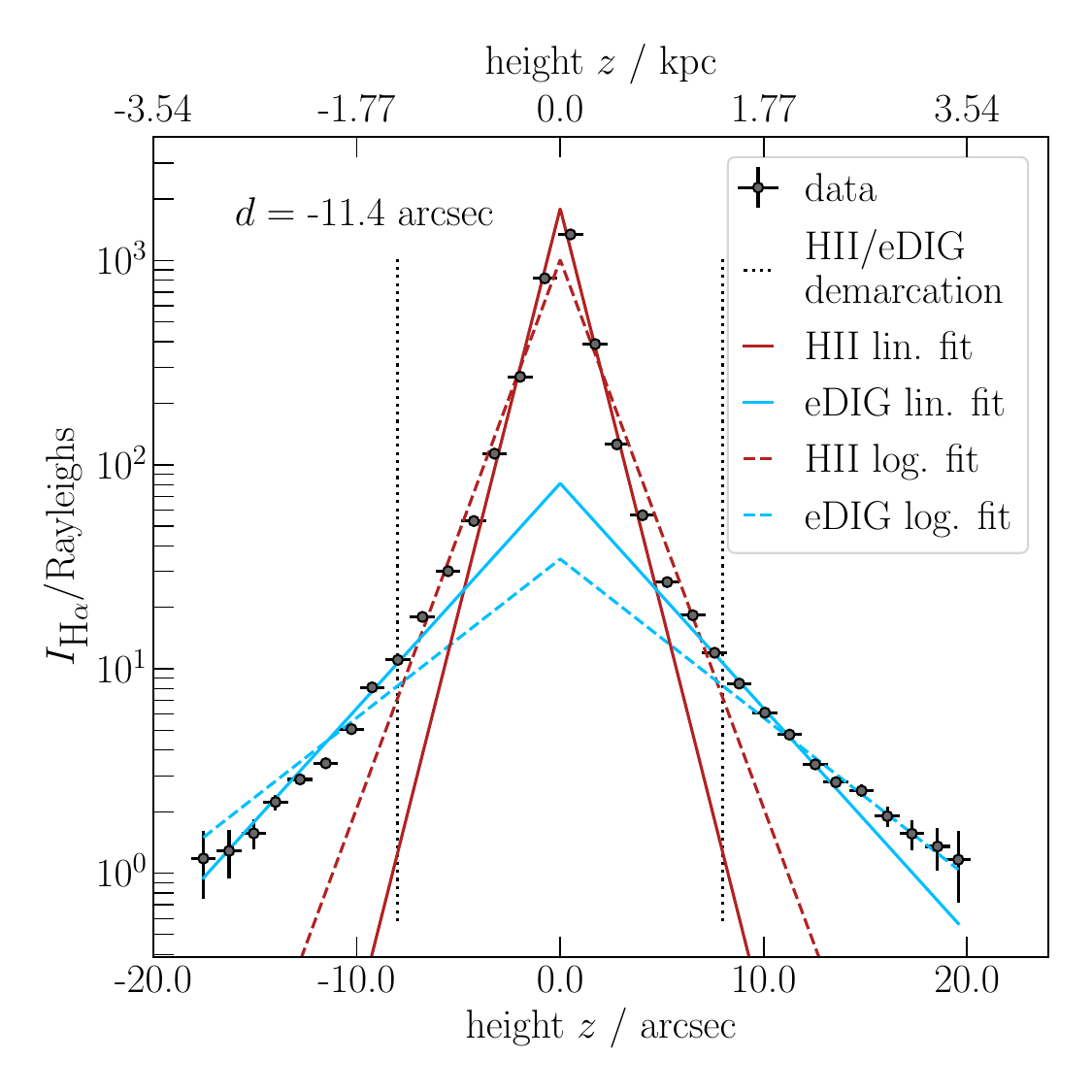}
    \end{subfigure}\\
        \begin{subfigure}{.24\linewidth}
        \includegraphics[width=\hsize]{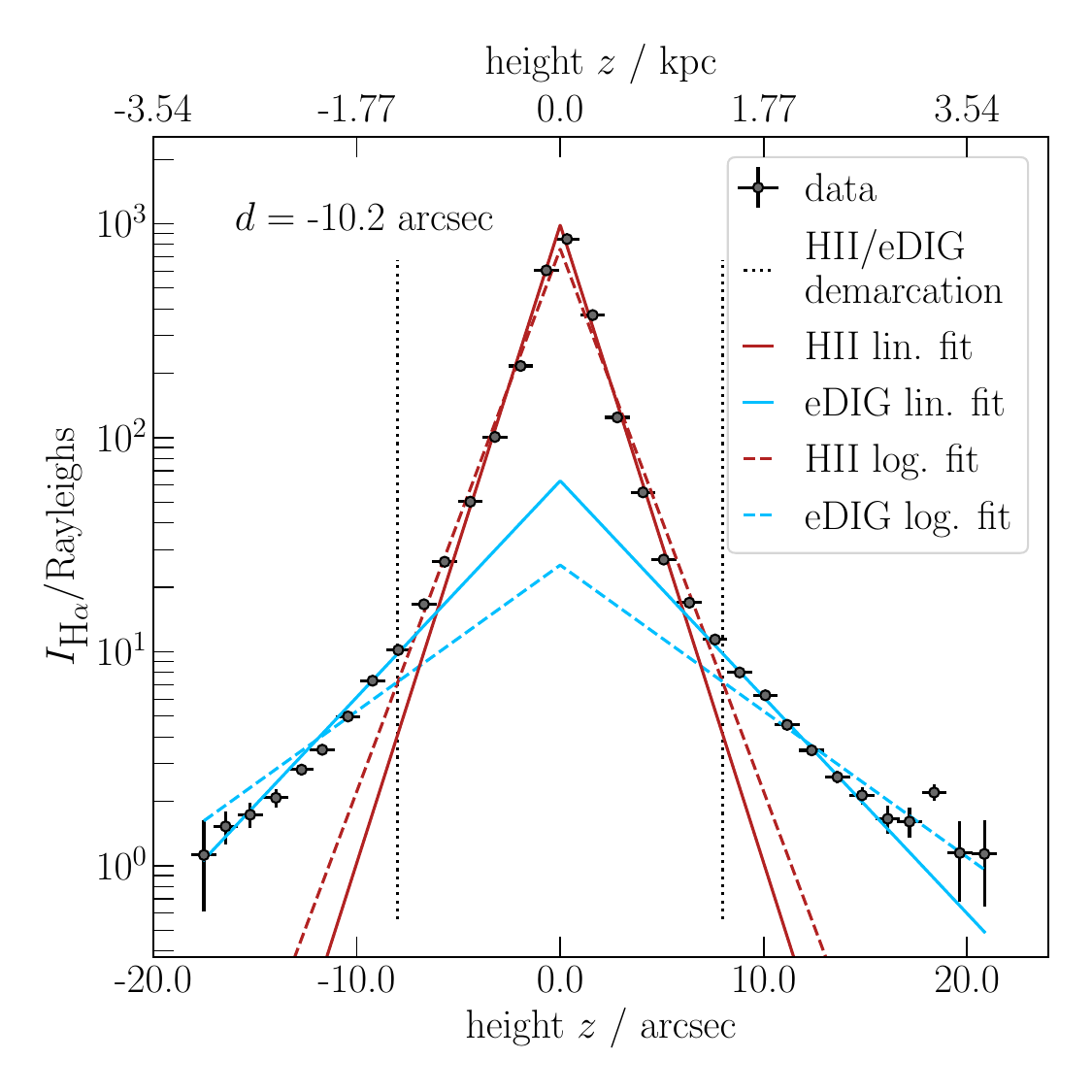}
    \end{subfigure}
        \begin{subfigure}{.24\linewidth}
        \includegraphics[width=\hsize]{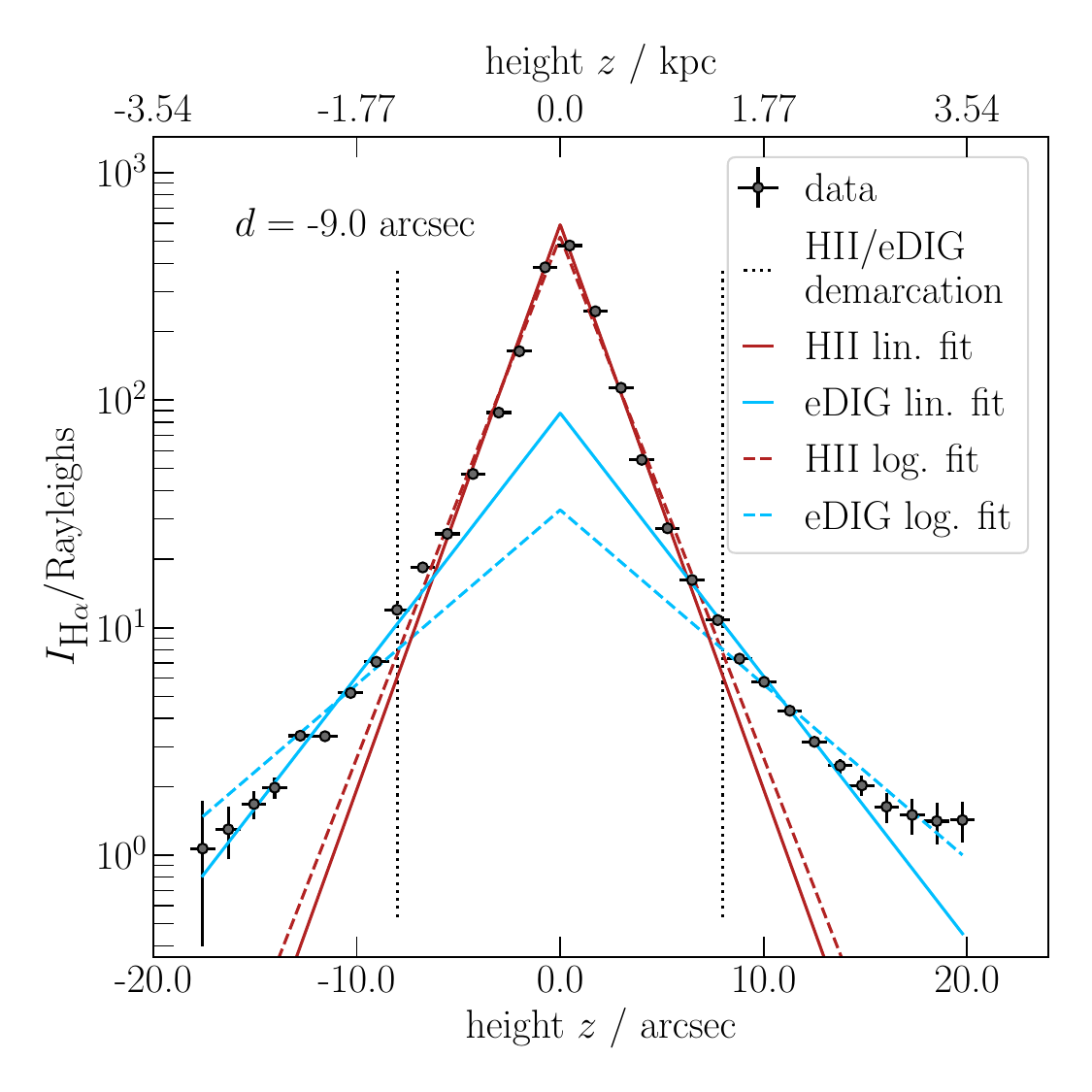}
    \end{subfigure}
        \begin{subfigure}{.24\linewidth}
        \includegraphics[width=\hsize]{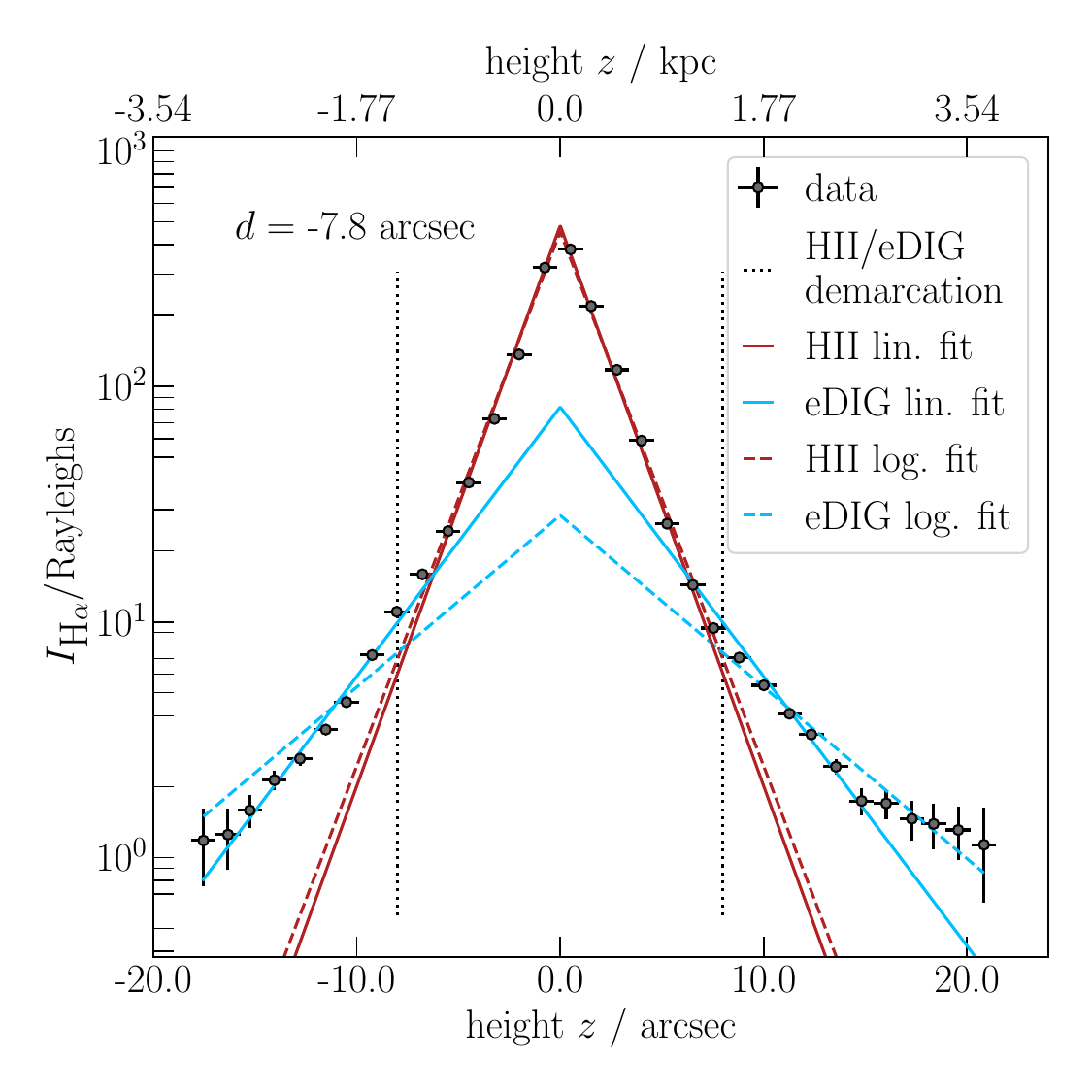}
    \end{subfigure}
        \begin{subfigure}{.24\linewidth}
        \includegraphics[width=\hsize]{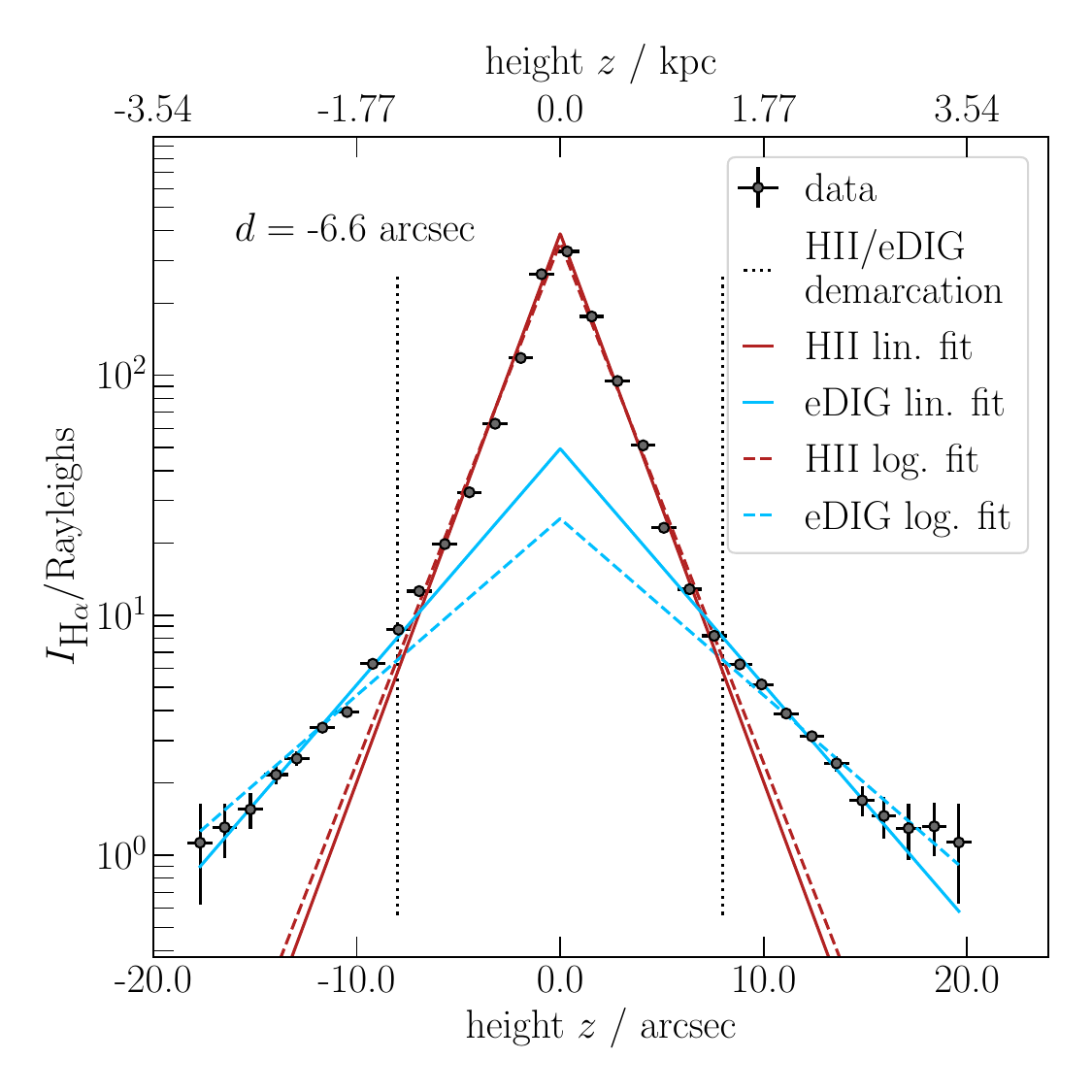}
    \end{subfigure}\\
    \caption{Visualization of the H$\alpha$ scale height fit analogous to the left panel of Fig. \ref{fig:scaleheights}, shown for all projected radial positions, $d$.}\label{fig:shfitting_full}
    
\end{figure*}

Figure \ref{fig:shfitting_full} visualizes the H$\alpha$ scale height fits presented in Sect. \ref{sect:Ha_dist} for all projected radial positions, analogous to the left panel of Fig. \ref{fig:scaleheights}.

    \begin{figure*}
        \begin{subfigure}{.24\linewidth}
        \includegraphics[width=\hsize]{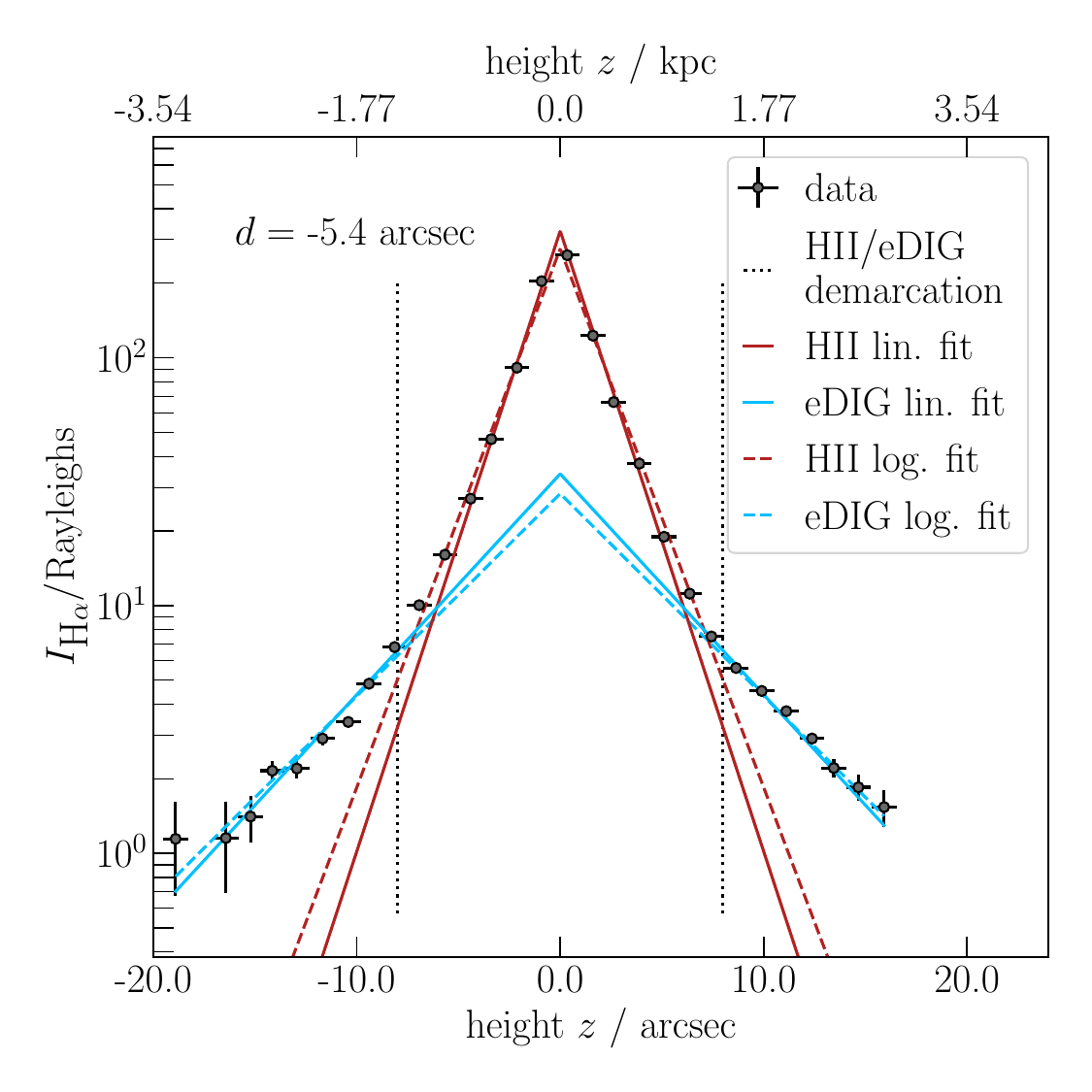}
    \end{subfigure}
        \begin{subfigure}{.24\linewidth}
        \includegraphics[width=\hsize]{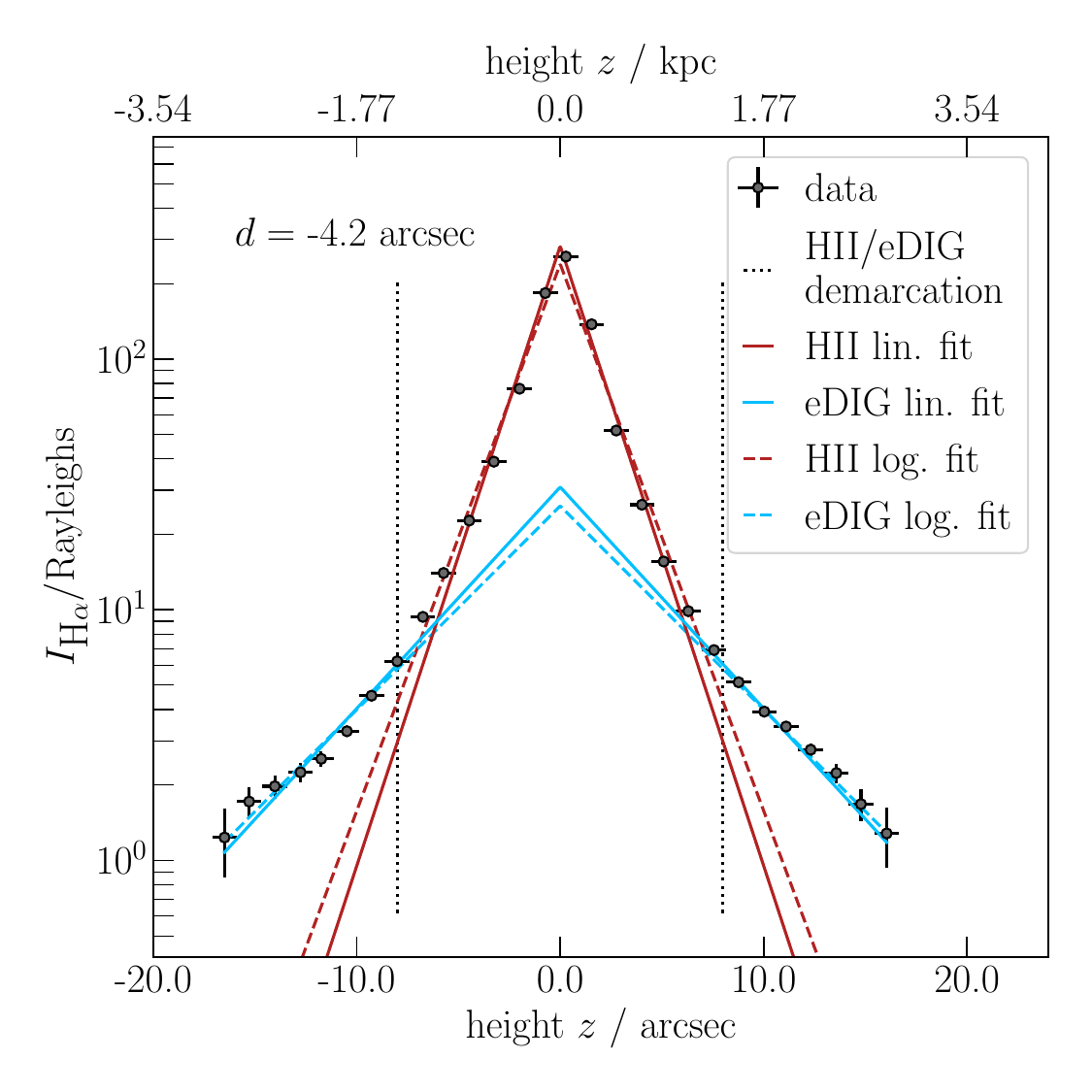}
    \end{subfigure}
        \begin{subfigure}{.24\linewidth}
        \includegraphics[width=\hsize]{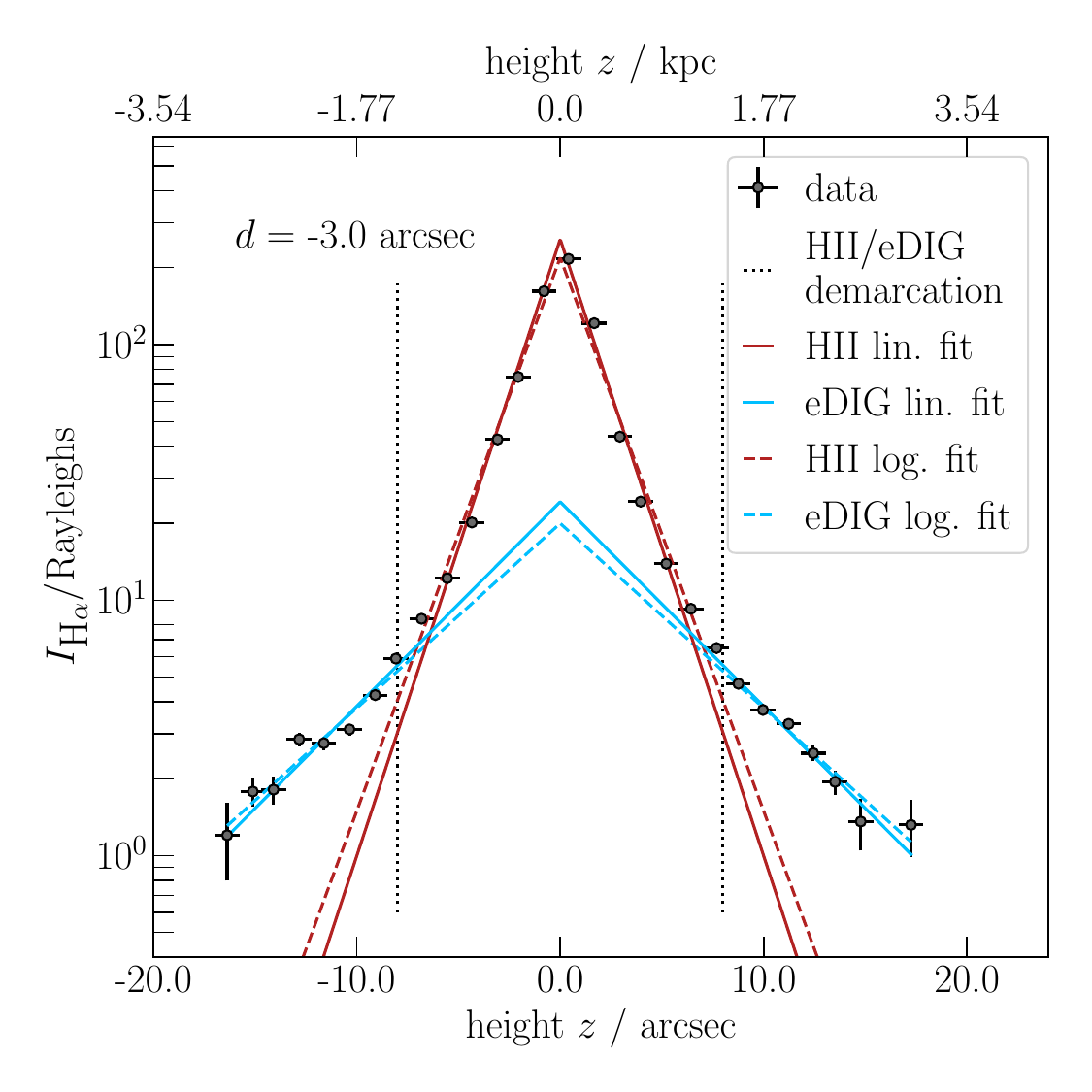}
    \end{subfigure}
        \begin{subfigure}{.24\linewidth}
        \includegraphics[width=\hsize]{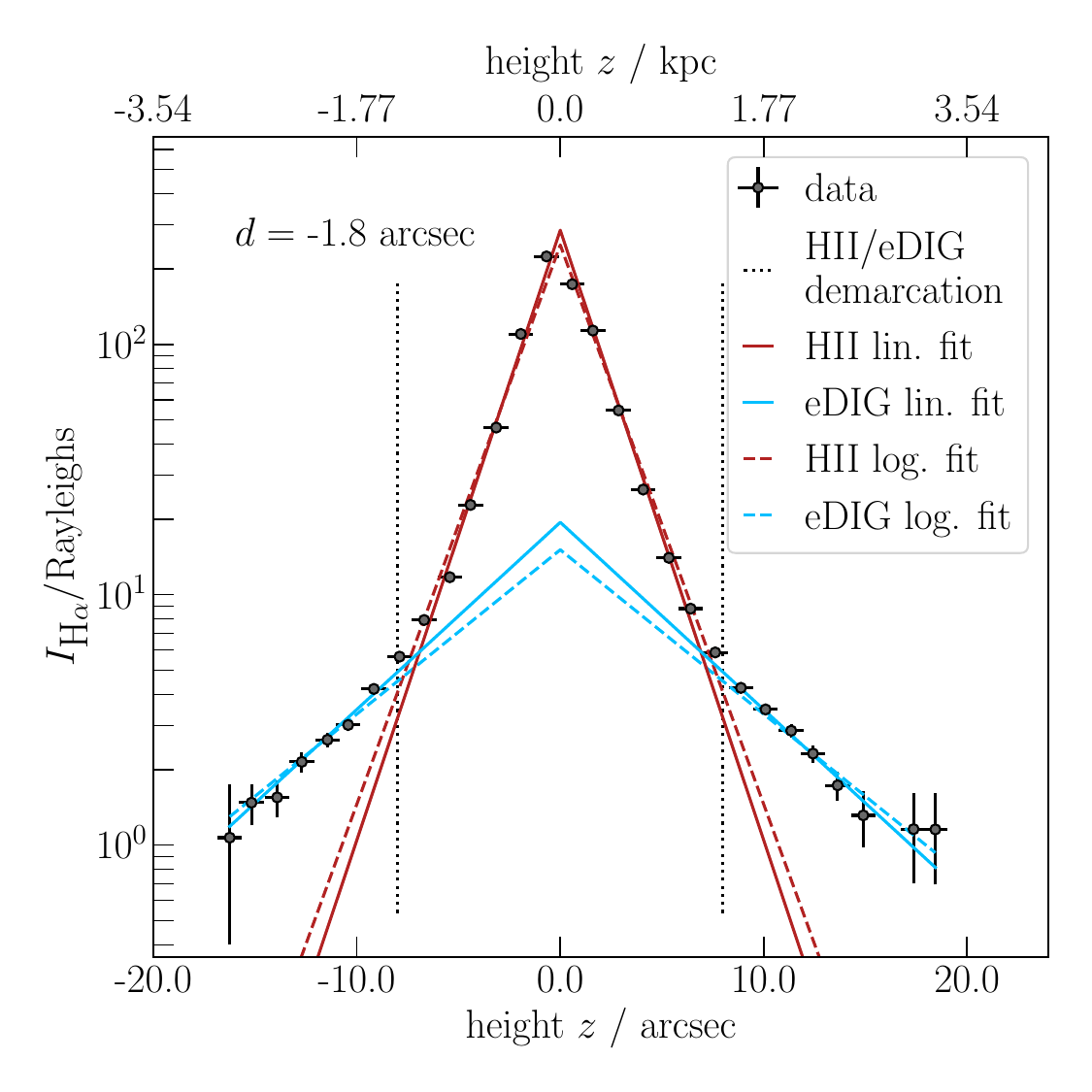}
    \end{subfigure}\\
        \begin{subfigure}{.24\linewidth}
        \includegraphics[width=\hsize]{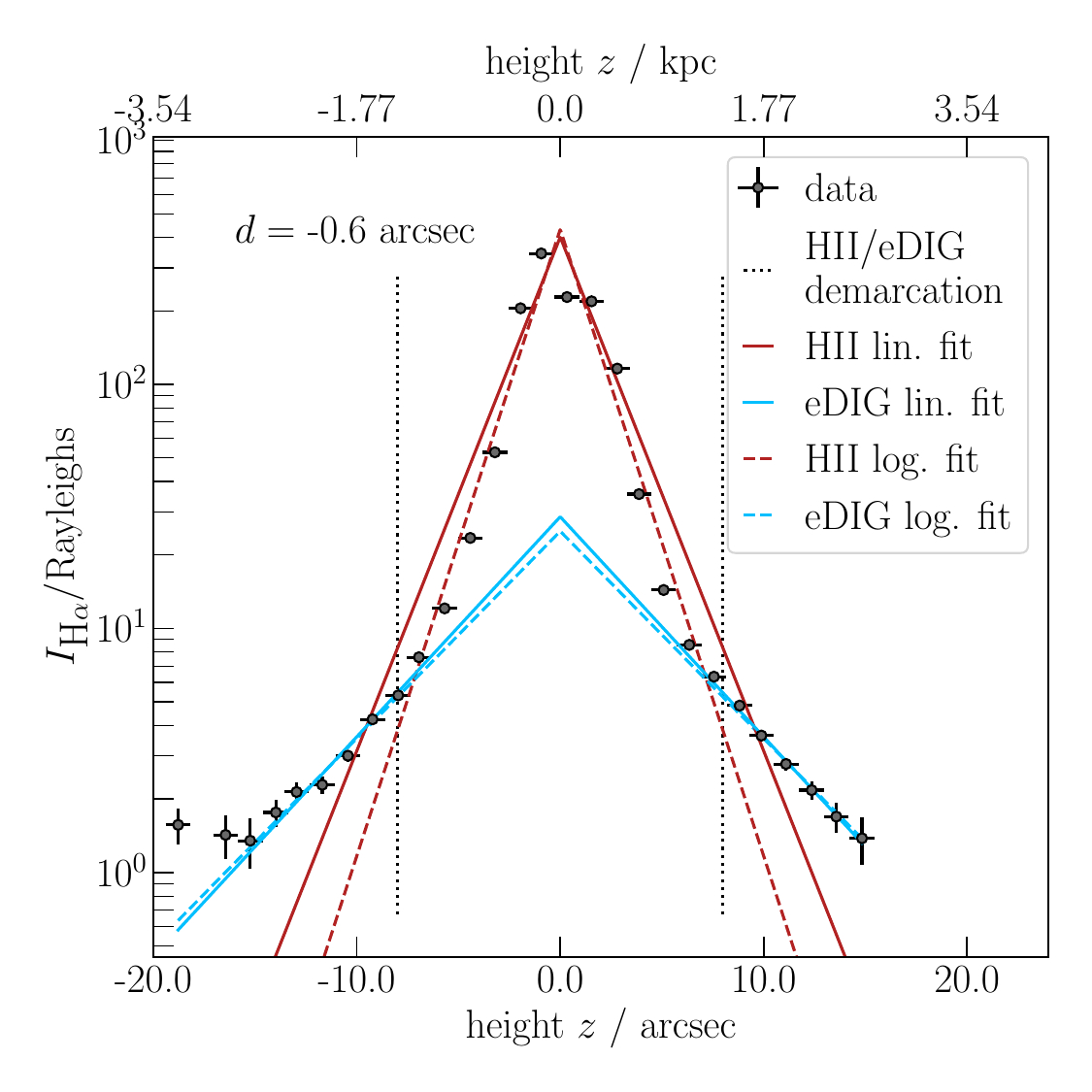}
    \end{subfigure}
        \begin{subfigure}{.24\linewidth}
        \includegraphics[width=\hsize]{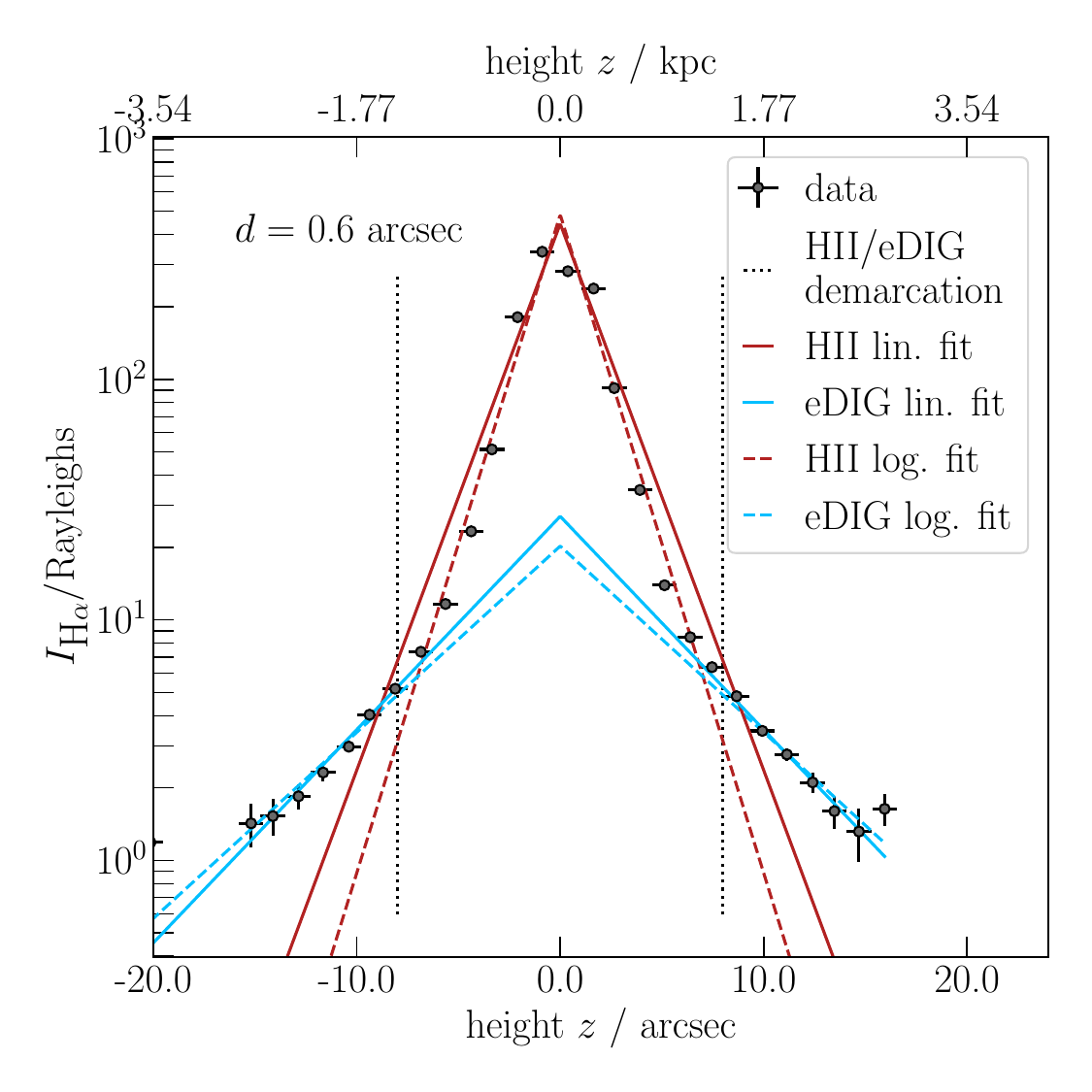}
    \end{subfigure}
        \begin{subfigure}{.24\linewidth}
        \includegraphics[width=\hsize]{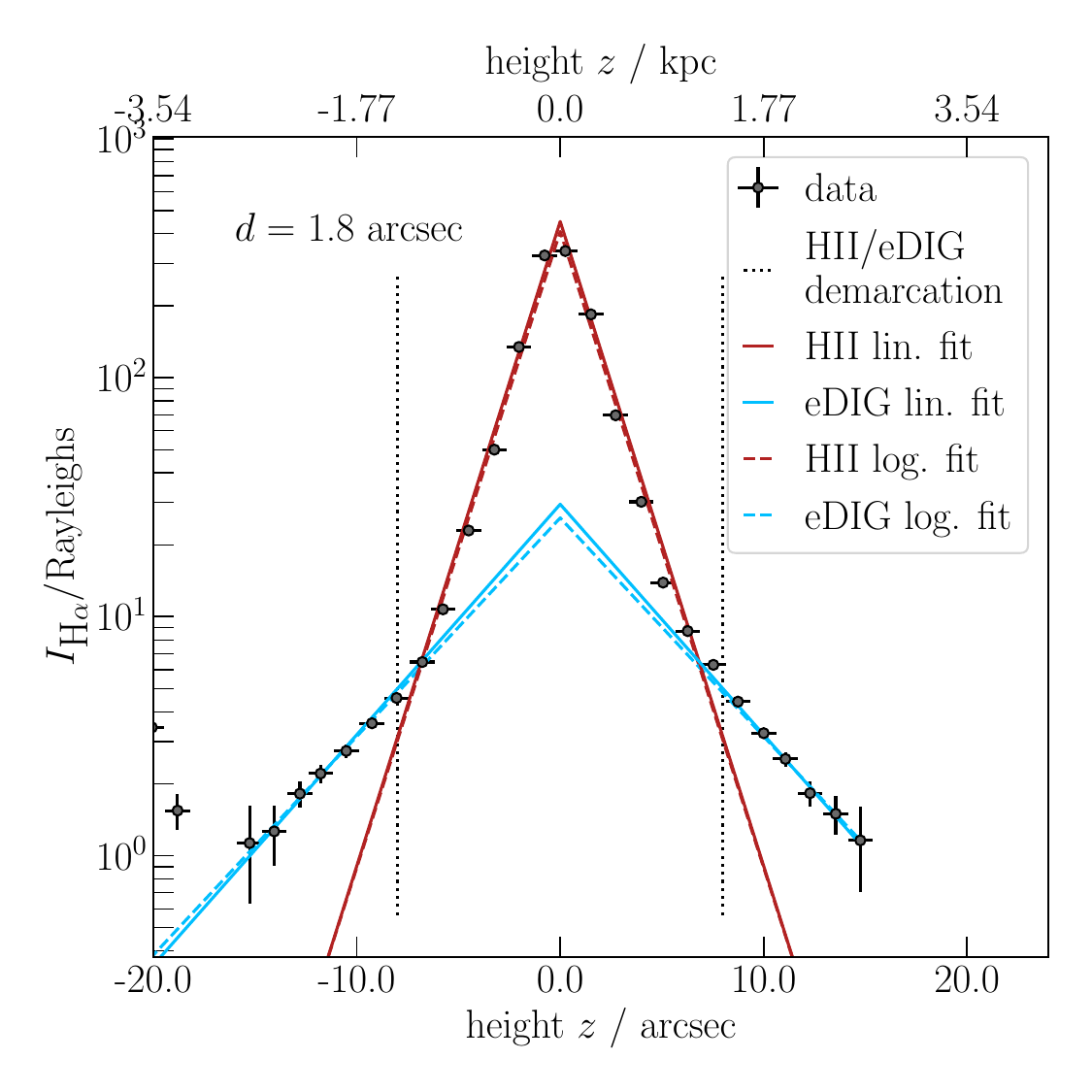}
    \end{subfigure}
        \begin{subfigure}{.24\linewidth}
        \includegraphics[width=\hsize]{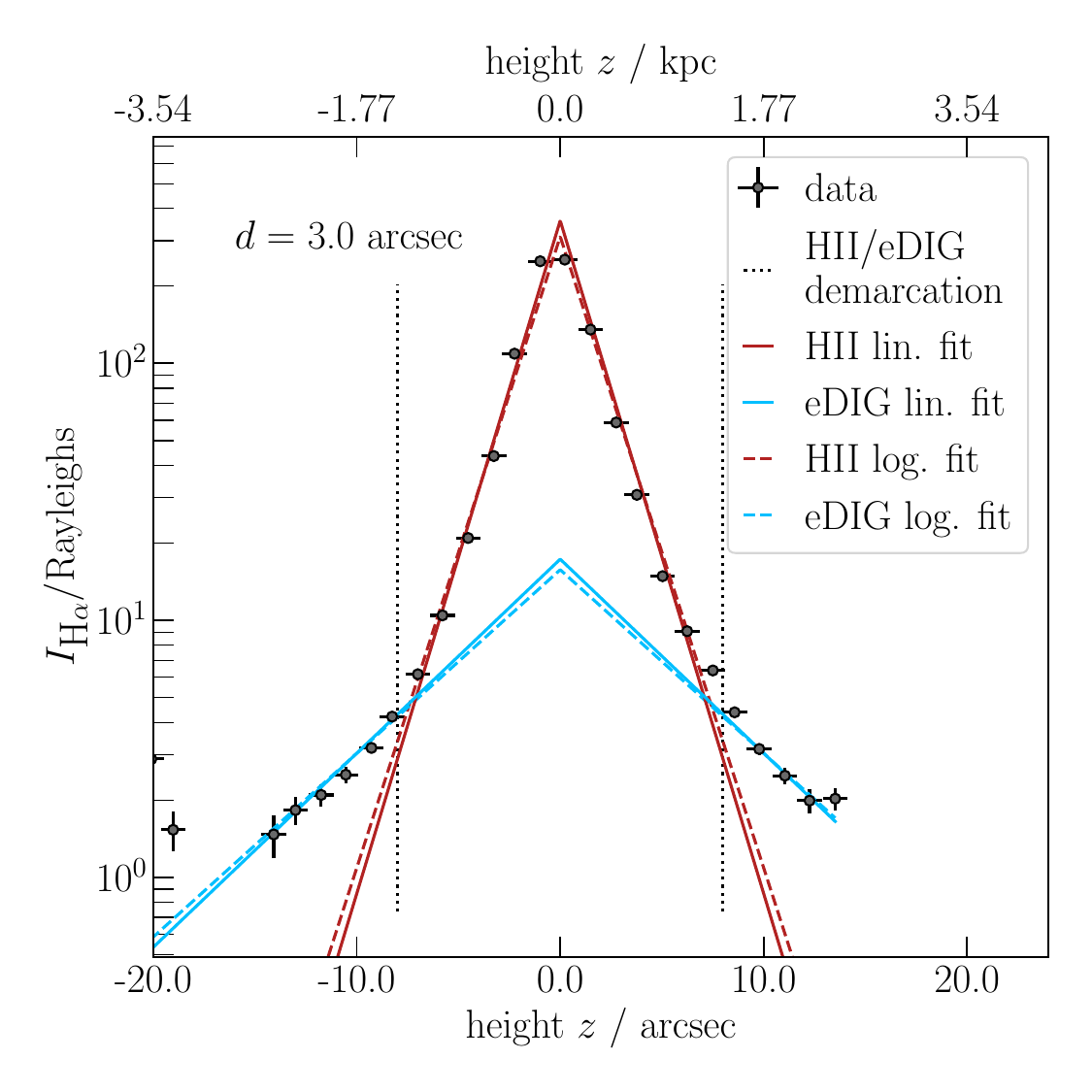}
    \end{subfigure}\\
        \begin{subfigure}{.24\linewidth}
        \includegraphics[width=\hsize]{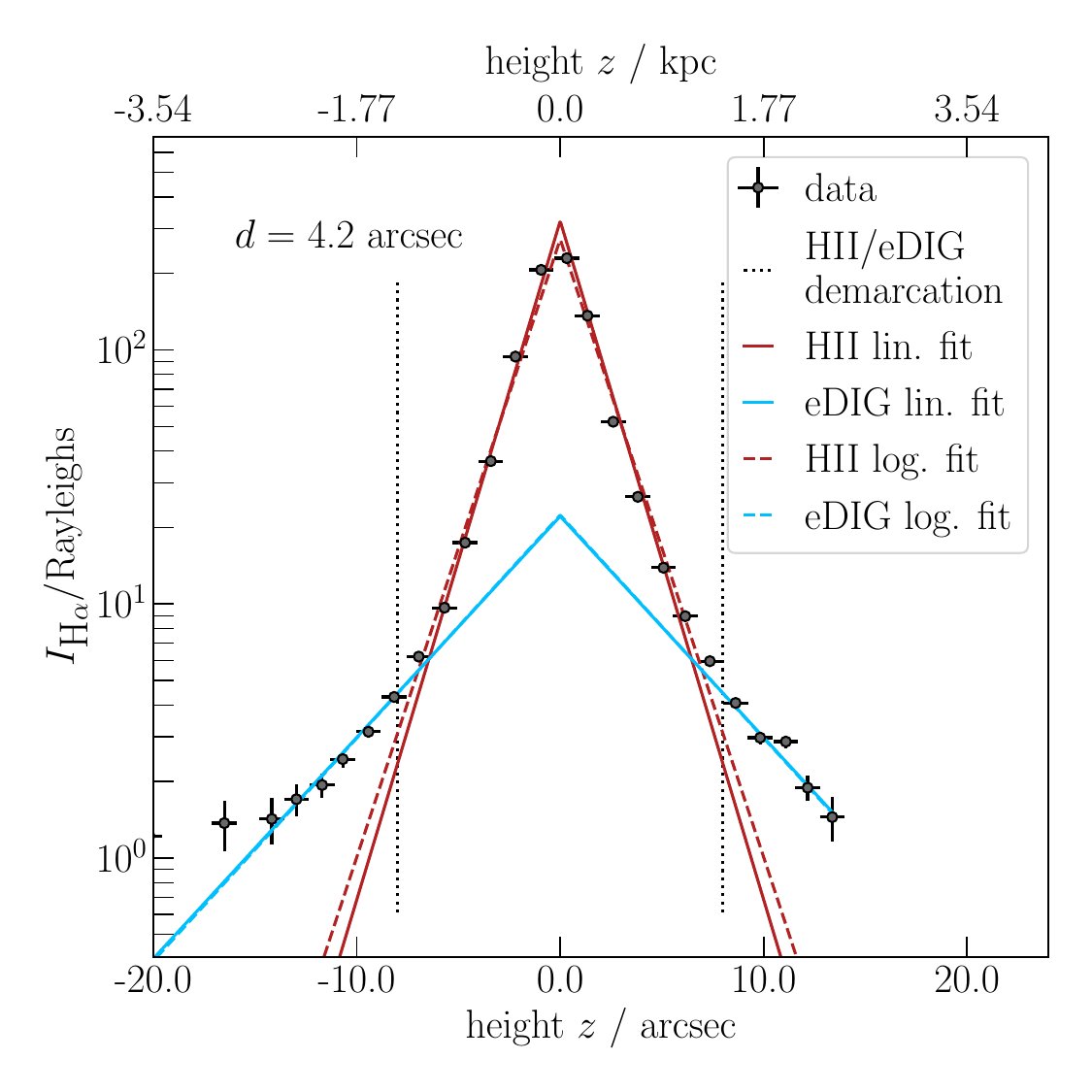}
    \end{subfigure}
        \begin{subfigure}{.24\linewidth}
        \includegraphics[width=\hsize]{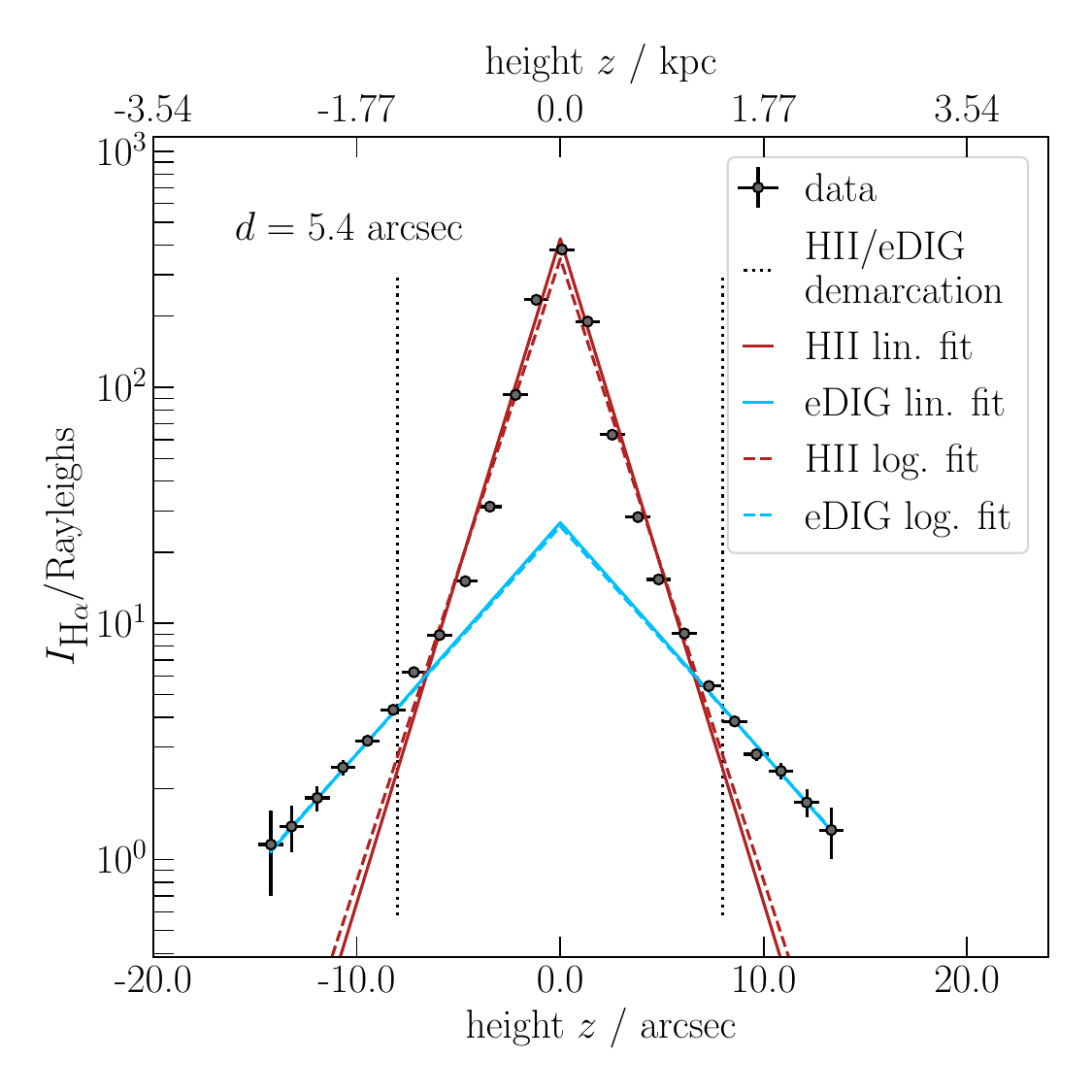}
    \end{subfigure}
        \begin{subfigure}{.24\linewidth}
        \includegraphics[width=\hsize]{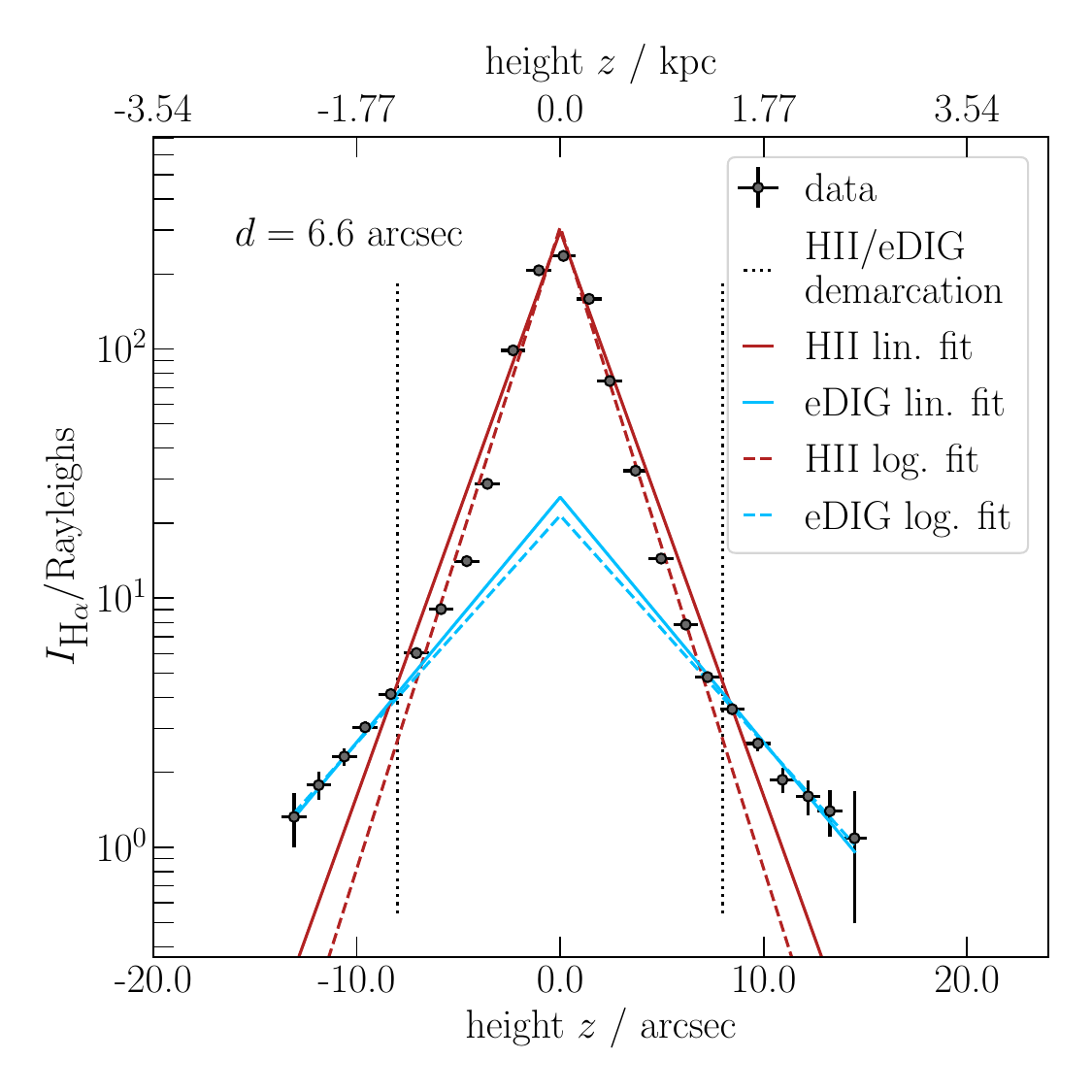}
    \end{subfigure}
        \begin{subfigure}{.24\linewidth}
        \includegraphics[width=\hsize]{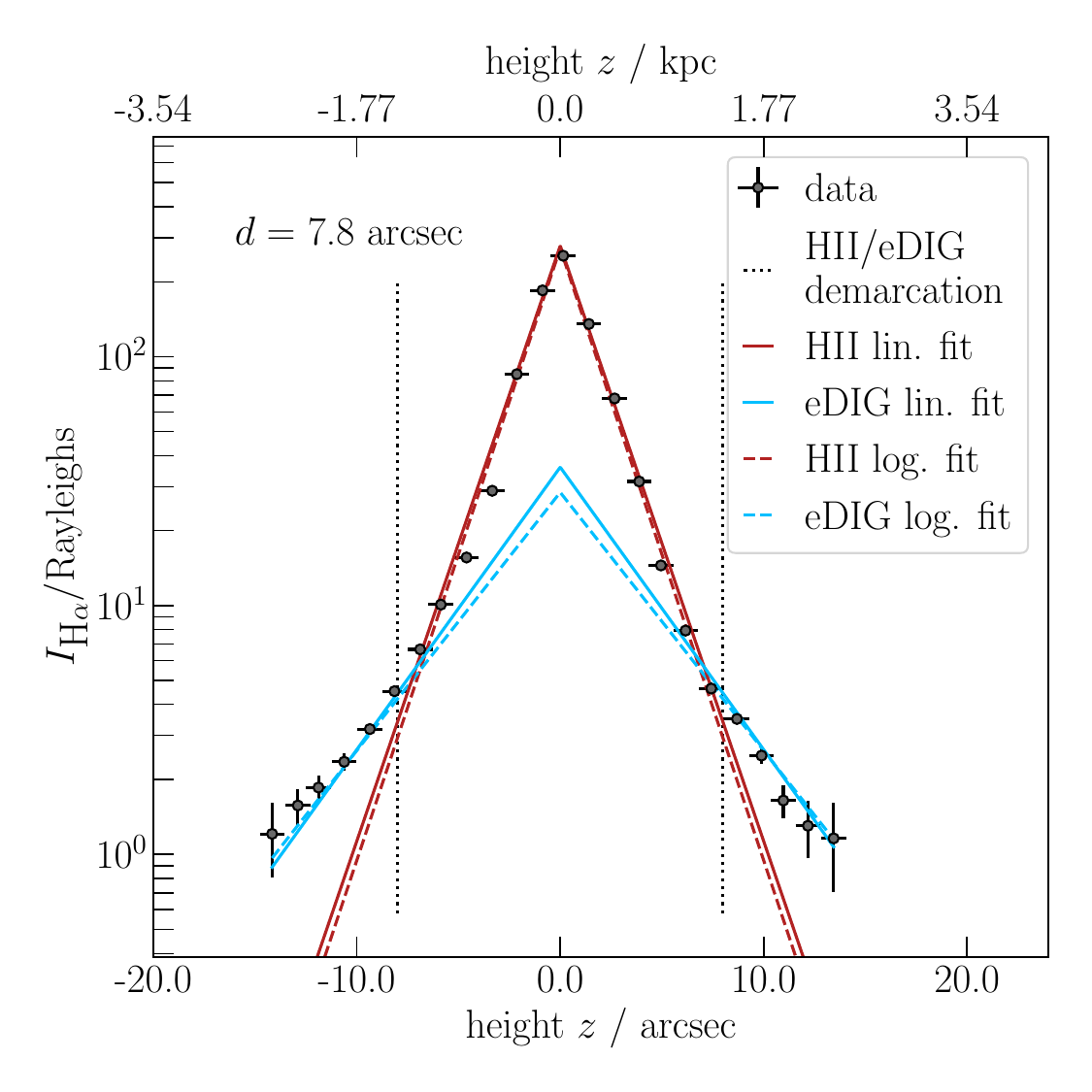}
    \end{subfigure}\\
        \begin{subfigure}{.24\linewidth}
        \includegraphics[width=\hsize]{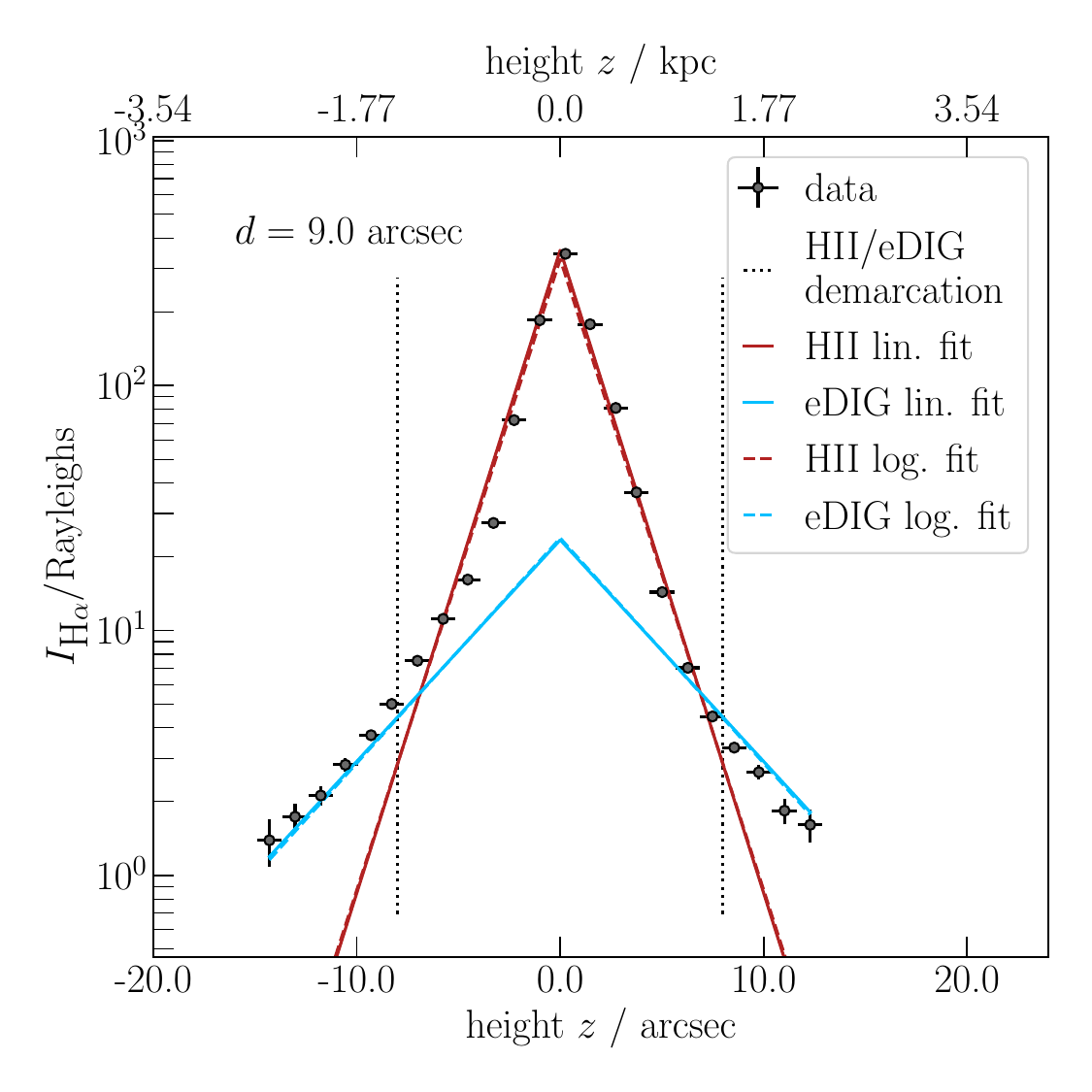}
    \end{subfigure}
        \begin{subfigure}{.24\linewidth}
        \includegraphics[width=\hsize]{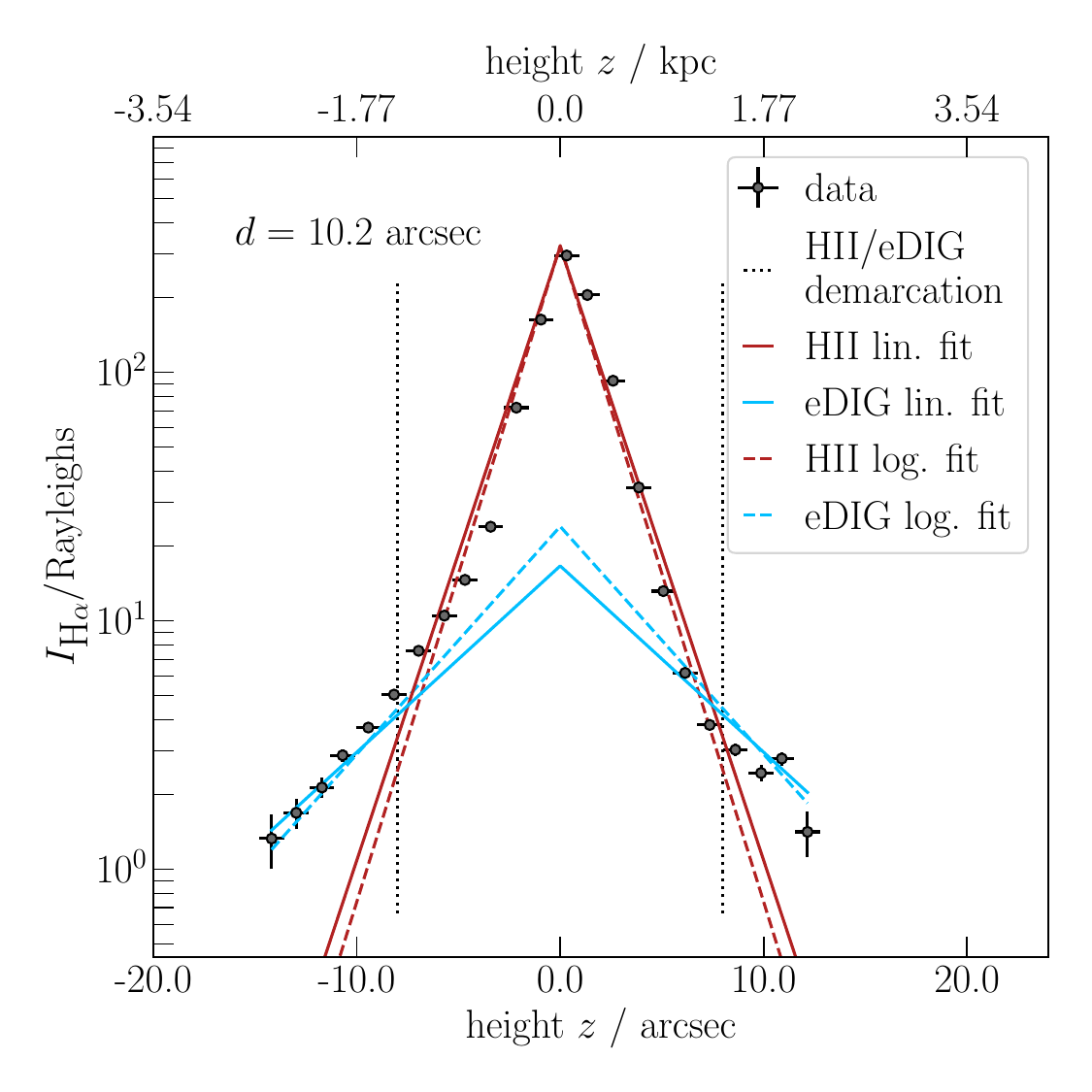}
    \end{subfigure}
        \begin{subfigure}{.24\linewidth}
        \includegraphics[width=\hsize]{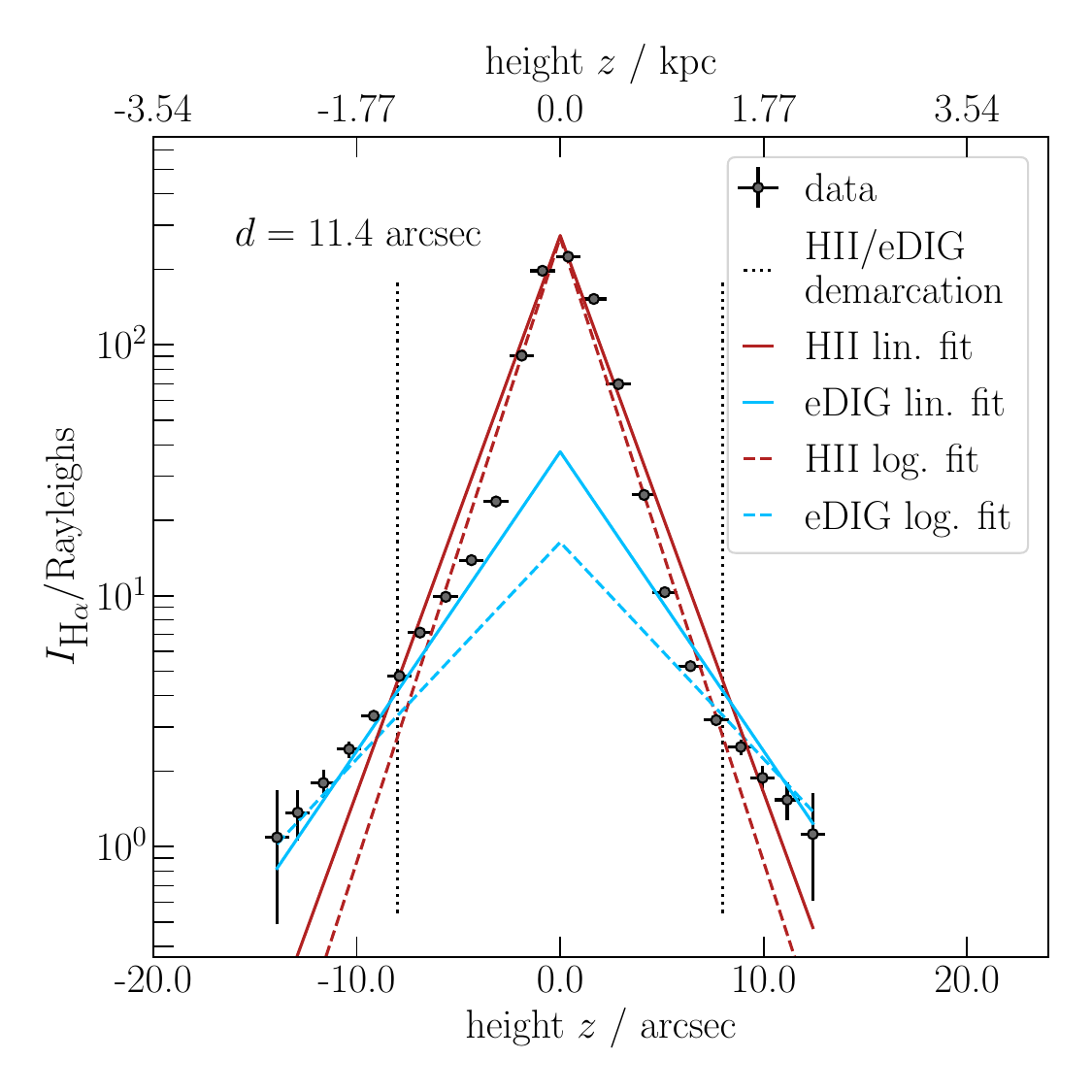}
    \end{subfigure}
        \begin{subfigure}{.24\linewidth}
        \includegraphics[width=\hsize]{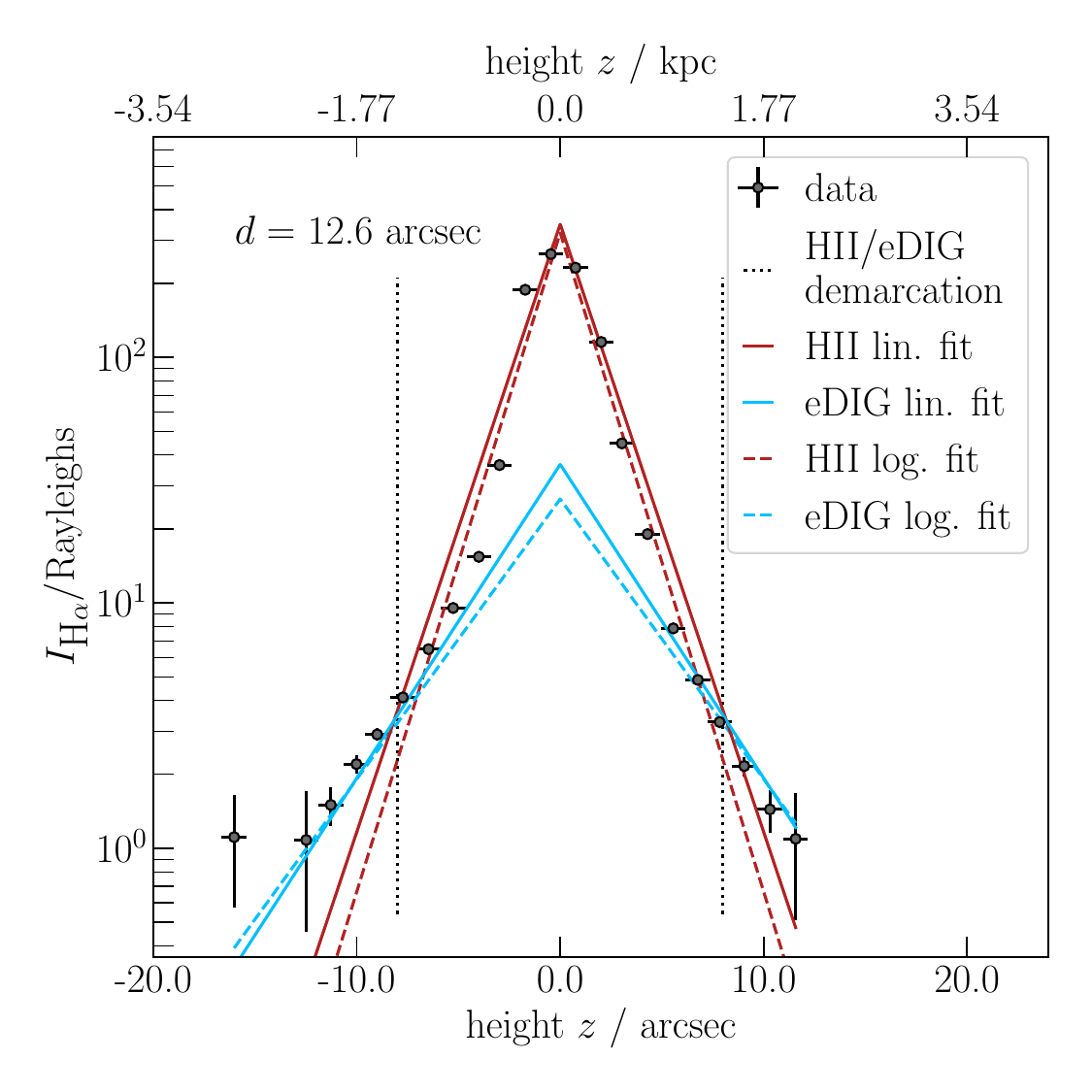}
    \end{subfigure}\\
        \begin{subfigure}{.24\linewidth}
        \includegraphics[width=\hsize]{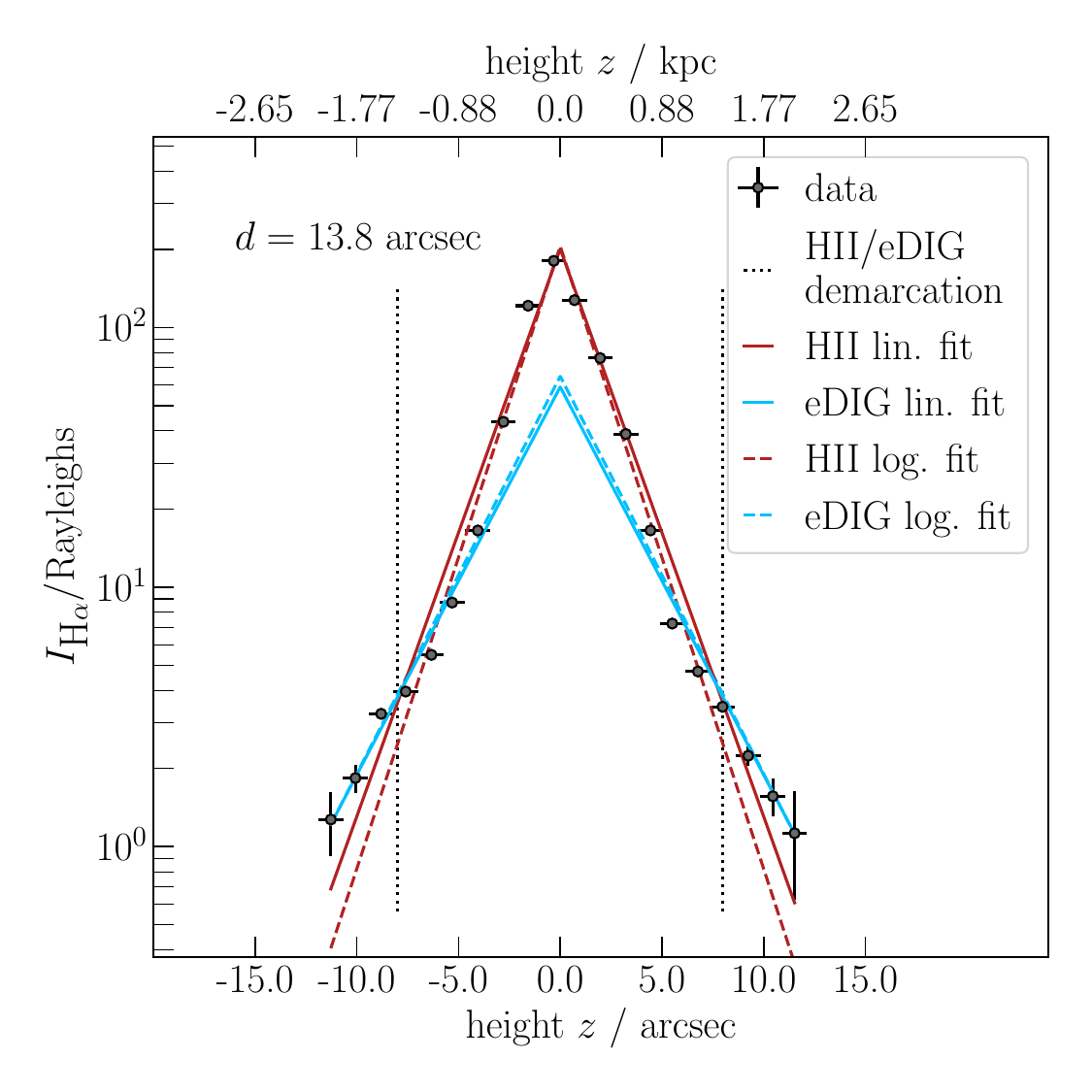}
    \end{subfigure}
        \begin{subfigure}{.24\linewidth}
        \includegraphics[width=\hsize]{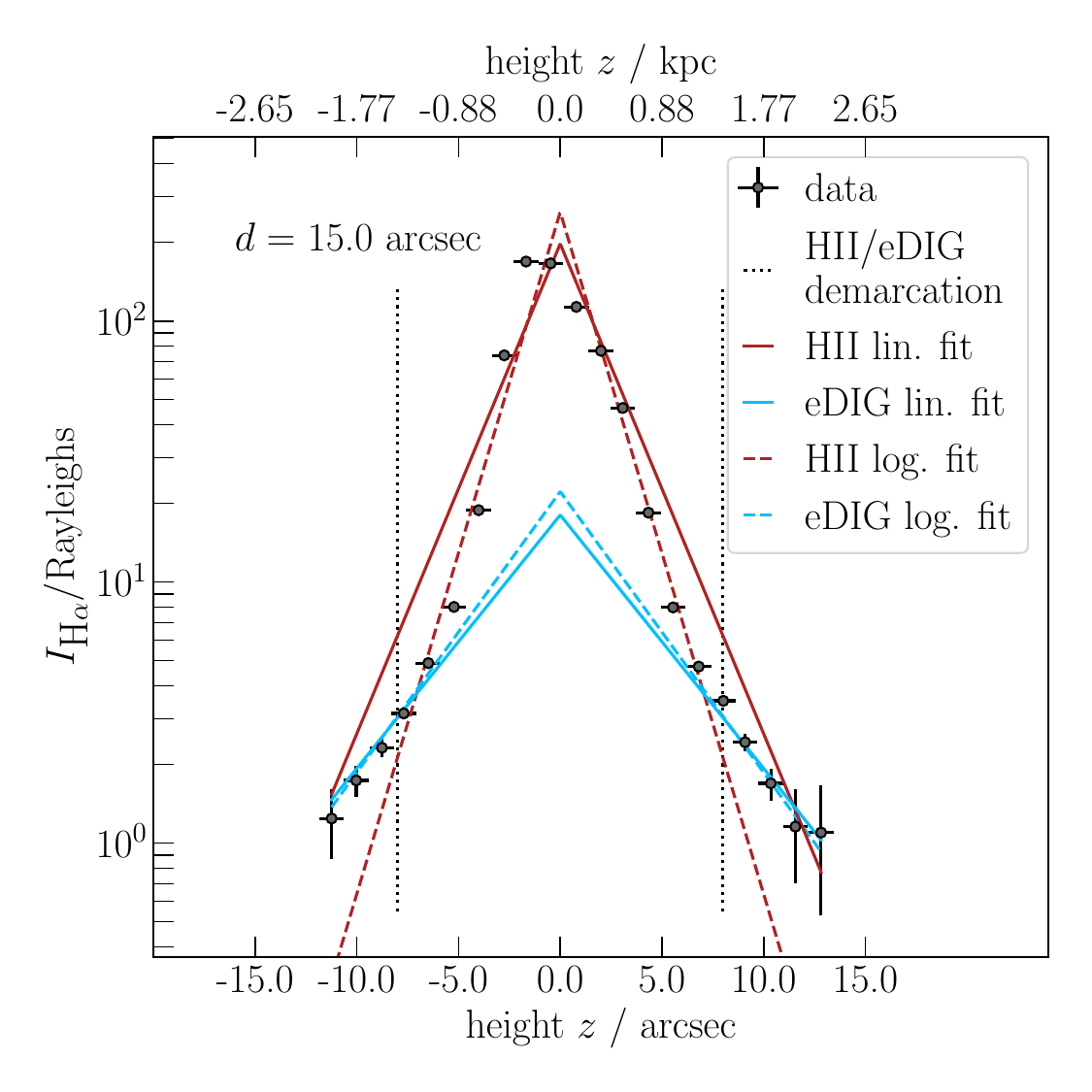}
    \end{subfigure}
        \begin{subfigure}{.24\linewidth}
        \includegraphics[width=\hsize]{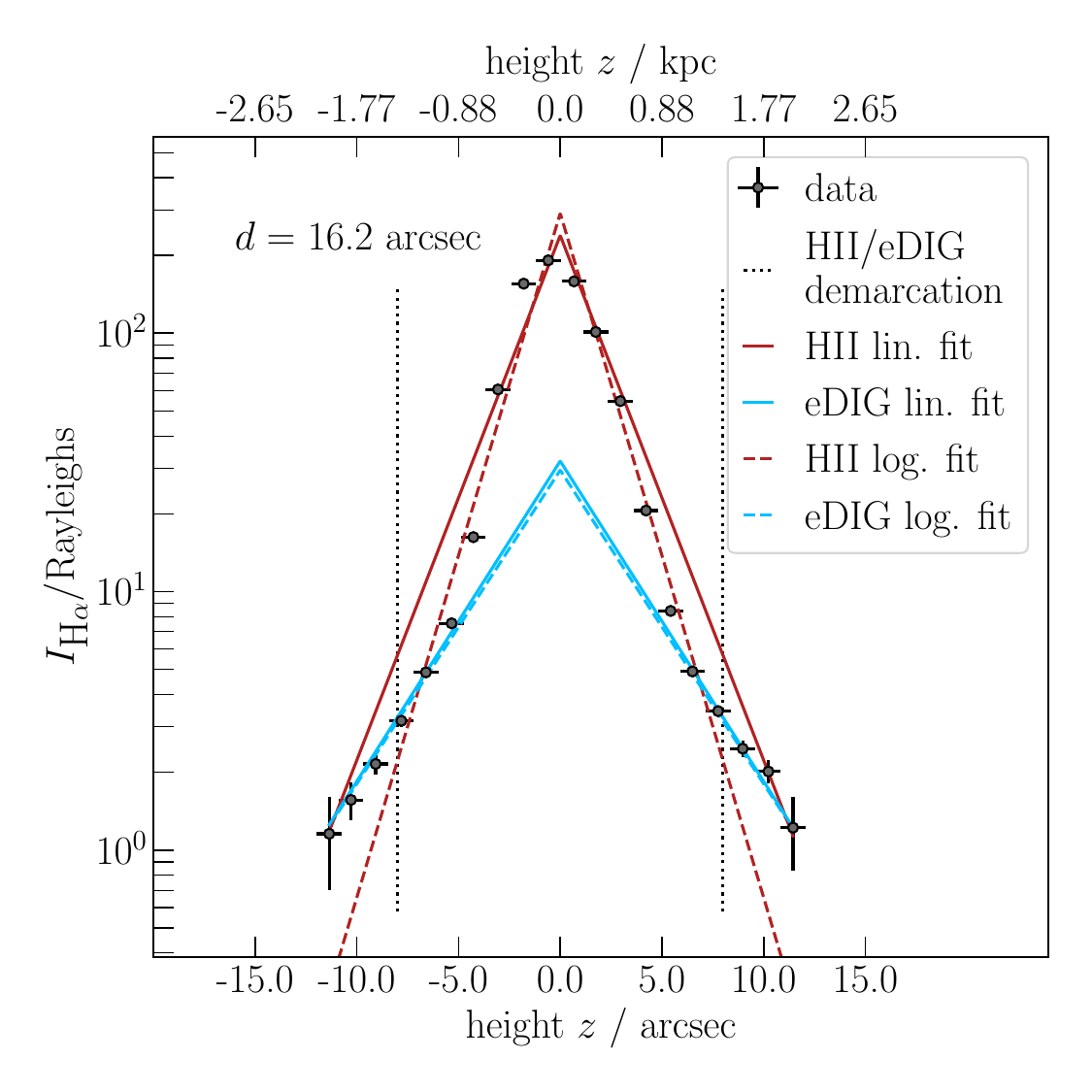}
    \end{subfigure}
        \begin{subfigure}{.24\linewidth}
        \includegraphics[width=\hsize]{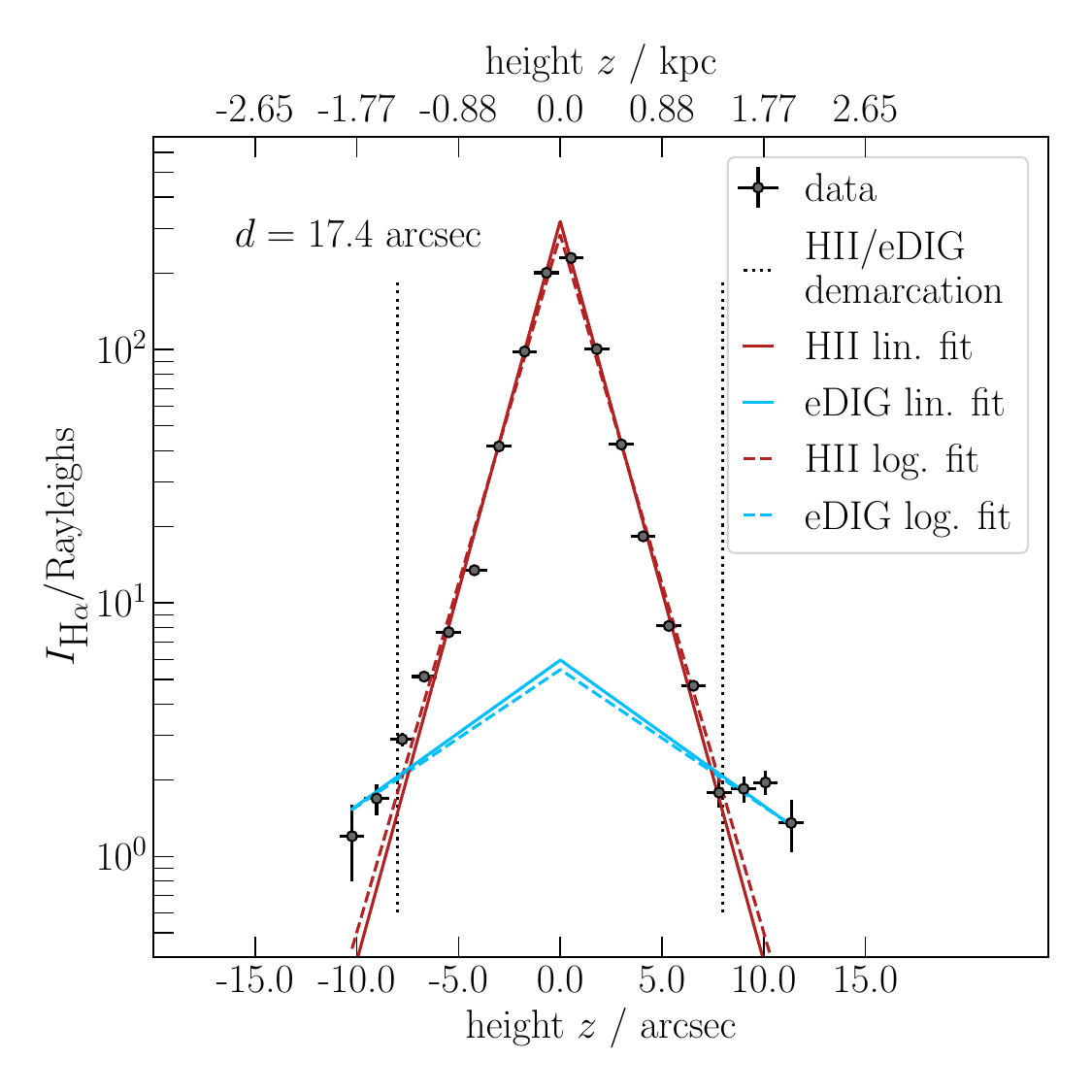}
    \end{subfigure}
    
    \caption*{Fig. \ref{fig:shfitting_full}. Continued.}
\end{figure*}

\end{appendix}

\end{document}